\documentclass[ip,revtex4]{emulateapj}
\usepackage{amsmath}
\usepackage{ifthen}
\newcommand{\forloop}[5][1]%
{%
\setcounter{#2}{#3}%
\ifthenelse{#4}%
	{%
	#5%
	\addtocounter{#2}{#1}%
	\forloop[#1]{#2}{\value{#2}}{#4}{#5}%
	}%
	{%
	}%
}% 

\newcommand{\ctbd}[1]{}

\newcommand{\band}[1]{\ensuremath{#1}~band}

\newcommand{\kms}{\ensuremath{\rm km\,s^{-1}}}

\newcommand{\gcmc}{\ensuremath{\rm g\,cm^{-3}}}

\newcommand{\logg}{\ensuremath{\log{g}}}

\newcommand{\rsun}{\ensuremath{R_\sun}}
\newcommand{\msun}{\ensuremath{M_\sun}}

\newcommand{\rsunnom}{\hbox{\ensuremath{\mathcal{R}^{\rm N}_\odot}}}
\newcommand{\msunnom}{\hbox{\ensuremath{\mathcal{M}^{\rm N}_\odot}}}
\newcommand{\lsunnom}{\hbox{\ensuremath{\mathcal{L}^{\rm N}_\odot}}}
\newcommand{\loglsunnom}{\hbox{\ensuremath{\log\mathcal{L}^{\rm N}_\odot}}}

\newcommand{\reffigl}[1]{Figure~\ref{fig:#1}}
\newcommand{\refsecl}[1]{\mbox{Section \ref{sec:#1}}}

\newcommand{\reftabl}[1]{Table~\ref{tab:#1}}

\newcommand{\sysname}{HAT-TR-318-007}

\newcommand{\tmassid}{2MASS~J08503296+1208239}
\newcommand{\fbcid}{Cl\*~NGC~2682~FBC~6558}
\newcommand{\gscid}{GSC2.3~N8X5006832}
\newcommand{\sdssid}{SDSS~J085032.94+120822.8}
\newcommand{\kicid}{KIC~6651}
\newcommand{\wgaid}{1WGA~J0850.5+1208}

\newcommand{\sysra}{\ensuremath{08^{\mathrm h}50^{\mathrm m}32.9578{\mathrm s}}}
\newcommand{\sysdec}{\ensuremath{+12{\arcdeg}08{\arcmin}23.644{\arcsec}}}

\newcommand{\ppmra}{\ensuremath{-21.0\pm2.4}}
\newcommand{\ppmdec}{\ensuremath{-131.6\pm2.5}}

\newcommand{\velU}{\ensuremath{-1.9\pm1.2}}
\newcommand{\velV}{\ensuremath{-81.0\pm2.4}}
\newcommand{\velW}{\ensuremath{-20.9\pm1.7}}

\newcommand{\distance}{\ensuremath{123.7\pm3.8}}
\newcommand{\parallax}{\ensuremath{8.09\pm0.25}}

\newcommand{\tmassmagK}{\ensuremath{11.131\pm0.020}}

\newcommand{\periodshort}{\ensuremath{3.344}}

\newcommand{\jdend}{\ensuremath{2457214.941784\pm0.000027}}
\newcommand{\period}{\ensuremath{3.34395390\pm0.00000020}}
\newcommand{\RVKA}{\ensuremath{48.22\pm0.17}}
\newcommand{\RVKB}{\ensuremath{79.41^{+0.89}_{-0.90}}}
\newcommand{\RVgammaB}{\ensuremath{140.489^{+0.031}_{-0.030}}}
\newcommand{\RVgammaBcorr}{\ensuremath{30.07\pm0.18}}
\newcommand{\massstarA}{\ensuremath{0.448\pm0.011}}
\newcommand{\massstarB}{\ensuremath{0.2721^{+0.0041}_{-0.0042}}}
\newcommand{\hRVeccen}{\ensuremath{-0.117\pm0.011}}
\newcommand{\kRVeccen}{\ensuremath{-0.00041^{+0.00015}_{-0.00016}}}
\newcommand{\RVeccen}{\ensuremath{0.0136\pm0.0026}}
\newcommand{\RVomega}{\ensuremath{-90.202^{+0.086}_{-0.093}}}
\newcommand{\bimpact}{\ensuremath{0.0850^{+0.0090}_{-0.0096}}}

\newcommand{\inclination}{\ensuremath{89.566^{+0.050}_{-0.047}}}
\newcommand{\rBoverrA}{\ensuremath{0.6405^{+0.0036}_{-0.0034}}}
\newcommand{\rAplusrBovera}{\ensuremath{0.08844\pm0.00022}}
\newcommand{\radiusstarA}{\ensuremath{0.4548^{+0.0035}_{-0.0036}}}
\newcommand{\radiusstarB}{\ensuremath{0.2913^{+0.0023}_{-0.0024}}}
\newcommand{\loggstarA}{\ensuremath{4.7740^{+0.0056}_{-0.0058}}}
\newcommand{\loggstarB}{\ensuremath{4.9442^{+0.0038}_{-0.0039}}}
\newcommand{\densitystarA}{\ensuremath{6.720^{+0.074}_{-0.072}}}
\newcommand{\densitystarB}{\ensuremath{15.53\pm0.22}}
\newcommand{\semimajoraxis}{\ensuremath{0.03923\pm0.00028}}
\newcommand{\jBoverjArfilter}{\ensuremath{0.590^{+0.057}_{-0.058}}}
\newcommand{\deltamagrfilter}{\ensuremath{1.54^{+0.11}_{-0.10}}}
\newcommand{\jBoverjAifilter}{\ensuremath{0.637^{+0.030}_{-0.026}}}
\newcommand{\deltamagifilter}{\ensuremath{1.1593^{+0.0089}_{-0.0085}}}
\newcommand{\jBoverjApfilter}{\ensuremath{0.624\pm0.022}}
\newcommand{\deltamagpfilter}{\ensuremath{1.2600\pm0.0028}}
\newcommand{\starAldcoeffArfilter}{\ensuremath{0.7834}}
\newcommand{\starAldcoeffAifilter}{\ensuremath{0.839\pm0.035}}
\newcommand{\starAldcoeffApfilter}{\ensuremath{0.795^{+0.032}_{-0.031}}}
\newcommand{\starBldcoeffArfilter}{\ensuremath{0.7834}}
\newcommand{\starBldcoeffAifilter}{\ensuremath{0.16^{+0.13}_{-0.12}}}
\newcommand{\starBldcoeffApfilter}{\ensuremath{0.303^{+0.095}_{-0.096}}}

\newcommand{\starAootvperiod}{\ensuremath{3.41315^{+0.00030}_{-0.00032}}}

\newcommand{\starBootvperiod}{\ensuremath{3.28498^{+0.00035}_{-0.00034}}}

\newcommand{\starAldcoeffAadjsqrtifilter}{\ensuremath{0.771^{+0.027}_{-0.028}}}
\newcommand{\starAldcoeffBadjsqrtifilter}{\ensuremath{0.24^{+1.19}_{-0.66}}}
\newcommand{\starBldcoeffAadjsqrtifilter}{\ensuremath{0.198^{+0.076}_{-0.108}}}
\newcommand{\starBldcoeffBadjsqrtifilter}{\ensuremath{-2.2^{+1.3}_{-1.1}}}
\newcommand{\starAldcoeffAadjsqrtpfilter}{\ensuremath{0.731^{+0.024}_{-0.025}}}
\newcommand{\starAldcoeffBadjsqrtpfilter}{\ensuremath{-0.29^{+0.68}_{-0.50}}}
\newcommand{\starBldcoeffAadjsqrtpfilter}{\ensuremath{0.407^{+0.057}_{-0.069}}}
\newcommand{\starBldcoeffBadjsqrtpfilter}{\ensuremath{1.66^{+0.77}_{-1.22}}}

\newcommand{\spectypestarA}{\ensuremath{{\rm M}3.7\pm0.7}}
\newcommand{\spectypestarAtotal}{\ensuremath{{\rm M}3.6\pm0.7}}
\newcommand{\spectypestarB}{\ensuremath{{\rm M}5.0\pm0.7}}

\newcommand{\teffstarbtsettlA}{\ensuremath{3190\pm110}}

\newcommand{\teffstarbtsettlB}{\ensuremath{3100\pm110}}

\newcommand{\fehstarbtsettlA}{\ensuremath{+0.25\pm0.13}}

\newcommand{\fehstarbtsettlB}{\ensuremath{+0.09\pm0.15}}
\newcommand{\fehsystembtsettl}{\ensuremath{+0.18\pm0.10}}

\newcommand{\teffstarbesselA}{\ensuremath{3200\pm120}}

\newcommand{\teffstarbesselB}{\ensuremath{3000\pm130}}

\newcommand{\teffstarA}{\teffstarbtsettlA}

\newcommand{\teffstarB}{\teffstarbtsettlB}

\newcommand{\fehstarA}{\ensuremath{+0.40\pm0.11}}
\newcommand{\fehstarB}{\ensuremath{+0.20\pm0.11}}
\newcommand{\fehsystem}{\ensuremath{+0.298\pm0.080}}

\newcommand{\fehterrienKstarAfullerr}{\ensuremath{0.49\pm0.14}}
\newcommand{\fehterrienHstarAfullerr}{\ensuremath{0.23\pm0.19}}
\newcommand{\fehterrienKstarBfullerr}{\ensuremath{0.07\pm0.14}}
\newcommand{\fehterrienHstarBfullerr}{\ensuremath{0.43\pm0.19}}
\newcommand{\fehterrienKstarAtotalfullerr}{\ensuremath{0.49\pm0.14}}
\newcommand{\fehterrienHstarAtotalfullerr}{\ensuremath{0.39\pm0.16}}

\newcommand{\fehrojasstarAfullerr}{\ensuremath{0.55\pm0.17}}
\newcommand{\mhrojasstarAfullerr}{\ensuremath{0.39\pm0.14}}
\newcommand{\spectyperojasstarAfullerr}{\ensuremath{3.71\pm0.69}}
\newcommand{\fehrojasstarBfullerr}{\ensuremath{0.08\pm0.21}}
\newcommand{\mhrojasstarBfullerr}{\ensuremath{0.06\pm0.15}}
\newcommand{\spectyperojasstarBfullerr}{\ensuremath{5.01\pm0.73}}
\newcommand{\fehrojasstarAtotalfullerr}{\ensuremath{0.63\pm0.16}}
\newcommand{\mhrojasstarAtotalfullerr}{\ensuremath{0.45\pm0.13}}

\newcommand{\fehterrienKsyserr}{\ensuremath{0.12}}
\newcommand{\fehterrienHsyserr}{\ensuremath{0.12}}

\newcommand{\fehrojassyserr}{\ensuremath{0.14}}
\newcommand{\mhrojassyserr}{\ensuremath{0.10}}

\newcommand{\luminositystarA}{\ensuremath{(1.93\pm0.27)\times10^{-2}}}
\newcommand{\luminositystarB}{\ensuremath{(7.1\pm1.0)\times10^{-3}}}

\newcommand{\logluminositystarA}{\ensuremath{-1.715\pm0.060}}
\newcommand{\logluminositystarB}{\ensuremath{-2.151\pm0.062}}

\newcommand{\bolmagA}{\ensuremath{9.03\pm0.15}}
\newcommand{\bolmagB}{\ensuremath{10.12\pm0.16}}

\newcommand{\teffsunval}{5772}
\newcommand{\mbolsunval}{4.740}

\newcommand{\periodlttcorr}{\ensuremath{3.34361825\pm0.00000020}}

\newboolean{emulateapj}
\setboolean{emulateapj}{true}

\newboolean{astroph}
\setboolean{astroph}{true}
\shortauthors{Hartman et al.}
\shorttitle{\sysname}
\ifthenelse{\boolean{emulateapj}}{
    \newcommand{\titledag}{$\dagger$}
}{
    \newcommand{\titledag}{\dagger}
}

\begin{document}

%% Titlepage
\title{\sysname{}: a double-lined M-dwarf binary with total
  secondary eclipses discovered by HATNet and observed by K2\altaffilmark{$\star$},\altaffilmark{\titledag}}

%% Authors
\author{
        J.~D.~Hartman\altaffilmark{1},
        S.~N.~Quinn\altaffilmark{2},
        G.~\'A.~Bakos\altaffilmark{1,+},
        G.~Torres\altaffilmark{2},
        G.~Kov\'acs\altaffilmark{3},
        D.~W.~Latham\altaffilmark{2},
        R.~W.~Noyes\altaffilmark{2},
        A.~Shporer\altaffilmark{4},
        B.~J.~Fulton\altaffilmark{5},
        G.~A.~Esquerdo\altaffilmark{2},
        M.~E.~Everett\altaffilmark{6},
        K.~Penev\altaffilmark{7},
        W.~Bhatti\altaffilmark{1},
        Z.~Csubry\altaffilmark{1}
}
\altaffiltext{1}{Princeton University, Department of Astrophysical Sciences,
	Princeton, NJ; email: jhartman@astro.princeton.edu}

\altaffiltext{2}{Harvard-Smithsonian Center for Astrophysics, Cambridge, MA 02138, USA}

\altaffiltext{3}{Konkoly Observatory of the Hungarian Academy of Sciences, Budapest, Hungary}

\altaffiltext{4}{Jet Propulsion Laboratory, California Insitute of Technology, 4800 Oak Grove Drive, Pasadena, CA 91109, USA}

\altaffiltext{5}{Institute for Astronomy, University of Hawaii,
	Honolulu, HI 96822; NSF Postdoctoral Fellow}

\altaffiltext{6}{National Optical Astronomy Observatory, Tucson, AZ, USA}

\altaffiltext{7}{Department of Physics, The University of Texas at Dallas, Richardson, TX, USA}

\altaffiltext{+}{Packard Fellow}

\altaffiltext{$\star$}{
	Based in part on observations made with the Nordic Optical
        Telescope, operated on the island of La Palma jointly by
        Denmark, Finland, Iceland, Norway, and Sweden, in the Spanish
        Observatorio del Roque de los Muchachos of the Instituto de
        Astrofisica de Canarias.
}
\altaffiltext{$\dagger$}{
	This paper includes data gathered with the 6.5 meter Magellan
        Telescopes located at Las Campanas Observatory, Chile.
}

%% EOF authors

% #####################################################################
%% abstract
\begin{abstract}
%++++++++++++++++++++++++++++++++++++++++++++++++++++++++++++++++++++++
%++++++++++++++++++++++++++++++++++++++++++++++++++++++++++++++++++++++

We report the discovery by the HATNet survey of \sysname{}, a $P =
\period{}$\,d period detached double-lined M-dwarf binary with total
secondary eclipses. We combine radial velocity (RV) measurements from
TRES/FLWO~1.5\,m, and time-series photometry from HATNet, FLWO~1.2\,m,
BOS~0.8\,m and NASA {\em K2} Campaign 5, to determine the masses and
radii of the component stars: $M_{A} = \massstarA{}$\,\msunnom, $M_{B}
= \massstarB{}$\,\msunnom, $R_{A} = \radiusstarA{}$\,\rsunnom, and
$R_{B} = \radiusstarB{}$\,\rsunnom. We obtained a FIRE/Magellan
near-infrared spectrum of the primary star during a total secondary
eclipse, and use this to obtain disentangled spectra of both
components.  We determine spectral types of ${\rm ST}_{A} = {\rm
  M}\spectyperojasstarAfullerr$ and ${\rm ST}_{B} = {\rm
  M}\spectyperojasstarBfullerr$, and effective temperatures of T$_{\rm
  eff,A} = \teffstarA$\,K and T$_{\rm eff,B} = \teffstarB$\,K, for the
primary and secondary star, respectively. We also measure a
metallicity of [Fe/H]$=\fehsystem$ for the system.  We find that the
system has a small, but significant, non-zero eccentricity of
$\RVeccen$. The {\em K2} light curve shows a coherent variation at a
period of $\starAootvperiod$\,d, which is slightly longer than the
orbital period, and which we demonstrate comes from the primary
star. We interpret this as the rotation period of the primary.  We
perform a quantitative comparison between the Dartmouth stellar
evolution models and the seven systems, including \sysname{}, that
contain M dwarfs with $0.2\,\msunnom < M < 0.5\,\msunnom$, have
metallicity measurements, and have masses and radii determined to
better than 5\% precision. Discrepancies between the predicted and
observed masses and radii are found for three of the systems.
\end{abstract}

% #####################################################################
%% keywords

% ####################################################################
%% space velocity
\section{Introduction}

\setcounter{footnote}{0}

Detached double-lined eclipsing binary systems are fundamental to our
understanding of stellar evolution. By applying simple geometry and
orbital mechanics it is possible to measure the masses and radii of
the component stars in such a system. Assuming both stars are the same
age, one can then test a theoretical model by checking whether there
is an age at which the model would predict that two stars with the
measured masses would have the measured radii. These models also
predict the temperatures (or luminosities) of the stars, and depend on
the composition of the stars. A sharper test of the model can be
performed if the temperatures and metallicities (a.k.a.\ atmospheric
parameters) of the component stars can be measured
spectroscopically. 

The double-lined nature of the spectrum, which enables the measurement
of the masses of both components of a binary, complicates the
measurement of the stellar atmospheric parameters. Although algorithms
exist to disentangle the composite spectra into spectra of the
individual component stars \citep{simon:1994,hadrava:1995},
determining the continuum level of each component is difficult,
leading to systematic uncertainties in the depths of the absorption
lines, and hence in the atmospheric parameters. In rare cases the
primary or secondary eclipse may be total in which case one can obtain
an uncontaminated spectrum of one component star during total eclipse,
providing an opportunity to cleanly measure the atmospheric parameters
of the star. This spectrum also serves as an ideal template for
disentangling the out-of-eclipse composite spectra, allowing one to
obtain a spectrum for the totally eclipsed star as well. Such a
technique was applied by \citet{terrien:2012:CMDra} to the late
M-dwarf eclipsing binary system CM~Dra. Although this object does not
exhibit total eclipses, the eclipses are close enough to totality that
it is possible to obtain an effectively uncontaminated spectrum of a
single component during eclipse. \citet{terrien:2012:CMDra} measured a
subsolar metallicity of [Fe/H]$= -0.30 \pm 0.12$ for CM~Dra based on
these observations. \citet{feiden:2014}, however, conclude that CM~Dra
may have a 0.2\,dex enhancement in $\alpha$-elements compared to the
Sun, in which case its [Fe/H] metallicity is close to solar.

Another advantage of binaries with total eclipses is that the radii of
the component stars can often be measured with greater precision than
can be done for grazing systems due to the additional information
provided by the times of second and third contact (end of ingress and
start of egress, respectively). The contact points can be measured
quite precisely due to the sharp change in light curve morphology at
these times. Moreover, the determination of the times of contact is
relatively insensitive to the presence of spots on one or both of the
stars, which may be a significant source of systematic errors in
determining the radii of M dwarf stars in grazing eclipsing binaries
\citep{morales:2010,windmiller:2010}.

Over the past decade there has been a significant observational effort
to determine the fundamental parameters of M dwarf stars, which, due
to their faintness, have not been well-studied until recently. The
first precise determinations of the fundamental parameters for early
and mid M dwarfs in the eclipsing systems YY~Gem and CU~Cnc indicated
that the components of these systems have radii that are 10--20\%
larger than predicted by theoretical models
\citep{torres:2002,ribas:2003}, whereas the late M dwarfs in CM~Dra
appeared to be in agreement with theoretical models
\citep{metcalfe:1996}. Subsequent M dwarf binary studies appear to
have corroborated this effect for mid and early M dwarfs while finding
that the late M dwarfs may also be inflated at the few percent level
\citep[e.g.,][and references
  therein]{torres:2013,zhou:2013,dittmann:2017}. These stars also
appear to have a temperature discrepancy, whereby the measured
effective temperatures are lower than predicted by the models. While
it has been speculated that these discrepencies may be due to strong
magnetic fields inhibiting convection in the atmospheres of these
stars \citep[e.g.][]{chabrier:2007,macdonald:2012}, no clear
observational distinction in radii has been found between stars
expected to have strong magnetic activity and stars expected to be
less active\footnote{One caveat is that the magnetic activity is not
  measured directly, but rather is assumed to correlate with the
  rotation period of the star. The latter is measured in some cases,
  or else assumed to be tidally synchronized to the orbital period for
  short period systems.}  \citep[e.g.][]{irwin:2011,doyle:2011}. Other
suggestions include the possibility that the stars have a range of
compositions \citep[e.g.,][who consider the case of brown dwarfs and
  very low mass stars]{burrows:2011}, that a proper equation of state
is not used in generating the models \citep[e.g.,][]{dotter:2008}, or
that there are systematic errors in the measurements due, for example,
to spots on the stars \citep{morales:2010,windmiller:2010}.

In this paper we report the discovery by the HATNet survey
\citep{bakos:2004:hatnet} of a detached, double-lined M-dwarf binary
with total eclipses. As an M-dwarf system this object is useful for
testing theoretical models of low-mass stars. Moreover, we take
advantage of the total eclipses to obtain an uncontaminated
near-infrared spectrum of the primary star, and a disentangled
spectrum of the secondary star, which we use to measure their
atmospheric parameters. The eclipse totality also allows us to obtain
accurate mass and radius measurements for the system, with systematic
errors due to unaccounted-for starspots that are lower than if the
system were grazing. Finally, because the system was observed by the
NASA {\em K2} mission, we are able to determine precise photometric
parameters for the binary system, including a direct measurement of
the rotation period for the primary star, which we find to be close
to, but clearly not synchronized with, the orbital period of the
system.

\section{Observations and Reductions}

% ####################################################################
\subsection{Photometric Detection}
\label{sec:detection}

%% ----------------
\begin{figure*}[!ht]
\plotone{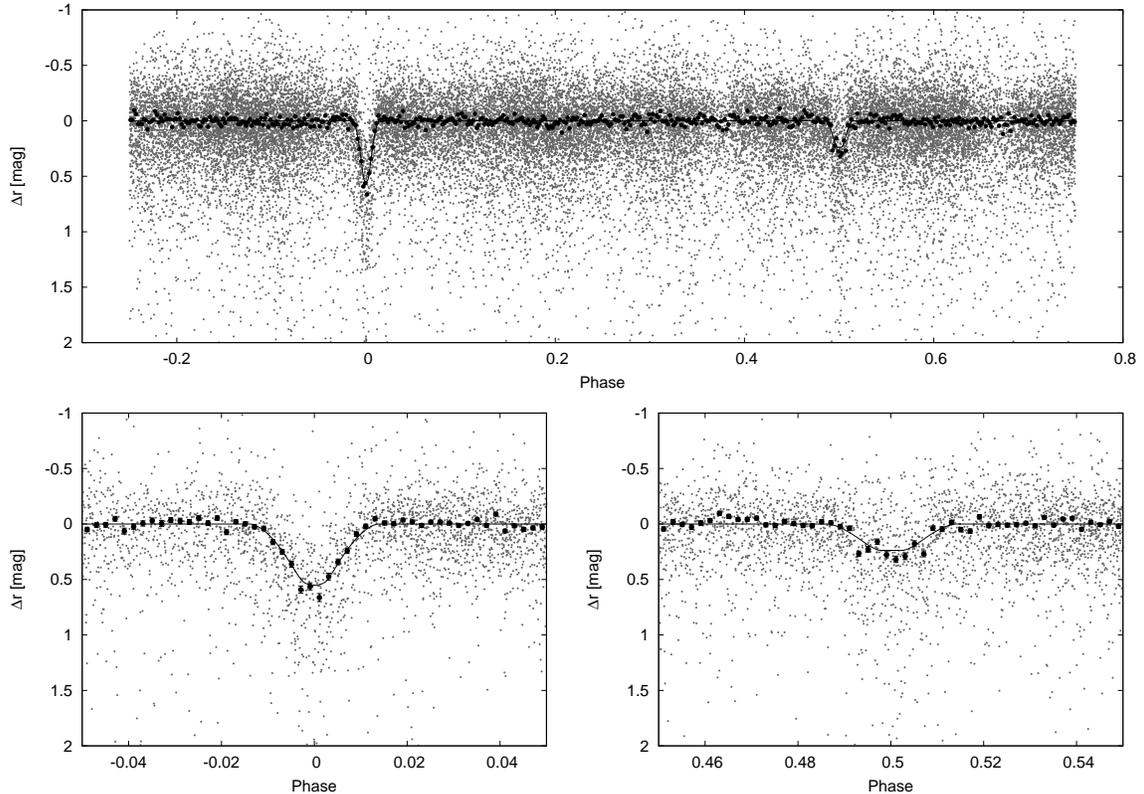}
\caption{ Phase-folded HATNet light curve for \sysname{} after
  correcting for the artificial reduction in the eclipse depth due to
  an incorrect reference flux and application of EPD and TFA in
  signal-search mode. The light filled points are the individual
  measurements after application of EPD and TFA, and with the
  fitted-for dilution removed.  The solid line shows the best-fit
  model. The dark filled circles show the light curve binned in phase,
  with a binsize of 0.002. The upper panel shows the full phased light
  curve, the bottom left and bottom right panels show the light curve
  zoomed-in on the primary and secondary eclipses, respectively. We
  show the depth-corrected light curves so that all of the HATNet data
  can be presented together on the same scale. The fitting itself uses
  the original EPD- and TFA-corrected light curves, with the
  additional depth correction determined as part of the fit. Note that
  the procedure of depth-correcting the light curves for display in
  this figure results in assymetric magnitude uncertainties which
  distorts slightly the binned light curve, and results in a slight
  discrepancy between the model and binned values in the figure.
\label{fig:lcHN}}
\end{figure*}
%% ----------------

\sysname{}\footnote{Following the convention established in
  \citet{beatty:2007} we adopt the HAT transit candidate
  identification as the name for this object. Here 318 indicates that
  the object falls in HAT field 318, while 007 indicates that this was
  the seventh transiting planet candidate identified in this field.}
(also known as \tmassid{}, \fbcid{}, \gscid{}, \sdssid{}, \kicid{},
\wgaid{}; $\alpha = \sysra{}$, $\delta = \sysdec{}$; J2000;
$K=\tmassmagK$\,mag) was initially detected as a candidate transiting
planet system by the HATNet survey. The available HATNet observations
of this system are summarized in \reftabl{photobs}, while the combined
phase-folded light curve is shown in Figure~\ref{fig:lcHN}. The data
are provided in \reftabl{photdata}. 

The HATNet images were processed
and reduced to trend-filtered light curves following the procedure
described by \cite{bakos:2010:hat11}. Candidate transits were
identified using the Box-fitting Least-Squares \citep[BLS;][]{kovacs:2002:BLS}
procedure. Typically eclipsing binary systems which show unequal
primary and secondary eclipse depths are automatically rejected,
however due to the faintness of \sysname{}, the difference in eclipse
depths for this star could not be clearly detected in the HATNet
observations. Moreover, an inaccurate reference flux estimate for this
star in our image subtraction procedure caused us to grossly
underestimate the eclipse depths in some of our light
curves. Subsequent photometric follow-up observations discussed in
\refsecl{photfu} showed clearly that \sysname{} is not a transiting
planet system, but also revealed the total secondary eclipses, which
motivated us to continue studying this object.

Coincidentally \sysname{} has a projected separation on the sky of
23$\arcmin$ from the center of the open cluster M67. Although this is
within the projected tidal radius of the cluster, the apparent
magnitude of \sysname{} indicates that it is clearly a foreground
object, and not a cluster member. Nonetheless, its location on the sky
has resulted in this star having calibrated photometry from several
surveys of this very well studied open
cluster. Table~\ref{tab:catalogphotometry} lists the available photometric
measurements of the system from the literature.

\ifthenelse{\boolean{emulateapj}}{
  \begin{deluxetable*}{lllrrrr}
}{
  \begin{deluxetable}{lllrrrr}
}
\tablewidth{0pc}
\tabletypesize{\scriptsize}
\tablecaption{
  Summary of Time-Series Photometry for \sysname{}
  \label{tab:photobs}
}
\tablehead{
  \multicolumn{1}{c}{Facility} &
  \multicolumn{1}{c}{Date(s)} &
  \multicolumn{1}{c}{Event(s)} &
  \multicolumn{1}{c}{\# obs.} &
  \multicolumn{1}{c}{Filter} &
  \multicolumn{1}{c}{Median Cadence} &
  \multicolumn{1}{c}{RMS precision} \\
  &
  &
  &
  &
  &
  \multicolumn{1}{c}{(s)} &
  \multicolumn{1}{c}{(mag)}
}
\startdata
HAT-5/G317 & 2010 Nov.--2011 Apr. & Both & 4167 & $r$ & 236 & 0.166 \\
HAT-8/G317 & 2010 Nov.--2011 Apr. & Both & 4289 & $r$ & 239 & 0.166 \\
HAT-6/G318 & 2008 Dec.--2009 May  & Both & 2972 & $r$ & 352 & 0.149 \\
HAT-7/G365 & 2010 Nov.--2011 May  & Both & 8115 & $r$ & 235 & 0.348 \\
HAT-8/G365 & 2011 Apr.--2011 May  & Both &  452 & $r$ & 232 & 0.348 \\
HAT-6/G366 & 2010 Nov.--2011 Apr. & Both & 4334 & $r$ & 236 & 0.142 \\
HAT-9/G366 & 2010 Nov.--2011 Apr. & Both & 6011 & $r$ & 230 & 0.142 \\
FLWO~1.2\,m & 2011 Mar. 30        & Secondary & 123 & $i$ & 133 & 0.005 \\
FLWO~1.2\,m & 2011 Dec. 29        & Secondary & 327 & $i$ &  59 & 0.006 \\
FLWO~1.2\,m & 2012 Jan. 03        & Primary   & 464 & $i$ &  44 & 0.008 \\
BOS~0.8\,m  & 2012 Jan. 03        & Primary   & 111 & $i$ & 140 & 0.013 \\
K2~C5 \tablenotemark{1}     & 2015 Apr--Jul       & Both & 3298 & $Kp$ & 1765 & 0.005 \\
\enddata
\tablenotetext{1}{For the K2 Campaign 5 data we list for the precision the point-to-point r.m.s.\ of the PDC light curve after removing the eclipse events and using a Fourier series to fit and remove the dominant rotational variability signal. The r.m.s.\ of the residuals is dominated by systematic variations on a time-scale of several hours, likely resulting from inaccuracies in removing the instrumental variations due to the spacecraft roll for a such large amplitude variable star.}
\ifthenelse{\boolean{emulateapj}}{
  \end{deluxetable*}
}{
  \end{deluxetable}
}

\ifthenelse{\boolean{emulateapj}}{
  \begin{deluxetable*}{lllrrrrrrrr}
}{
  \begin{deluxetable}{lllrrrrrrrr}
}
\tablewidth{0pc}
\tabletypesize{\scriptsize}
\tablecaption{
  Time-Series Photometry Data for \sysname{}
  \label{tab:photdata}
}
\tablehead{
  \multicolumn{1}{c}{Object\tablenotemark{1}} &
  \multicolumn{1}{c}{Facility\tablenotemark{2}} &
  \multicolumn{1}{c}{Filter} &
  \multicolumn{1}{c}{BJD$-2454000$\tablenotemark{3}} &
  \multicolumn{1}{c}{Raw Mag.\tablenotemark{4}} &
  \multicolumn{1}{c}{Err.~Mag.} &
  \multicolumn{1}{c}{Corr.~Mag.\tablenotemark{5}} &
  \multicolumn{1}{c}{$T-T_{c}$\tablenotemark{6}} &
  \multicolumn{1}{c}{$S$\tablenotemark{7}} &
  \multicolumn{1}{c}{$D$\tablenotemark{7}} &
  \multicolumn{1}{c}{$K$\tablenotemark{7}} \\
  &
  &
  &
  &
  &
  &
  &
  \multicolumn{1}{c}{[d]} &
  &
  &
}
\startdata
HTR318-007     & bos      & i & 1930.780201 & 14.3763 & 0.0146 & 14.3523 &  -0.08311 &   0.26320 &  -0.00660 &   0.00930 \\
HTR318-007     & bos      & i & 1930.781811 & 14.3679 & 0.0128 & 14.3544 &  -0.08150 &   0.29080 &   0.00310 &   0.00480 \\
HTR318-007     & bos      & i & 1930.783431 & 14.3831 & 0.0133 & 14.3657 &  -0.07988 &   0.27920 &  -0.00070 &   0.01190 \\
HTR318-007     & bos      & i & 1930.785041 & 14.3746 & 0.0135 & 14.3619 &  -0.07827 &   0.33410 &   0.00050 &   0.01330 \\
HTR318-007     & bos      & i & 1930.786671 & 14.3396 & 0.0147 & 14.3339 &  -0.07664 &   0.33640 &   0.00420 &   0.01530 \\
HTR318-007     & bos      & i & 1930.788321 & 14.3473 & 0.0132 & 14.3419 &  -0.07499 &   0.33640 &  -0.00140 &  -0.00080 \\
HTR318-007     & bos      & i & 1930.789941 & 14.3662 & 0.0144 & 14.3549 &  -0.07337 &   0.32710 &  -0.00920 &  -0.00830 \\
HTR318-007     & bos      & i & 1930.791561 & 14.3643 & 0.0142 & 14.3544 &  -0.07175 &   0.29770 &  -0.00650 &   0.01090 \\
HTR318-007     & bos      & i & 1930.793201 & 14.3588 & 0.0132 & 14.3453 &  -0.07011 &   0.38650 &  -0.01590 &   0.01060 \\
HTR318-007     & bos      & i & 1930.794831 & 14.3753 & 0.0119 & 14.3533 &  -0.06848 &   0.33170 &  -0.00660 &   0.01190 \\

\enddata
\tablenotetext{1}{Either ``HTR318-007'' to indicate that this is a measurement for \sysname{}, or ``TFA\_?\_?'' to indicate that this is a measurement for one of the TFA trend vectors. The first digit in names of the form ``TFA\_?\_?'' is either 0, 1, or 2 to indicate if it is a trend used for the KeplerCam observations, the BOS observations, or the {\em K2} observations respectively. The second digit indicates which trend vector this measurement is associated with (1-20 for KeplerCam, 1-5 for BOS, or 1-12 for {\em K2}). For the {\em K2} observations these trends are the $\sin$ and $\cos$ components of a harmonic series to sixth order with a period equal to the time-spanned by the full quarter.}
\tablenotetext{2}{Either ``kepcam'' for KeplerCam observations, ``bos'' for BOS observations, ``K2/Campaign5'' for {\em K2}, ``kepcam\_binned'' for the time-binned out-of-eclipse KeplerCam observations, or ``HAT/G???'' to indicate HATNet observations. In the latter case the last three digits in the name indicate the HATNet field from which these observations were obtained. The {\em K2} observations have been cleaned as described in Section~\ref{sec:k2phot}.}
\tablenotetext{3}{BJD is on the TDB system, and we have corrected the times for the light-travel time effect as described in Appendix~\ref{sec:appendixltt}. The uncorrected times can be obtained by request to the authors.}
\tablenotetext{4}{For KeplerCam, BOS and {\em K2} observations of \sysname{} this is the measured magnitude without application of EPD or TFA. For HATNet this is the magnitude after application of EPD and TFA run in signal-search mode. For TFA trend measurements this is the value of the TFA vector at the specified time.}
\tablenotetext{5}{For KeplerCam, BOS and {\em K2} observations of \sysname{} this is the magnitude after application of EPD and TFA. For other observations this value is undefined.}
\tablenotetext{6}{The time from eclipse center. This is used as an EPD term to second order for the KeplerCam and BOS observations of \sysname{}. For other observations this value is left undefined.}
\tablenotetext{7}{The parameters $S$, $D$ and $K$ describe the shape of the PSF, and are provided for KeplerCam and BOS observations of \sysname{}, for which they are used as EPD terms to first order. Here we assume an elliptical Gaussian PSF parameterized by the form: $\exp\{-\frac{1}{2}(S(x^{2}+y^{2})+D(x^{2}-y^{2})+K(2xy))\}$, with $x$ and $y$ being the distance in pixels from the PSF center.}
\tablecomments{
   This table is presented in its entirety in the electronic edition
   of the Journal. A portion is shown here for guidance regarding its
   form and content.
}
\ifthenelse{\boolean{emulateapj}}{
  \end{deluxetable*}
}{
  \end{deluxetable}
}

\ifthenelse{\boolean{emulateapj}}{
  \begin{deluxetable}{lrl}
}{
  \begin{deluxetable}{lrl}
}
\tablewidth{0pc}
\tabletypesize{\scriptsize}
\tablecaption{
  Photometric Measurements of \sysname{} From the Literature
  \label{tab:catalogphotometry}
}
\tablehead{
  \multicolumn{1}{c}{Filter} &
  \multicolumn{1}{c}{Measurement} &
  \multicolumn{1}{c}{Reference} \\
  &
  &
}
\startdata
$g$ & 17.262 & KIC \citep{brown:2011}\tablenotemark{a} \\
$r$ & 15.780 & KIC \\
$i$ & 14.417 & KIC \\
$z$ & 13.510 & KIC \\
DDO-51 & 16.881 & KIC \\
$J$ & 12.028$\pm$0.021 & 2MASS \citep{skrutskie:2006} \\
$H$ & 11.429$\pm$0.022 & 2MASS \\
$K_{S}$ & 11.131$\pm$0.020 & 2MASS \\
$u$ & 19.522$\pm$0.039 & SDSS DR9 \citep{sdss:dr9} \\
$g$ & 17.270$\pm$0.004 & SDSS \\
$r$ & 15.848$\pm$0.003 & SDSS \\
$i$ & 14.353$\pm$0.003 & SDSS \\
$z$ & 13.520$\pm$0.003 & SDSS \\
3890 & 18.845$\pm$0.085 & BATC \citep{fan:1996}\tablenotemark{b} \\
5795 & 16.093$\pm$0.016 & BATC \\
6075 & 16.118$\pm$0.017 & BATC \\
6660 & 15.163$\pm$0.017 & BATC \\
7215 & 14.626$\pm$0.015 & BATC \\
8020 & 13.928$\pm$0.010 & BATC \\
8480 & 13.784$\pm$0.015 & BATC \\
9190 & 13.257$\pm$0.014 & BATC \\
9745 & 13.109$\pm$0.010 & BATC \\
NUV  & 21.57$\pm$0.38 & GALEX DR5 \citep{bianchi:2011} \\
W1   & 11.042$\pm$0.023 & WISE \citep{wright:2010} \\
W2   & 10.866$\pm$0.021 & WISE \\
W3   & 10.919$\pm$0.106 & WISE \\
W4   & $< 8.853$ & WISE \\
$B$  & 17.450 & NOMAD \citep{zacharias:2004} \\
$V$  & 16.130 & NOMAD \\
$R$  & 15.260 & NOMAD \\
$G$  & $14.922 \pm 0.020$ & Gaia DR1 \citep{gaia:2016} \\
\enddata
\tablenotetext{a}{Observations of M67 were carried out as part of calibrating the KIC. \sysname{} was serendipitously included in these observations.}
\tablenotetext{b}{Data from the BATC color survey of M67. Values are AB magnitudes within narrow-band filters. The filtername is the center wavelength of the filter in Angstroms.}
\ifthenelse{\boolean{emulateapj}}{
  \end{deluxetable}
}{
  \end{deluxetable}
}

% ####################################################################
\subsection{Ground-Based Photometric Follow-up}
\label{sec:photfu}

%% ----------------
\begin{figure}[!ht]
\plotone{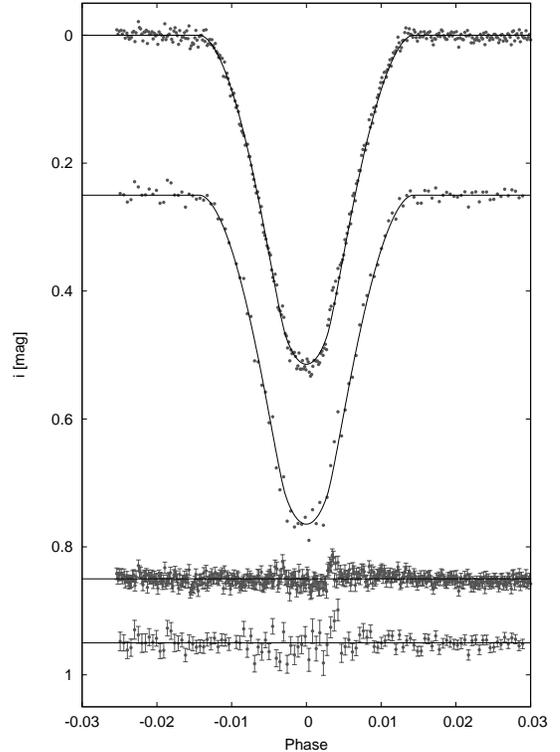}
\caption{ Primary eclipse light curves for \sysname{} offset
  vertically for clarity. The top light curve was obtained with
  KeplerCam on the night of 2012 January 3, while the second light
  curve was obtained with BOS on the night
  of 2012 January 3. The solid lines show the best-fit model light
  curves. Residuals from the best-fit models are shown at the bottom,
  in the same order as above.
\label{fig:lcpri}}
\end{figure}
%% ----------------
%% ----------------
\begin{figure}[!ht]
\plotone{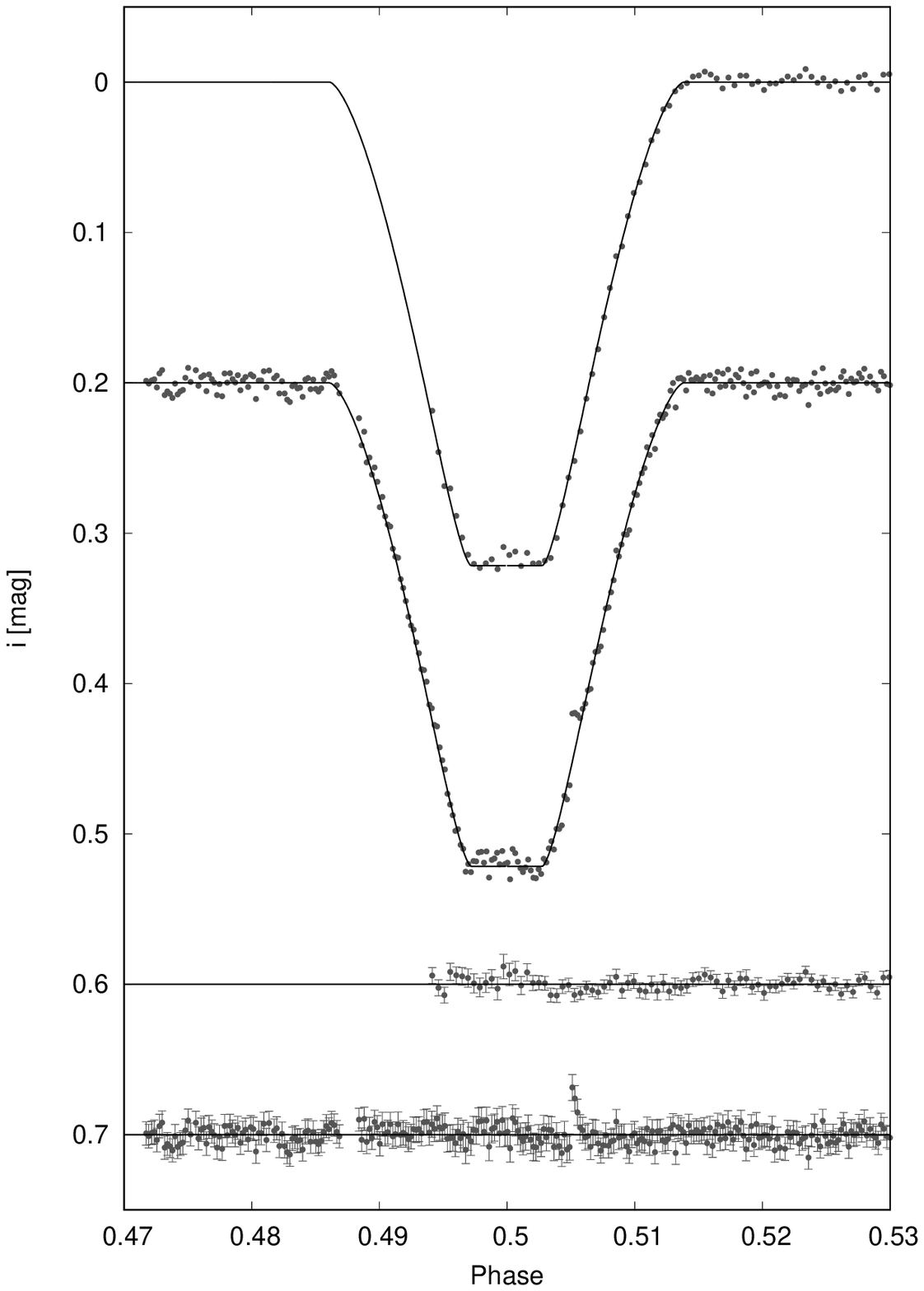}
\caption{ Secondary eclipse light curves for \sysname{} offset
  vertically for clarity. Both light curves were obtained with
  KeplerCam, the top on the night of 2011 March 30, the second on the
  night of 2011 December 29. Residuals from the best-fit models are
  shown at the bottom, in the same order as above.
\label{fig:lcsec}}
\end{figure}
%% ----------------

Additional photometric light curves of \sysname{} covering both
primary and secondary eclipses were obtained with the KeplerCam
instrument on the FLWO~1.2\,m telescope and the CCD imager on the
Byrne Observatory at Sedgewick, CA (BOS) 0.8\,m telescope. The dates,
number of observations gathered, exposure times, and filters used are
listed in \reftabl{photobs}. The light curves covering primary eclipse
are shown in Figure~\ref{fig:lcpri}, while those covering secondary
eclipse are shown in Figure~\ref{fig:lcsec}. These observations were
reduced to light curves via the aperture photometry procedure
described by \cite{bakos:2010:hat11}. As discussed in
\refsecl{jointfit} we applied trend-filtering to the light curves
simultaneously with the fitting procedure.

As seen in Figure~\ref{fig:lcpri}, we obtained two light curves covering primary eclipse on the same night with different facilities. Both of these light curves show a slight brightening in the residuals at the same phase during egress. This is either due to a stellar flare, or spots on the primary star. A similar feature is seen during egress of a secondary eclipse observed on the night of 2011 December 29 (Figure~\ref{fig:lcsec}), but no such feature is seen during a secondary eclipse observed 274 days earlier. Note that if these features are due to spots we would not expect them to reappear at the same phase after 274 days due to the slight difference between the stellar rotation period(s) and the orbital period of the system (Section~\ref{sec:stellarrotation}).

% ####################################################################
\subsection{K2 Photometry}
\label{sec:k2phot}

%% ----------------
\begin{figure*}[!ht]
\plotone{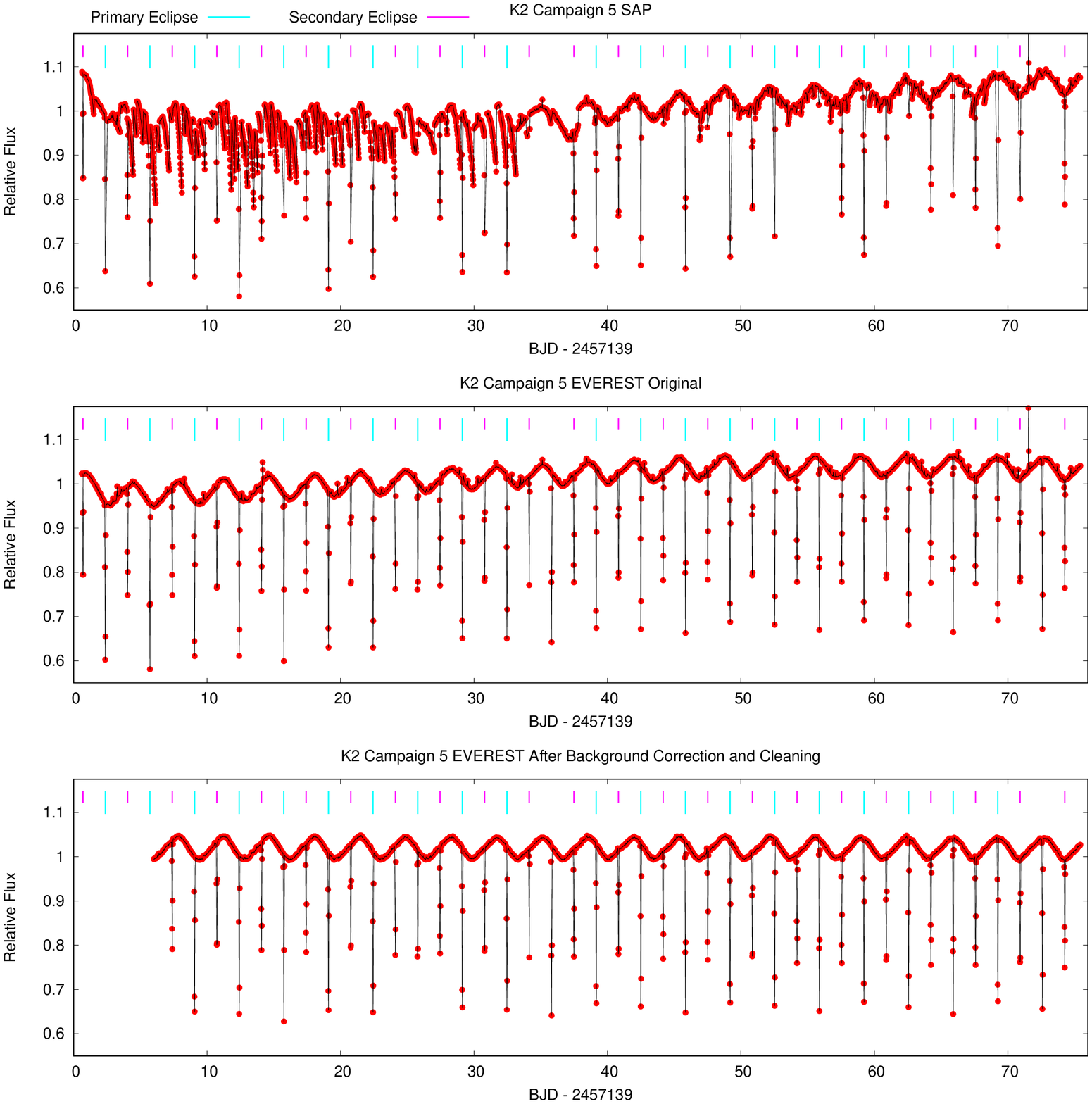}
\caption{K2 Campaign 5 light curve of \sysname{} (EPIC~211432946) showing ({\em top}) the original SAP light
  curve in the top panel, ({\em middle}) the EVEREST light curve from \citet{luger:2017} which applies a pixel-level decorrelation against the spacecraft roll, and ({\em bottom}) the EVEREST light curve after correcting for an increase in the brightness which we attribute to a slight error in the background correction over the course of the campaign, and filtering of bright outliers (likely due to stellar flares) and exclusion of points from the start of the campaign. This is the light curve that we analyze in Section~\ref{sec:jointfit}. In all cases the overplotted line is not a model, but simply connects observed points in time. Observed primary and secondary elcipse events are marked in each panel.
\label{fig:k2lcfulleverestflux}}
\end{figure*}
%% ----------------

%% ----------------
\begin{figure*}[!ht]
\plotone{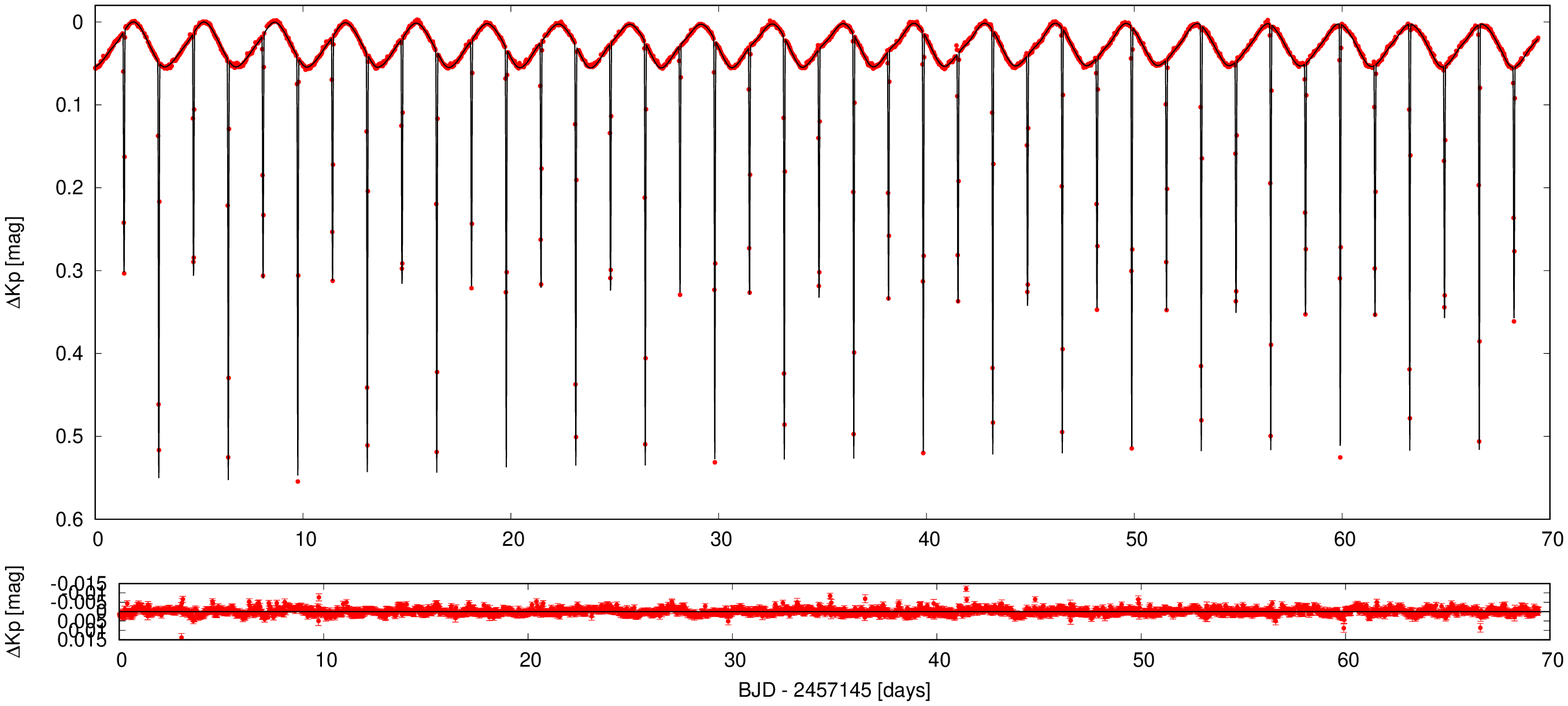}
\plotone{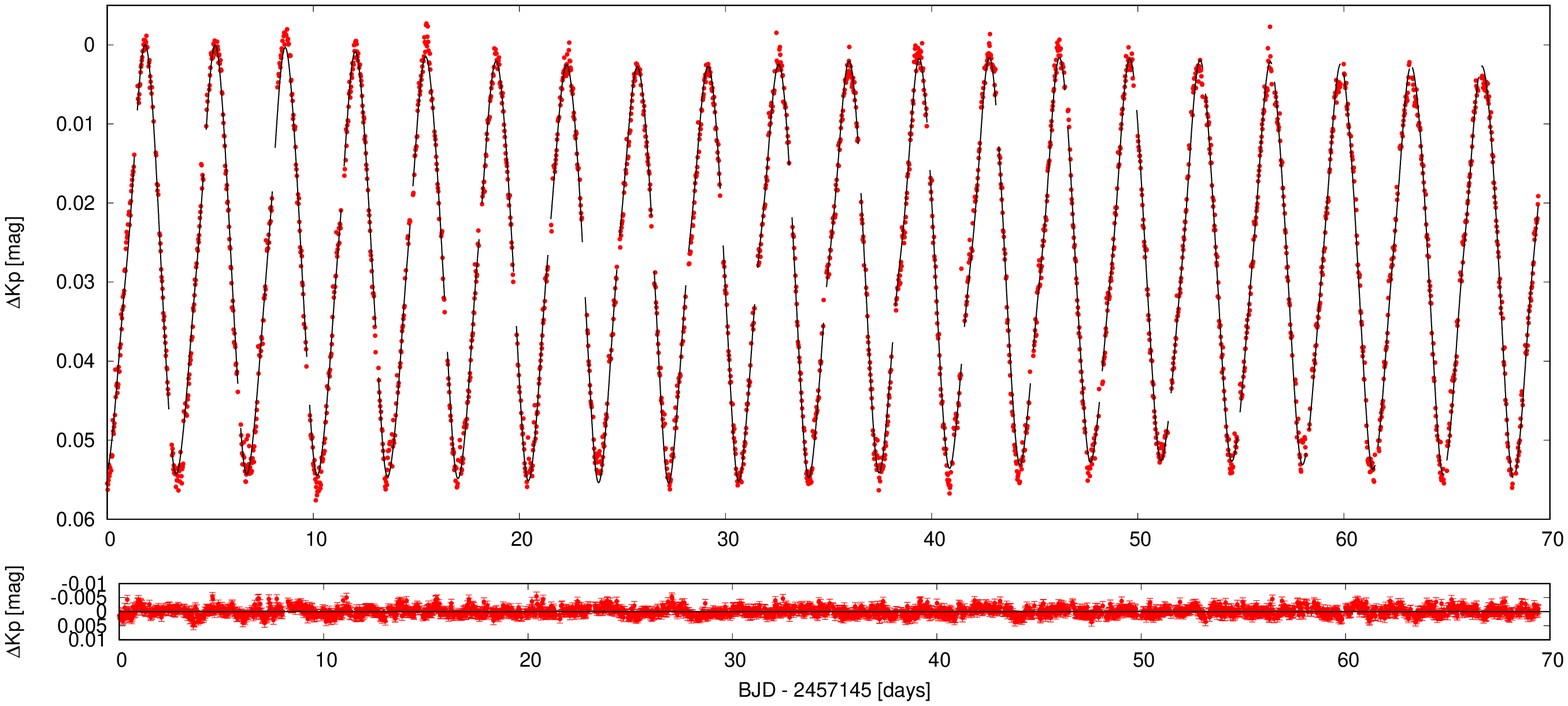}
\caption{{\em Top}: K2 Campaign 5 light curve of \sysname{} showing
  the EVEREST light curve after correcting for a systematic variation
  in the background flux, and filtering outliers. Overplotted is our
  best-fit model including primary and secondary eclipses, ellipsoidal
  variability (which is included through the JKTEBOP model), and the $P = \starAootvperiod$\,d and $P = \starBootvperiod$\,d rotational variation
  attributed in this modeling to the primary and secondary components, respectively. The
  residuals from the best-fit model are shown immediately below the
  full light curve. {\em Bottom}: Same as top, here we exclude
  points within $0.02$ in phase from the primary and secondary eclipse
  centers to highlight the out-of-eclipse variation, and its phase
  coherence over the time-span of the observations.
\label{fig:k2lcfulleverestmag}}
\end{figure*}
%% ----------------

%% ----------------
\begin{figure*}[!ht]
\plottwo{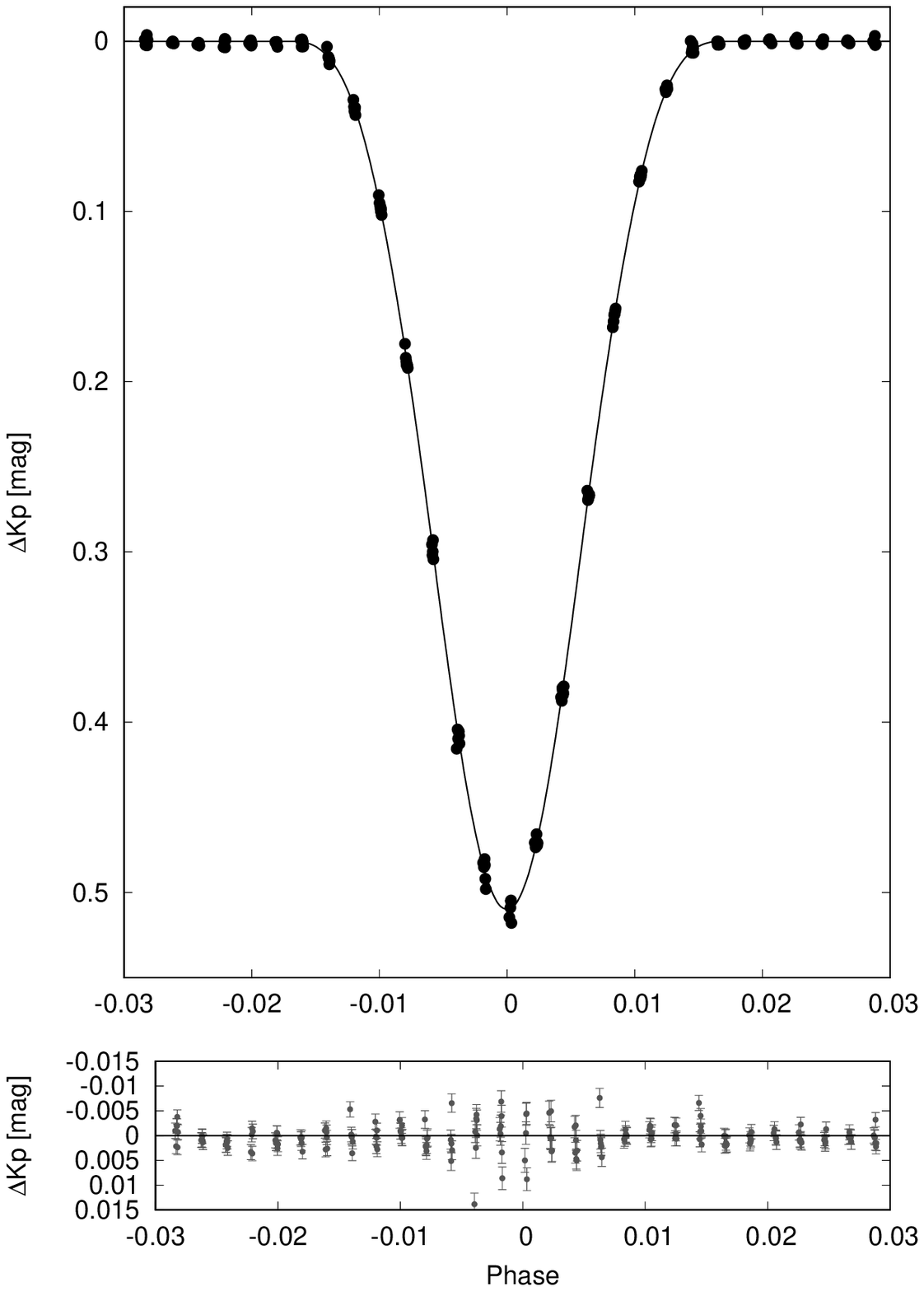}{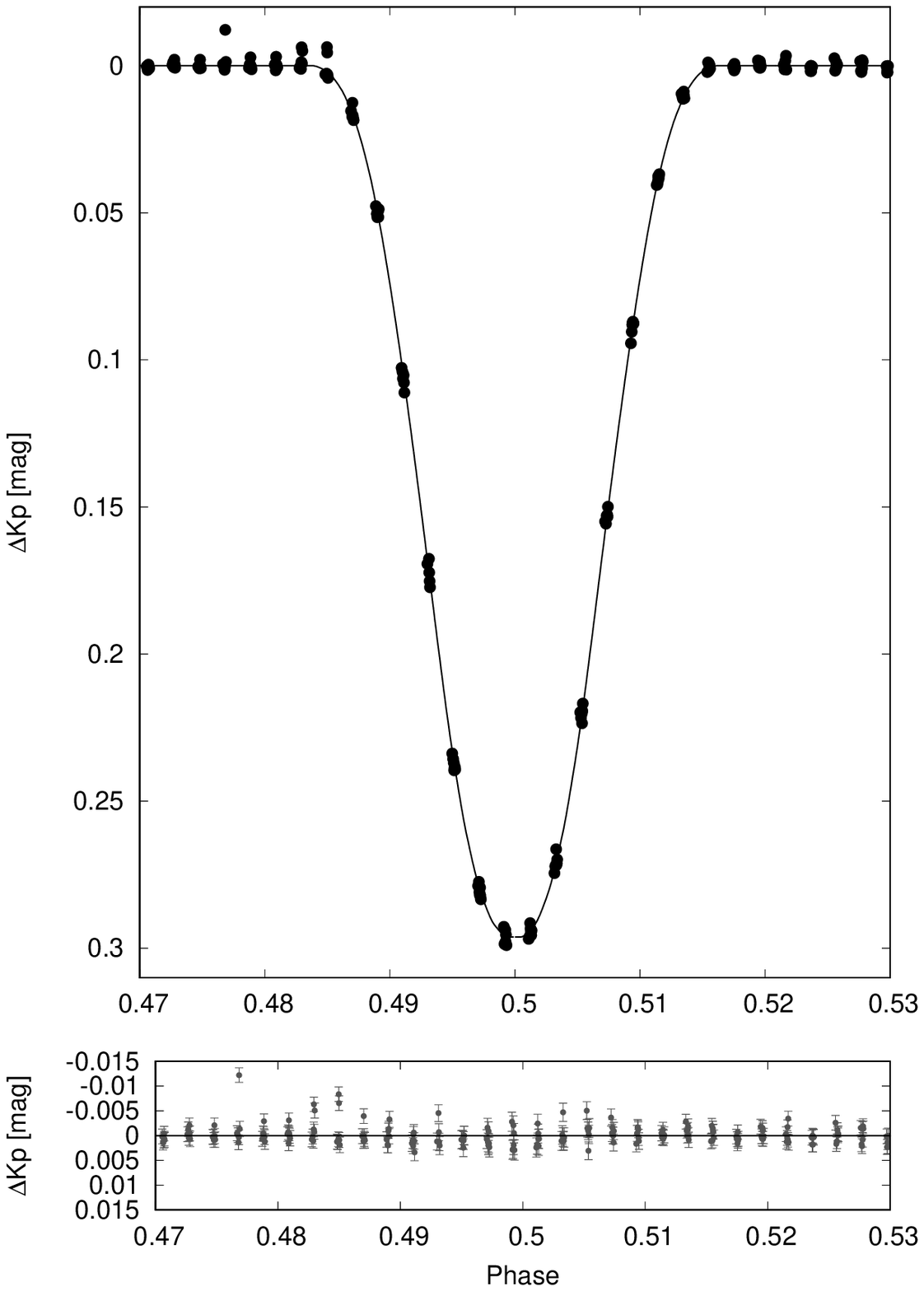}
\caption{{\em Left}: Phase-folded K2 Campaign 5 EVEREST light curve of \sysname{} showing
  primary eclipses, together with our best-fit model, after subtracting our model for the rotational variability (Figure~\ref{fig:k2lcfulleverestmag} right). The phase-folded residuals from the full model are plotted below the light curve. {\em Right}: same as at left, here we show the secondary eclipses. Note the different vertical scale used in each of the panels.
\label{fig:k2lceclipses}}
\end{figure*}
%% ----------------

Due to its propitious location in the field of the open cluster M67,
\sysname{} was observed by the NASA {\em K2} mission
\citep{howell:2014} during Campaign 5. A total of 3298 long cadence
(29.4\,minute) photometric time series observations were collected by
the satellite for this source (EPIC~211432946). The observations span
74.8 days from UT 2015 April 26 to UT 2015 July 10, and cover 20
primary eclipse events and 23 secondary events. 

Based on the {\em K2} observations \citet{barros:2016} have also
independently identified EPIC~211432946 as an eclipsing binary
system. They included it in a catalog of eclipsing binary candidates
providing the period, epoch of eclipse, primary eclipse depth, eclipse
and ingress durations, and flags indicating and that there is a
definite secondary eclipse present in the data. There is no additional
analysis or discussion of this particular object in that publication,
and we did not become aware of this independent identification until
very shortly before submitting this paper for publication.

We considered three different publicly available reductions of the
{\em K2} long cadence observations of \sysname{}. These include (1)
the Data Release 7 {\em K2} long cadence light curve for \sysname{}
from the Mikulski Archive for Space Telescopes (MAST); (2) the EVEREST
pixel-level decorrelated light curve from \citet{luger:2017}; and (3)
the decorrelated light curve produced by the method of
\citet{vanderburg:2014}. We found that the EVEREST reduction of
\sysname{} had substantially lower scatter around the astrophysical
signal and fewer residual systematic variations associated with the
six hour spacecraft roll than the other reductions, and so we adopt
the EVEREST light curve for the remainder of the analysis.

Figure~\ref{fig:k2lcfulleverestflux} shows the original EVEREST light
curve of \sysname{} in units of flux. In addition to the numerous
primary and secondary eclipses visible, there is a periodic variation
in the out-of-eclipse light curve that we associate with the rotation
of one of the binary components. There is also a significant
increasing trend in the flux over the quarter. The trend is correlated
with the change in the zodiacal background flux over the quarter, and
likely indicates a slight systematic error in subtracting this
background from the observations of this faint source. Because both
the primary and secondary eclipse depths remain constant in flux over
the quarter, this is clearly a change in the background flux level,
rather than a variation in the brightness of one or more components in
the system. To correct for this we fit the out-of-eclipse points in
the flux light curve as a combination of two harmonic series. The
first is a sequence of six harmonics with a fundamental period of $P =
3.4164940808$\,days, the highest peak in the Lomb-Scargle periodogram
(we discuss the photometric rotational variation seen in the {\em K2}
light curve in greater detail in
Section~\ref{sec:stellarrotation}). This series captures the variation
due to the periodic rotation of one of the binary components. The
second series that we used was a series of six harmonics with a
fundamental period of $P = 93$\,days which we use to capture the
variation in the background flux. We subtract the latter model from
the full light curve and convert the corrected fluxes to
magnitudes. The choice of $P = 93$\,days for the background
  harmonic series is effectively arbitrary. This value is longer than
  the time-span of the {\em K2} data and was determined by optimizing
  a fit during a preliminary analysis of the light curve, but was kept
  fixed during our final analysis.

Upon further inspection we found that there are numerous
short-time-scale brightenings in the light curve that may either be
due to flaring activity, or result from imperfect removal of
systematic artifacts associated with the spacecraft roll for this
large amplitude variable star. To deal with these we clipped
out-of-eclipse points with residuals less than
-0.005\,mag from a first iteration model fit to the full light
curve. We also restrict the analysis to observations obtained after
$BJD > 2457145$ to avoid large systematic variations seen in the residuals
at the start of the time series.

Figure~\ref{fig:k2lcfulleverestmag} shows the resulting corrected {\em
  K2} light curve together with our best-fit physical model
(Section~\ref{sec:jointfit}). Figure~\ref{fig:k2lceclipses} show the
phase-folded {\em K2} light curve with our full model for the
rotational signals of the two component stars subtracted, and zoomed
in on the primary and secondary eclipses, respectively. In these same
figures we overplot our best-fit model for the eclipse signals,
accounting for the 1625.35\,sec {\em K2} integration time.

% ####################################################################
\subsection{Spectroscopic Observations}

\subsubsection{Optical}
\label{sec:optspec}

%% ----------------
\begin{figure}[!ht]
\plotone{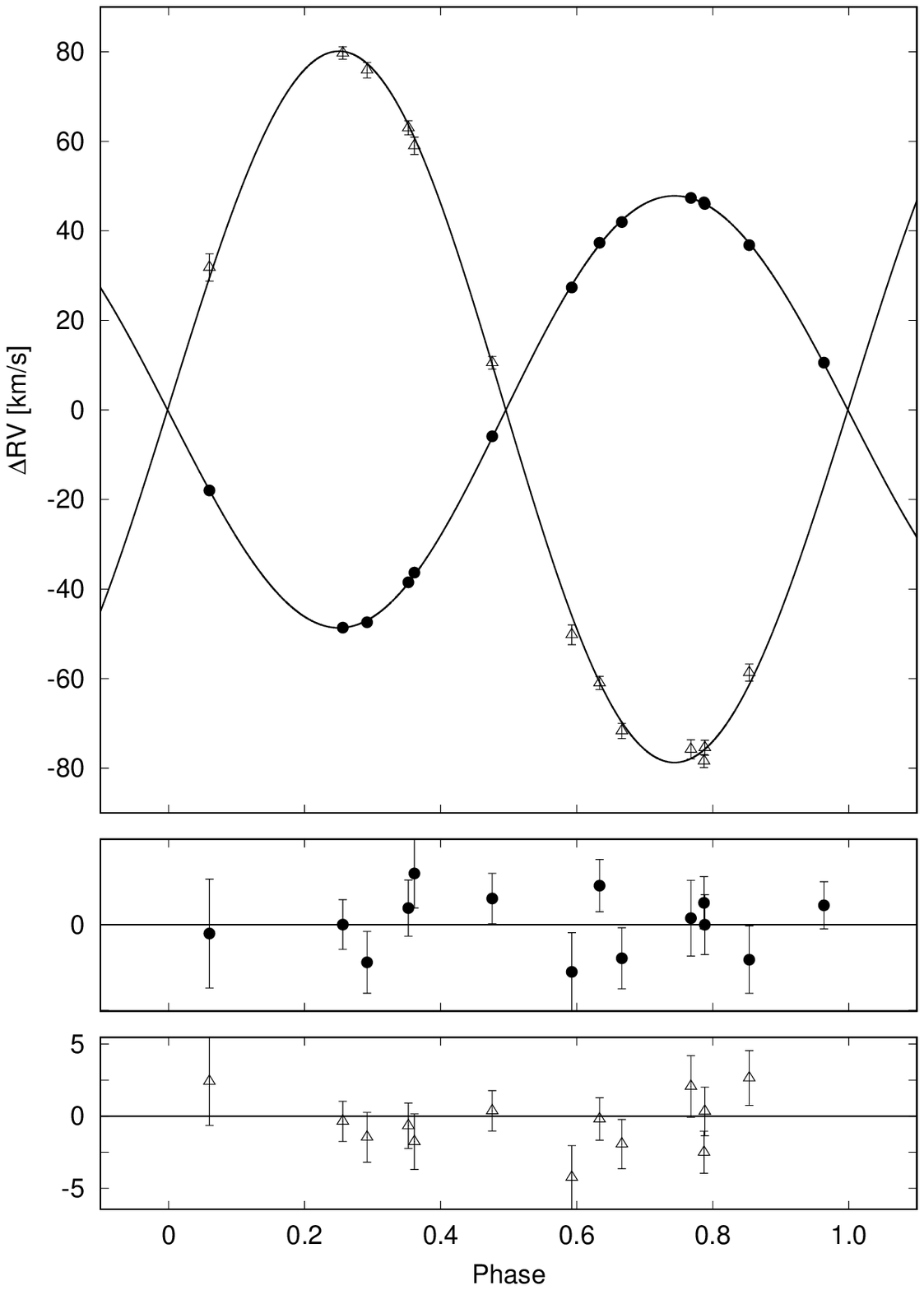}
\caption{ Phase-folded RVs for \sysname{} measured with TRES on the
  FLWO~1.5\,m together with the best-fit orbit (solid lines).  Filled
  circles show the RVs for the primary star, while open triangles show
  the RVs for the secondary star. The bottom two panels show the
  residuals for each component from the best-fit model. The systemic
  $\gamma$ velocity has been subtracted from the RVs.
\label{fig:RVs}}
\end{figure}
%% ----------------

We obtained optical spectra of \sysname{} using the Tillinghast
Reflector Echelle Spectrograph (TRES), with the medium-resolution
fiber, on the 1.5\,m Tillinghast Reflector at FLWO
\citep{furesz:2008}. This instrument and configuration delivers
multi-order spectra with a resolution of $\lambda/\Delta\lambda
\approx $44\,000 and a wavelength coverage of $\sim
3900$--$8900$\,\AA. A total of 14 spectra were obtained between
2011~Mar.~28 and 2011~Dec.~9. The spectra were extracted and
wavelength calibrated using the pipeline of
\cite{buchhave:2010:hat16}. While the star is too faint for much of
the optical spectra to be useful, we clearly detected double H$\alpha$
emission lines moving in phase with the photometric ephemeris, and
also detect TiO absorption bands from both components. The strength of the H$\alpha$ emission line for the primary star appears to vary by $\sim 20$\% between observations. We used these
observations to measure RVs for both of the components with the TwO
Dimensional CORrelation \citep[TODCOR;][]{zucker:1994:todcor}
algorithm, using a TRES spectrum of Barnard's star as a template for
both components. The correlation was done over a single order
containing TiO lines spanning 7063--7201\,\AA. In doing this we
determine an optical spectroscopic light ratio of $L_{B}/L_{A} =
0.350\pm 0.014$ from the highest S/N exposures with unblended lines, and
used this value in measuring the velocities from all of the
spectra. This is similar to the light ratio of $L_{B,i}/L_{A,i} =
0.300\pm 0.005$ determined from the $i$-band light curves. Because the
$i$ bandpass is broader than the order over which the spectroscopic
light ratio is determined, we do not expect these two estimates to be
equal. The resulting RVs, measured relative to Barnard's star, are
listed in Table~\ref{tab:tresrvs}, and plotted in
Figure~\ref{fig:RVs}.

\ifthenelse{\boolean{emulateapj}}{
  \begin{deluxetable*}{rrrrrr}
}{
  \begin{deluxetable}{rrrrrr}
}
\tablewidth{0pc}
\tabletypesize{\scriptsize}
\tablecaption{
  Relative Radial Velocities for \sysname{} from FLWO~1.5\,m/TRES
  \label{tab:tresrvs}
}
\tablehead{
  \multicolumn{1}{c}{BJD} &
  \multicolumn{1}{c}{RV$_{\rm A}$\tablenotemark{a}} &
  \multicolumn{1}{c}{$\sigma$RV$_{\rm A}$\tablenotemark{b}} &
  \multicolumn{1}{c}{RV$_{\rm B}$\tablenotemark{a}} &
  \multicolumn{1}{c}{$\sigma$RV$_{\rm B}$\tablenotemark{c}} &
  \multicolumn{1}{c}{C\tablenotemark{d}} \\
  &
  \multicolumn{1}{c}{\kms} &
  \multicolumn{1}{c}{\kms} &
  \multicolumn{1}{c}{\kms} &
  \multicolumn{1}{c}{\kms} &
}
\startdata
2455648.7452 & 177.85 &  0.66 &  79.52 &  1.77 & 0.714 \\
2455662.6386 & 186.50 &  0.76 &  65.06 &  2.03 & 0.623 \\
2455667.6664 &  93.06 &  0.79 & 216.39 &  2.08 & 0.608 \\
2455668.6728 & 167.85 &  1.00 &  90.28 &  2.66 & 0.476 \\
2455672.6665 & 186.85 &  0.66 &  62.07 &  1.76 & 0.718 \\
2455699.6407 & 177.31 &  0.86 &  81.83 &  2.28 & 0.554 \\
2455704.6528 & 102.01 &  0.71 & 203.50 &  1.90 & 0.668 \\
2455888.9826 & 134.62 &  0.64 & 151.02 &  1.69 & 0.747 \\
2455899.9911 & 187.85 &  0.96 &  64.69 &  2.58 & 0.492 \\
2455900.9678 & 122.51 &  1.39 & 172.33 &  3.69 & 0.343 \\
2455901.9759 & 104.17 &  0.87 & 199.50 &  2.32 & 0.545 \\
2455902.9955 & 182.46 &  0.78 &  68.80 &  2.06 & 0.614 \\
\tablenotemark{e}2455903.9885 & 151.06 &  0.60 & $\cdots$ & $\cdots$ & 0.792 \\
2455904.9681 &  91.86 &  0.63 & 220.20 &  1.68 & 0.750 \\
\enddata
\tablenotetext{a}{RVs are measured relative to Barnard's star.}
\tablenotetext{b}{Primary star RV uncertainties have been scaled by a factor of $0.81$ as determined in Section~\ref{sec:jointfit}.}
\tablenotetext{c}{Secondary star RV uncertainties have been scaled by a factor of $0.96$ as determined in Section~\ref{sec:jointfit}.}
\tablenotetext{d}{Normalized Cross-Correlation Peak Height}
\tablenotetext{e}{This observation was obtained near eclipse. A separate velocity for the secondary component could not be resolved from the primary velocity.}
\ifthenelse{\boolean{emulateapj}}{
  \end{deluxetable*}
}{
  \end{deluxetable}
}

\subsubsection{Near-IR}
\label{sec:nirspec}

%% ----------------
\begin{figure*}[!ht]
\plottwo{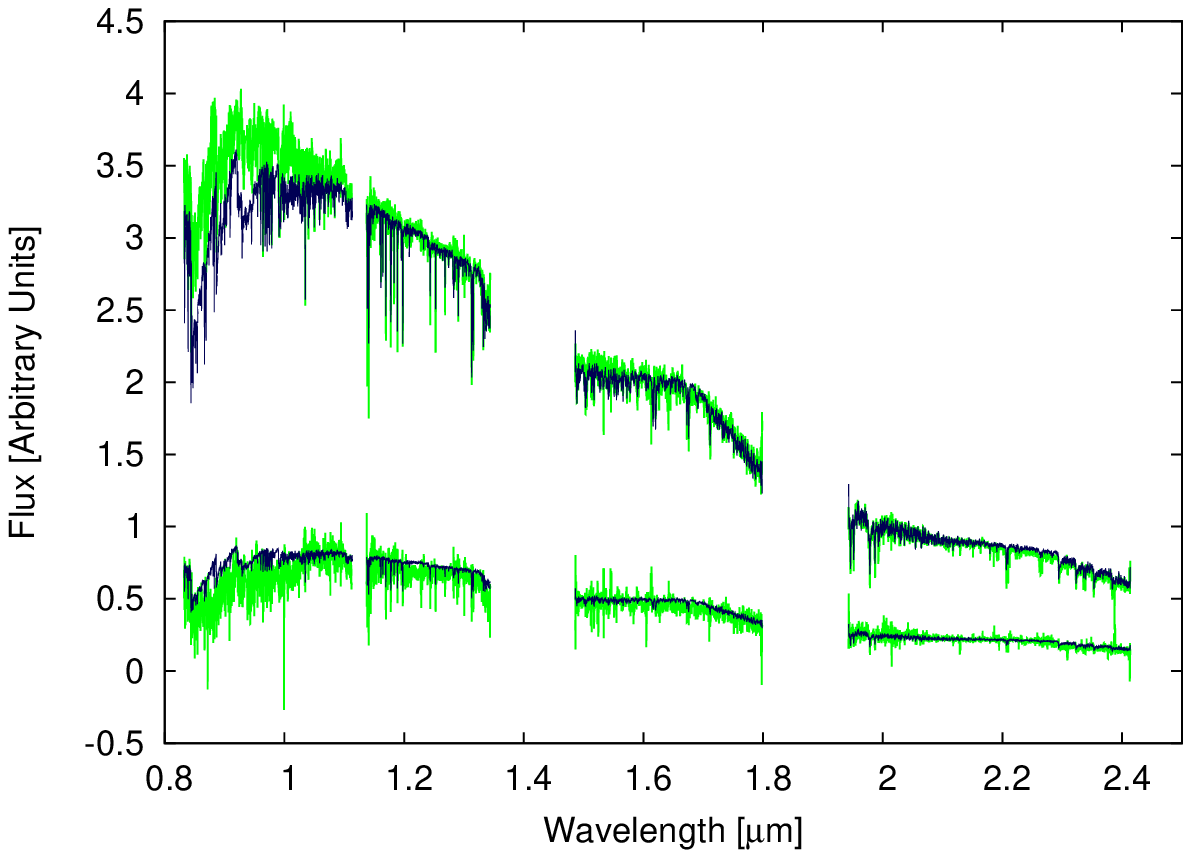}{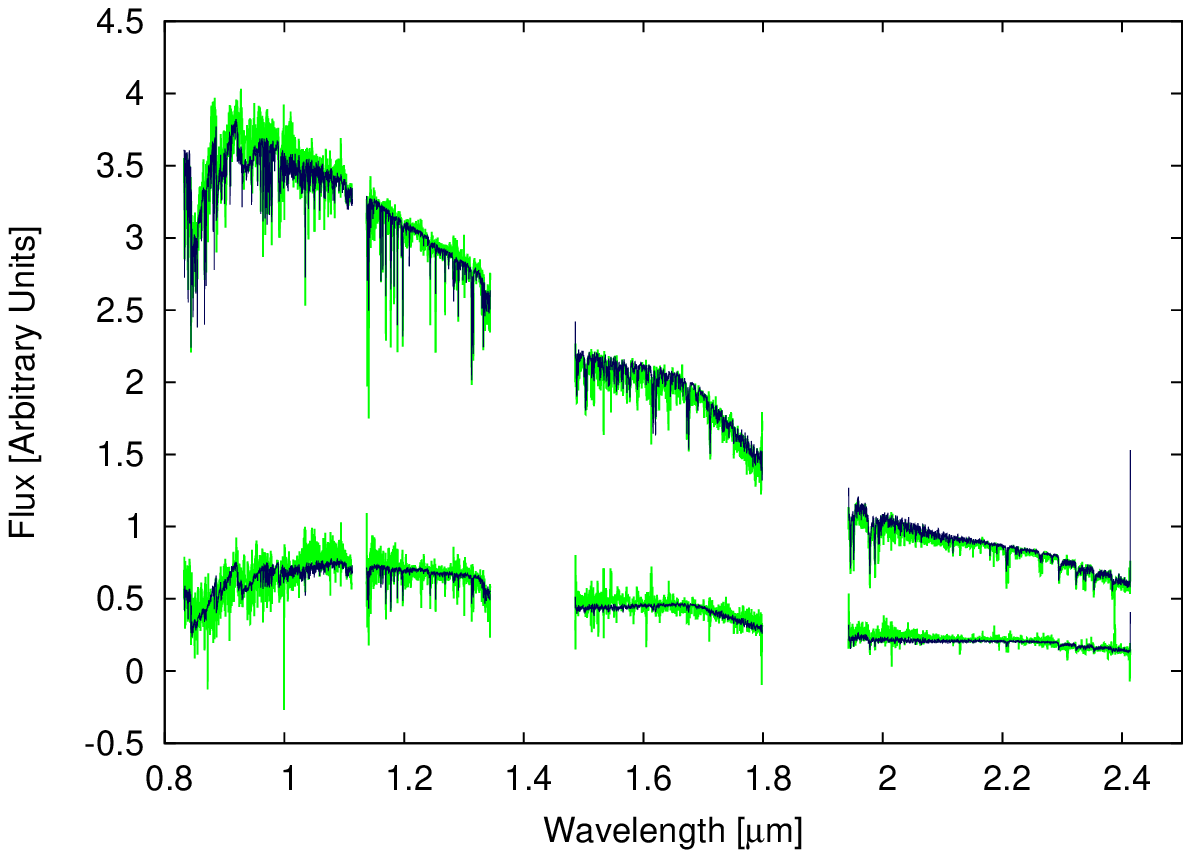}
\caption{ Full NIR spectra for the components of \sysname{} from Magellan/FIRE (green lines). In each panel the primary star spectrum is at the top, the secondary at the bottom. In the left panel we overlay the BT-Settl synthetic templates that provide the highest cross-correlation (blue lines). For the primary star this is a template with T$_{\rm eff} = 3200$\,K and [Fe/H]$=0.3$, while for the secondary star the template has T$_{\rm eff} = 3100$\,K and [Fe/H]$=0.3$. On the right-hand side we overlay templates (blue lines) that provide somewhat better matches to the bluest ($Z+Y$) band of the spectrum. For the primary this template has T$_{\rm eff} = 3400$\,K and [Fe/H]$=0.3$, while for the secondary this has T$_{\rm eff} = 2900$\,K and [Fe/H]$=0.3$. 
\label{fig:fullNIRspec}}
\end{figure*}
%% ----------------

In order to determine the atmospheric parameters for the individual
components of \sysname{} we obtained medium-resolution near-IR spectra
using the Folded-port InfraRed Echellette (FIRE) spectrograph
\citep{simcoe:2013} on the 6.5\,m Magellan Baade telescope at Las
Campanas Observatory in Chile. Observations were conducted during the
last $\sim$4 hours before twilight on the UT nights of 09 December
2011, 10 December 2011 and 11 December 2011, with a total secondary
eclipse occurring during the night of 10 December 2011. We observed
\sysname{} continuously over an 83 minute period
encompassing the secondary eclipse, and on each of the nights before
and after the eclipse. For calibration we also observed a number of M
dwarf standard stars. Our observations are summarized in
Table~\ref{tab:fireobslog}. In addition to the stars listed therein
we also observed GJ~273 and GJ~382, however due to a poor telluric
correction the data for these two stars proved to be not useable.

Observations were performed in echelle mode using a 0\farcs6 slit
width (7\arcsec\ length) with readout performed in the Fowler~1
mode. This setup provides spectra with a resolution of $\lambda/\Delta
\lambda = 6000$ over the $0.82$--2.51\,$\mu$m wavelength range. To facilitate
sky subtraction and flux calibration we performed A--B nodding and
observed telluric standards near in time and airmass to each of the
targets. To determine the wavelength calibration we obtained ThAr lamp
spectra before or after a set of observations for a given science
target. We also obtained quartz lamp spectra and observations of the
twilight sky to use in tracing the echelle orders and creating a
flat-field.

The observations were reduced to flux-calibrated spectra using the
FIRE reduction pipeline \citep{simcoe:2013} downloaded May 2012. The
reduction was performed using the boxcar extraction mode, with
apertures determined automatically, and with the closest associated B
(A) nod used to determine the sky subtraction for a given A (B) nod
observation.

To extract separate spectra for the primary and secondary star
components of \sysname{} we performed Fourier-based spectral
disentangling using version 3 of the FDBinary program
\citep{ilijic:2004:fdbinary}. We fixed the orbital parameters to those determined from fitting the TRES observations as the FIRE observations had poor phase coverage leading to a poor constraint on the parameters. We interpolated all spectra of
\sysname{} to a common wavelength grid uniformly sampled in $\log
\lambda$, linearly interpolating over the wavelength ranges
$0.9332$--$0.9378$\,$\mu$m, $1.1125$--$1.1371$\,$\mu$m,
$1.343967$--$1.48614$\,$\mu$m, $1.7990619$--$1.9427$\,$\mu$m,
$1.9667$--$1.97$\,$\mu$m, $2.4364$--$2.4401$\,$\mu$m,
$2.4473$--$2.4601$\,$\mu$m, $2.4663$--$2.4695$\,$\mu$m, and
$2.4777$--$2.48$\,$\mu$m which have significant telluric
contamination. We performed the disentangling separately for the $Z+Y$,
$J$, $H$ and $K$-band regions of the spectra (for the purposes of this
paper, these regions correspond to wavelength ranges of
$0.832$--$1.112$\,$\mu$m, $1.137$--$1.344$\,$\mu$m,
$1.486$--$1.798$\,$\mu$m, and $1.943$--$2.414$\,$\mu$m,
respectively). We excluded spectra obtained during partial
eclipse. 

The resulting spectra for both the primary and secondary components
are shown in Figures~\ref{fig:fullNIRspec}, together with theoretical
templates, which are described further in
Appendix~\ref{sec:ccf}. These spectra are also shown at higher resolution in
Figures~\ref{fig:NIRprimaryspecbands}
and~\ref{fig:NIRsecondaryspecbands} in the Appendix. The data are
provided in Table~\ref{tab:nirspecdat}.

\ifthenelse{\boolean{emulateapj}}{
  \begin{deluxetable*}{llrlrrr}
}{
  \begin{deluxetable}{llrlrrr}
}
\tablewidth{0pc}
\tabletypesize{\scriptsize}
\tablecaption{
  Summary of FIRE/Magellan Observations
  \label{tab:fireobslog}
}
\tablehead{
  \multicolumn{1}{c}{Object} &
  \multicolumn{1}{c}{Date(s)} &
  \multicolumn{1}{c}{$K_{s}$ mag} &
  \multicolumn{1}{c}{Spectral Type} &
  \multicolumn{1}{c}{Nspectra} &
  \multicolumn{1}{c}{Mean} &
  \multicolumn{1}{c}{Total S/N} \\
  &
  \multicolumn{1}{c}{(2011-12-\#)} &
  &
  &
  &
  \multicolumn{1}{c}{Exp.~Time} &
}
\startdata
    \sysname & 09,10,11 & 11.13 & M3.6+M5.0 & 21 &  371.4 &  490.2 \\
       GJ~205 &       11 &  4.04 & M1.5 &  2 &    1.0 &  169.1 \\
      GJ~250~B &       09 &  5.72 & M2.5 &  2 &    1.5 &  204.4 \\
       GJ~352 &    10,11 &  5.51 & M3.0 &  3 &    1.0 &  191.9 \\
       GJ~285 &       09 &  5.70 & M4.0 &  2 &    1.0 &  224.8 \\
     GJ~3348~B &       11 &  8.79 & M4.5 &  2 &   60.0 &  146.1 \\
       GJ~283 &       10 &  9.29 & M6.0 &  2 &   60.0 &  161.4 \\
   NLTT~15867 &       11 & 10.31 & M6.0 &  2 &  225.0 &  178.6 \\
     LHS~2065 &       10 &  9.94 & M9.0 &  2 &  180.0 &  166.8 \\
\enddata
\ifthenelse{\boolean{emulateapj}}{
  \end{deluxetable*}
}{
  \end{deluxetable}
}

\ifthenelse{\boolean{emulateapj}}{
  \begin{deluxetable*}{rrrrr}
}{
  \begin{deluxetable}{rrrrr}
}
\tablewidth{0pc}
\tabletypesize{\scriptsize}
\tablecaption{
  Disentangled FIRE/Magellan Spectra of \sysname{}
  \label{tab:nirspecdat}
}
\tablehead{
  \multicolumn{1}{c}{Wavelength} &
  \multicolumn{1}{c}{Pri.~Flux\tablenotemark{a}} &
  \multicolumn{1}{c}{Err.~Pri.~Flux\tablenotemark{b}} &
  \multicolumn{1}{c}{Sec.~Flux} &
  \multicolumn{1}{c}{Err.~Sec.~Flux} \\
  \multicolumn{1}{c}{\AA} &
  &
  &
  &
}
\startdata
 8321.040 &   3.44391 &   0.05910 &   0.56731 &   0.04367 \\
 8321.390 &   3.45643 &   0.04704 &   0.50339 &   0.03266 \\
 8321.730 &   3.46005 &   0.04686 &   0.56625 &   0.02873 \\
 8322.080 &   3.48033 &   0.04144 &   0.67510 &   0.03531 \\
 8322.430 &   3.55268 &   0.04595 &   0.63200 &   0.03725 \\
 8322.780 &   3.47703 &   0.05433 &   0.67137 &   0.03735 \\
 8323.120 &   3.38021 &   0.04889 &   0.68143 &   0.07494 \\
 8323.470 &   3.40215 &   0.07842 &   0.70101 &   0.05821 \\
 8323.820 &   3.39484 &   0.10027 &   0.66955 &   0.06367 \\
 8324.160 &   3.36470 &   0.06003 &   0.57566 &   0.08180 \\

\multicolumn{5}{c}{$\ldots$} \\
\enddata
\tablenotetext{a}{
   The spectra have been flux-calibrated using observations of
   telluric standards. The fluxes have units of power per wavelength,
   but are on an arbitrary scale.
}
\tablenotetext{b}{
   Uncertainties based on photon-counting statistics for the individual observations of \sysname{}, propagated through
   the spectral disentangling procedure via a Monte Carlo simulation.
}
\tablecomments{
   This table is presented in its entirety in the electronic edition
   of the Journal. A portion is shown here for guidance regarding its
   form and content.
}
\ifthenelse{\boolean{emulateapj}}{
  \end{deluxetable*}
}{
  \end{deluxetable}
}

\section{Analysis}

\subsection{Stellar Rotation and Activity}
\label{sec:stellarrotation}

%% ----------------
\begin{figure*}[!ht]
\plotone{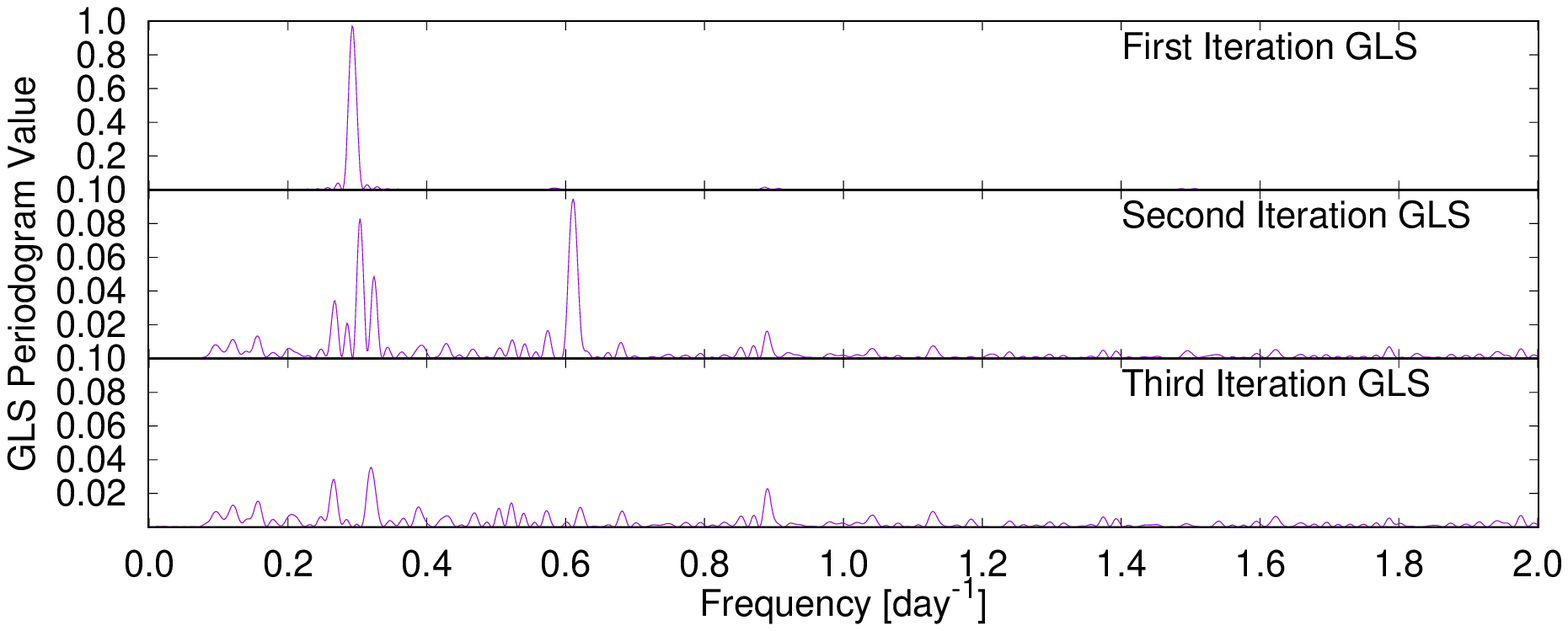}
\caption{Iterative application of the Generalized Lomb-Scargle
  Periodogram to the EVEREST {\em K2} Campaign 5 light curve of
  \sysname{} after removing observations taken during eclipses,
  removing high flux outliers, and correcting for a systematic
  variation in the background. The {\em top} panel shows the very
  significant detection of the $P = \starAootvperiod$\,d rotational
  period of the primary star (this value, with its uncertainty, comes
  from the analysis described in Section~\ref{sec:jointfit}, whereas
  the actual peak in the GLS periodogram is at $P =
  3.42005489$\,days). The {\em middle} panel shows the presence of an
  additional signal at $P = 1.63743194$\,days after fitting and
  subtracting a Fourier series to the light curve with the period
  fixed to $\starAootvperiod$\,d. We adopt double the GLS period in
  the analysis in Section~\ref{sec:jointfit}, yielding a period of $P
  = \starBootvperiod$\,d for this signal. The {\em bottom} panel shows
  that no additional periodic signals are detected at a significant
  level after subtracting a pair of Fourier series with periods of $P = \starAootvperiod$\,d and
  $\starBootvperiod$\,d. Note the different vertical scales between the three panels. Also note that the full periodogram is calculated up to the Nyquist Frequency, but we only display the region of the periodogram where significant peaks are present.
\label{fig:k2per}}
\end{figure*}
%% ----------------

We use the EVEREST {\em K2} light curve of \sysname{} to characterize
the photometric variability of the system due to stellar activity. The
light curve shows a clear sinusoidal variation with a period of
$\starAootvperiod$\,days, as detected by the Generalized Lomb-Scargle
(GLS) periodogram \citep{zechmeister:2009}. The value and uncertainty
listed are based on our full analysis of the system including
rotational variability as discussed in Section~\ref{sec:jointfit}. The
signal remains coherent over the full 74.8\,day time-span. Such a long spot coherence timescale is typical of M dwarf stars with large photometric amplitudes. Based on Figure~8 of \citet{giles:2017}, who used autocorrelation functions of {\em Kepler} light curves to determine the typical spot coherence lifetime as a function of spectral type and light curve r.m.s., an M dwarf star with a {\em Kepler} r.m.s.\ of $\sim 1$\% has an autocorrelation function that decays with an $e$-folding lifetime of $\sim 400$\,days. This is significantly longer than the time-span of the {\em K2} observations.
%, which is perhaps a surprisingly long time if the signal is due to the rotation of starspots on one of the two binary components. Nonetheless, lacking a better explanation for the signal, we interpret it as a rotation period. 
Figure~\ref{fig:k2per} shows the GLS
periodogram of the cleaned and background-corrected {\em K2} light
curve after removing points obtained during eclipses. 

In order to search for other periodic signals in the light curve, we
subtract the best-fit model of the $\starAootvperiod$\,d signal and
calculate the GLS periodogram of the residuals. We find a significant
peak in the periodogram at a period of $P = 1.63743194$\,days
(Fig.~\ref{fig:k2per}). Inspection of the residual light curve
phase-folded at both this period, and at double the period, indicates
that the double-period value is the correct period as the two minima
and maxima per cycle have noticeably different depths and heights.
Based on the full analysis of Section~\ref{sec:jointfit}, we find a
period and uncertainty of $P = \starBootvperiod$\,days for this
signal. 

A third iteration of GLS yields no additional significant periodic
signals (Fig.~\ref{fig:k2per}).

Although close to the orbital period of \periodshort\,d, the two most
signficant periods identified in the out-of-eclipse light curve are
definitely not the same as the orbital period, as can be seen by the
change in time in the rotational phase at which eclipses are
observed. In order to identify which binary star component is giving
rise to each of these signals, we make use of the {\em K2} observations
obtained near the base of the total secondary eclipses which enable us
to cleanly separate the contributions from each component to the total
integrated brightness of the system at those moments in time. We take
observations having a mid-exposure time within 11\,minutes of the
center time of a secondary eclipse as providing a measurement of the
integrated brightness of the primary star without any contribution
from the secondary. We then subtract this flux from the average flux
of nearby points that were obtained between 77\,minutes and
106\,minutes from the center of the eclipse to determine the
integrated brightness of the secondary star without any contribution
from the primary.

%% ----------------
\begin{figure*}[!ht]
\plottwo{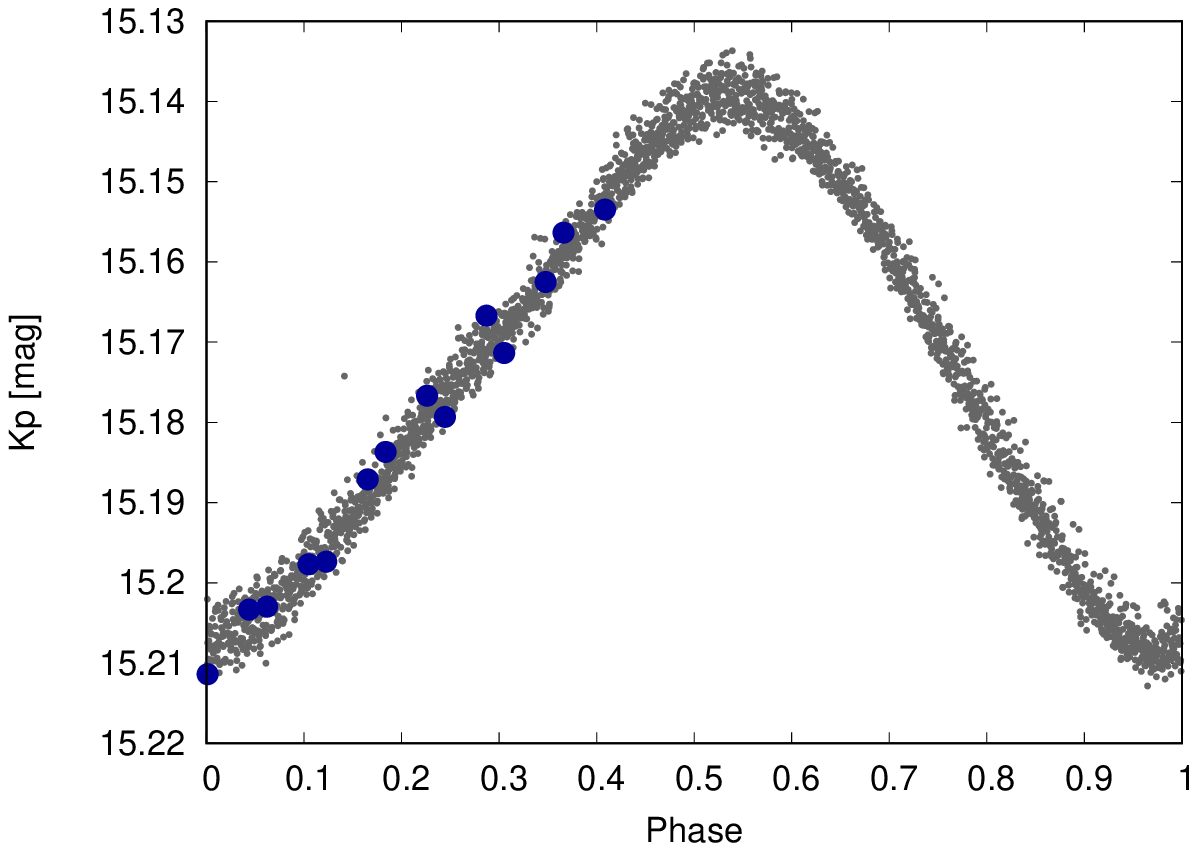}{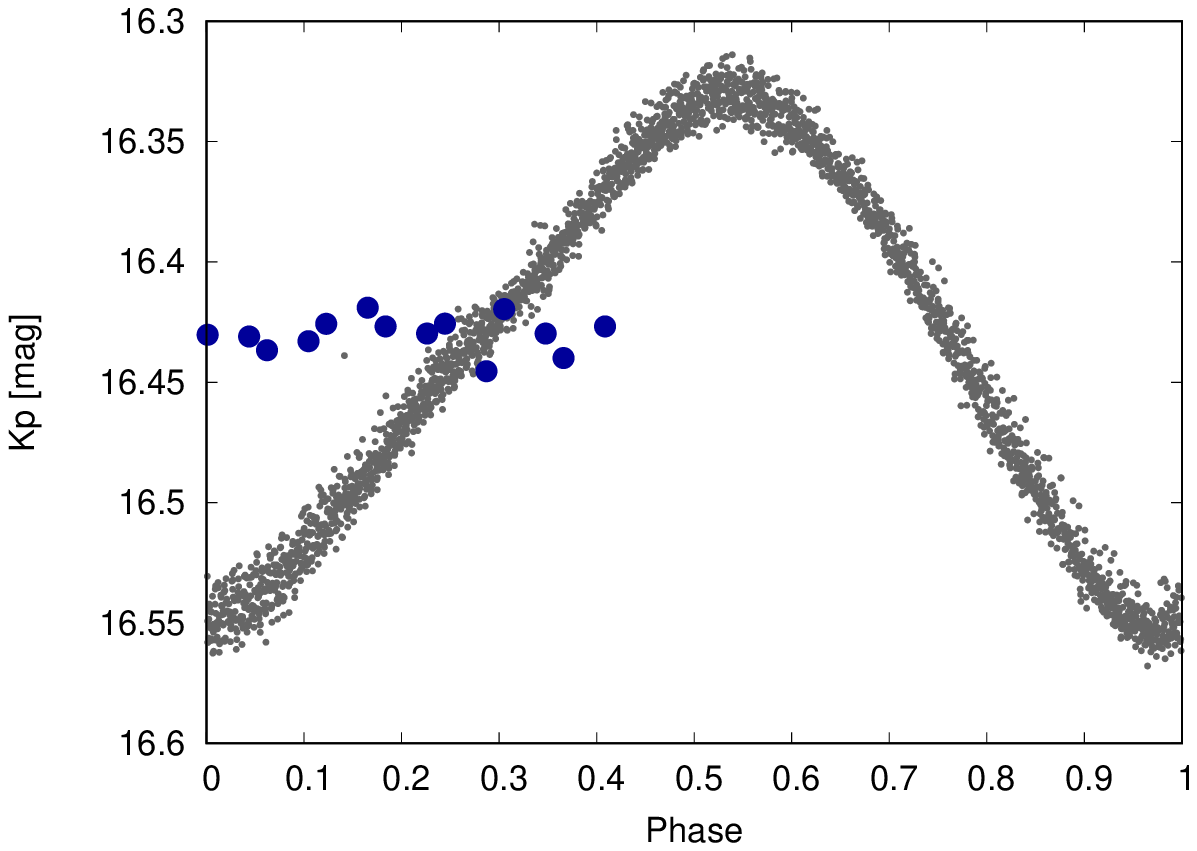}
\caption{{\em Left}: The out-of-eclipse {\em K2} observations of \sysname{} with a constant flux contribution from the secondary star subtracted and phase-folded at a period of $\starAootvperiod$\,days (small gray circles). We also show the brightness of the primary component directly measured from observations obtained during total secondary eclipses (larger blue circles), and phase-folded at the same period. {\em Right}: Same as at left, but here we subtract a constant flux contribution from the primary star (attributing the out-of-eclipse variability to the secondary star), while the larger blue circles show the secondary component brightness measured from the amount of light lost at each secondary eclipse. We conclude that the $P = \starAootvperiod$\,day signal originates on the primary star, and we interpret it as the rotation period of that component.
\label{fig:k2ootstar1}}
\end{figure*}
%% ----------------

%% ----------------
\begin{figure*}[!ht]
\plottwo{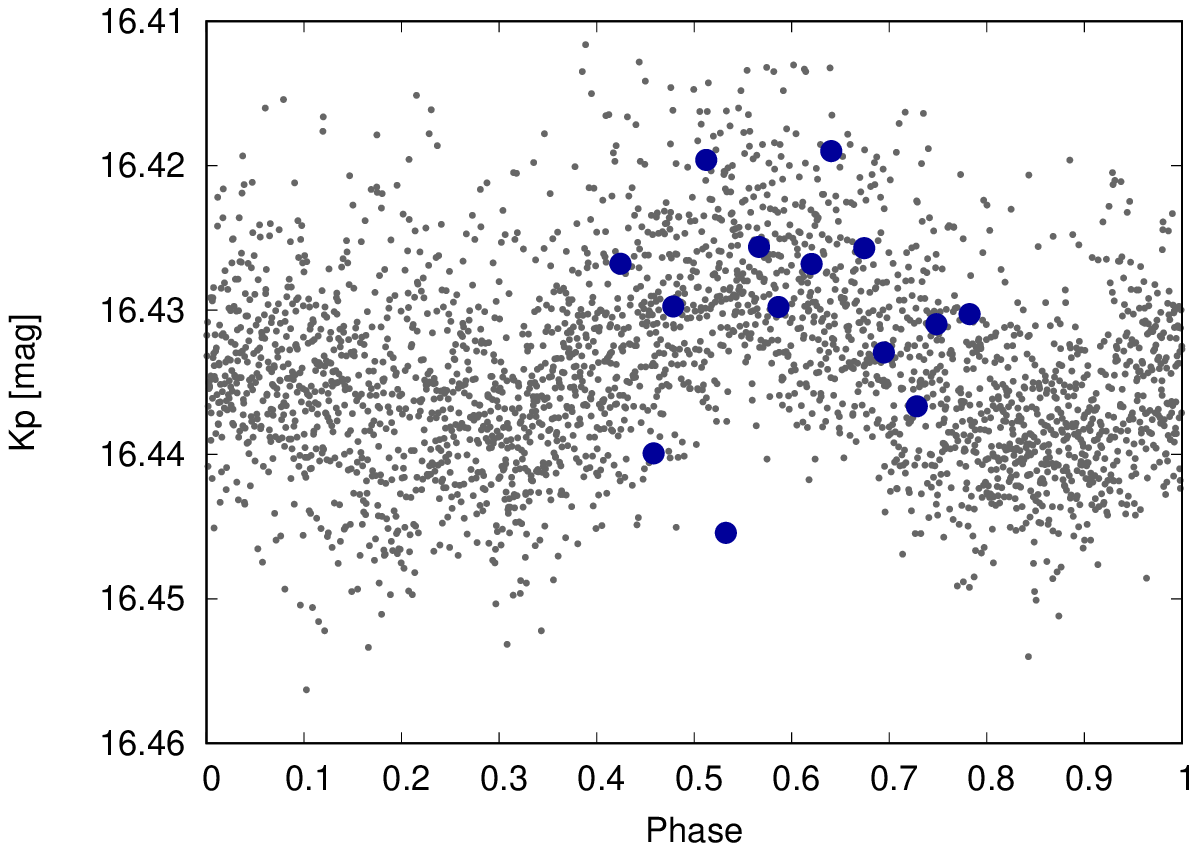}{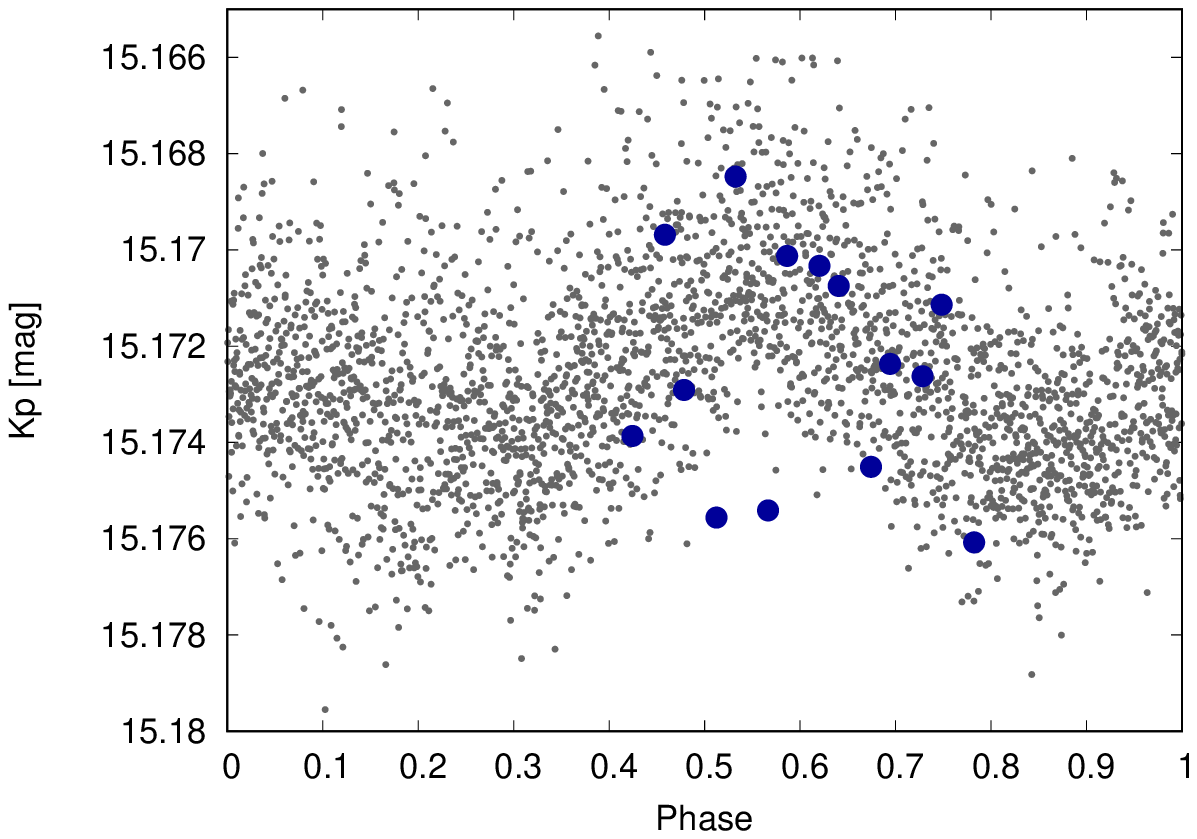}
\caption{Similar to Figure~\ref{fig:k2ootstar1}, here we show observations after subtracting the periodic signal shown in that figure from the flux of the primary star, and phase-fold the observations at a period of $\starBootvperiod$\,days. In this case the {\em left} panel shows the result when we attribute the variation to the secondary star, while the {\em right} panel shows the result when we attribute it to the primary star. In this case we cannot clearly determine which component of the binary system is giving rise to the observed variation.
\label{fig:k2ootstar2}}
\end{figure*}
%% ----------------

In Figure~\ref{fig:k2ootstar1} we show how the primary star brightness
values determined directly from the total secondary eclipse
observations phase up in excellent agreement with the larger amplitude
$P = \starAootvperiod$\,day out-of-eclipse modulation, where we have subtracted off
the average flux contribution from the secondary component in plotting
the out-of-eclipse values, and have converted everything to magnitudes
in the plot. By contrast, the secondary star brightness values
inferred from the secondary eclipse depths do not phase up at all with
this periodic out-of-eclipse variation when it is attributed to the
secondary star. We conclude that this variability arises from the primary star, and interpret the period as the photometric rotation period of this component.

Similarly in Figure~\ref{fig:k2ootstar2} we show how the secondary
star brightness values determined directly from the observed secondary
eclipse depths compare with the lower amplitude $P = \starBootvperiod$\,day
out-of-eclipse modulation, after subtracting the flux contribution
from the primary component, including the larger amplitude $P =
\starAootvperiod$\,day variation. We show the results when we attribute the
variation to the primary star, comparing the directly measured primary
star brightness (after subtracting the large amplitude $P = \starAootvperiod$\,day
variation) to the out-of-eclipse modulation after subtracting both the
large ampliutde $P = \starAootvperiod$\,day variation, and a constant flux
contribution from the secondary component. In this case the scatter in
the few observed eclipses is too large to clearly determine which
stellar component is giving rise to the $P = \starBootvperiod$\,day variability. In
carrying out the joint analysis (Section~\ref{sec:jointfit}) we model
the observations both ways, attributing the variability to the
secondary component, and attributing it to the primary component. We
find that the two scenarios are indistinguishable based on the
observations, with $\Delta \chi^2 < 1$ between the two best fit
models.

We also analyzed the HATNet and ground-based photometric follow-up
observations for stellar activity signals. We fit a harmonic signal to
the HATNet light curves from each of the four separate fields
containing \sysname{}. In each case we removed observations obtained
during eclipse before conducting the fit. Including the correction for
dilution, we find the following 95\% confidence upper-limits on the
semiamplitude of variability: HAT/G317$=0.017$\,mag,
HAT/G318$=0.026$\,mag, HAT/G365$=0.017$\,mag, HAT/G366$=0.013$\,mag.
A combined analysis of the four fields yields an upper limit of $\la 0.004$\,mag on the semiamplitude of variability at the {\em K2} period. 
We conclude that either the signal present during the {\em K2} observations was substantially lower during the time-period of the HATNet observations, and/or that the photometric variations must have a significantly lower amplitude
in $r$ than was observed by {\em K2} in the bluer {\em Kepler}
band-pass.

We also checked the HATNet light curves independently for periodic and
quasi-periodic variations using the Lomb-Scargle periodogram
\citep{lomb:1976,scargle:1982} and the discrete autocorrelation
function \citep{edelson:1988}. No significant signals are identified
by either method. When the combined HATNet light curve are searched for periodicity via a discrete Fourier transform analysis, there appears to be a peak in the power spectrum at a period of $27.87$\,days. The origin of this signal is unclear, though it may be due to modulation in the background at close to the lunar period.

%% ----------------
\begin{figure}[!ht]
\plotone{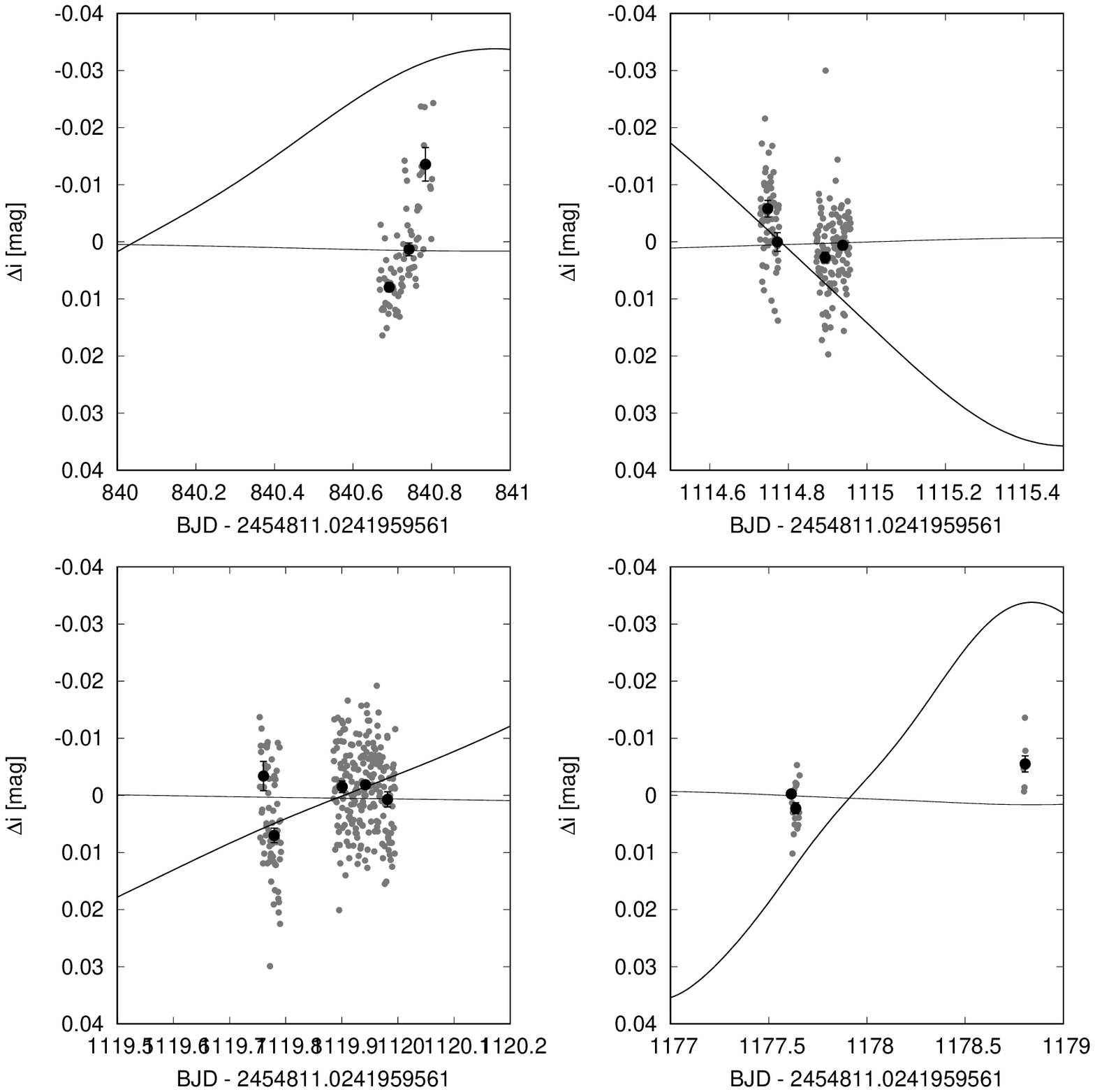}
\plotone{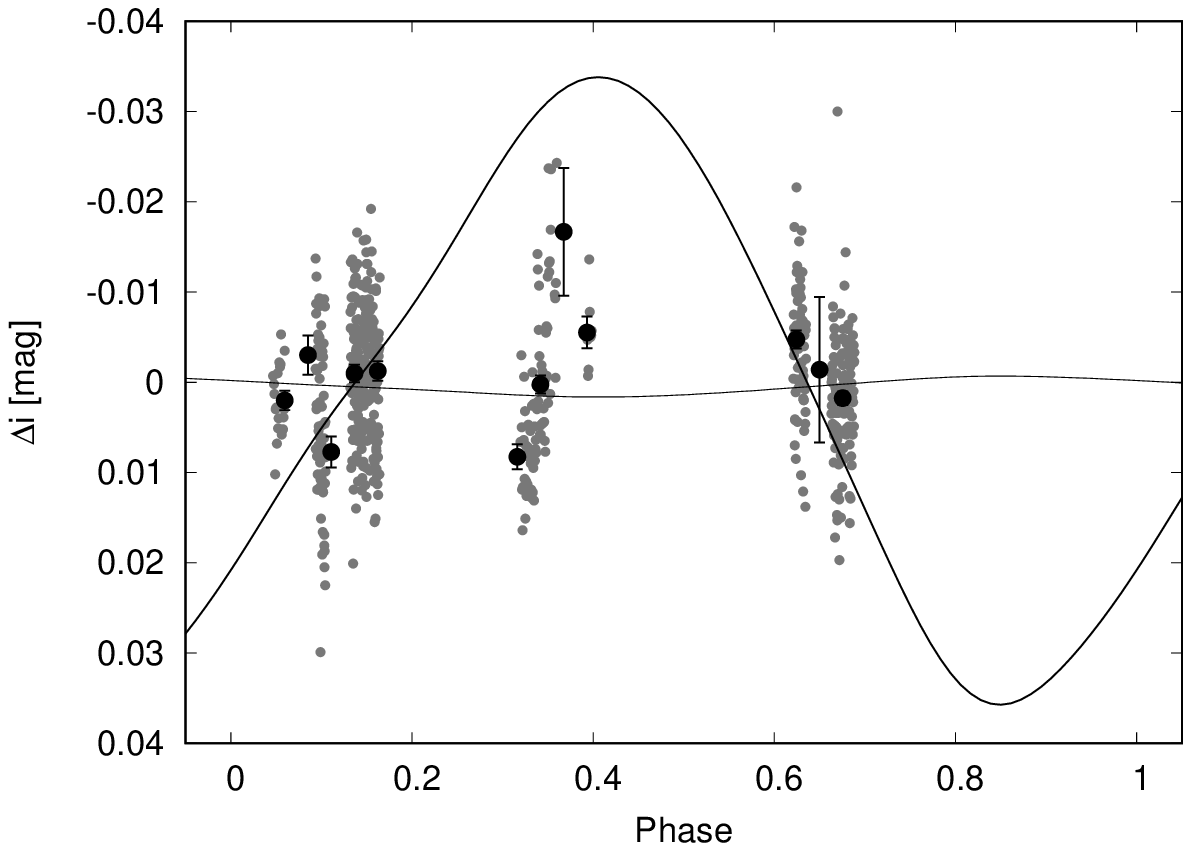}
\caption{ {\em Top:} Combined KeplerCam $i$-band light curve for \sysname{} after removing the primary and secondary eclipses. In order to better see the variability in the time-series, we split the figure into four groups of observations that were obtained near in time to each other. The light filled circles show the individual photometric measurements, while the dark filled circles show the measurements binned in time with a bin-size of 1.2\,hr. The large amplitude line is the best-fit $P = \starAootvperiod$\,day rotational signal from the {\em K2} light curve, while the lower amplitude line is the same signal with the amplitude fit to the {\em KeplerCam} observations. {\em Bottom:} Same as at top, here we phase-fold the observations at the rotational period. Overall the KeplerCam $i$-band observations, which were obtained several years before the {\em K2} data, show signficantly lower amplitude out-of-eclipse variability than was seen in the {\em Kepler} band-pass. The variability that is seen in the {\em KeplerCam} data does not phase up with the rotational period, and may be due to systematic errors in the photometry, which are fit for in modelling the individual KeplerCam eclipse observations, but which are not corrected-for in the combined light curve displayed here. We conclude that the rotational varaibility in the $i$-band has an amplitude $\la 0.01$\,mag.
\label{fig:kepcamrotation}}
\end{figure}
%% ----------------

While the combined KeplerCam $i$-band observations have much sparser
time coverage compared to the HATNet and {\em K2} observations, they
have sufficient precision to detect the photometric variability in the
$i$ band-pass. To do this we first perform a combined reduction of all
KeplerCam $i$-band observations, including out-of-eclipse observations
made on the nights of 2012-03-02 and 2012-03-03. This reduction
differs from the reduction used to make the light curves included in
our joint fit which was performed independently for each night. The
independent reduction produces higher precision light curves, but with
differing magnitude zero-points for each
night. Figure~\ref{fig:kepcamrotation} shows the out-of-eclipse
portion of the KeplerCam light curve, together with the rotational
signal as determined from the {\em K2} light curves. Overall the
KeplerCam $i$-band observations, which were obtained several years before the {\em K2} data, show significantly lower amplitude
out-of-eclipse variability than was seen in the {\em Kepler}
band-pass. The variability that is seen in the {\em KeplerCam} data
does not phase up with the rotational period, and may be due to
systematic errors in the photometry, which are fit for in modelling
the individual KeplerCam eclipse observations, but which are not
corrected-for in the combined light curve. We conclude
that the rotational varaibility in the $i$-band has an amplitude $\la
0.01$\,mag. Note that as the time separation of $\sim 1200$ days between the ground-based observations and the {\em K2} observations is much longer than the typical spot coherence time-scale of $\sim 400$\,days for an M dwarf like \sysname{}A \citep{giles:2017}, it is not surprising that a coherent signal seen in the {\em K2} data is not observed in the ground-based observations.

% ####################################################################
%% atmospheric parameters
\subsection{Joint Modeling of the Light Curves and Radial Velocity Curves}
\label{sec:jointfit}

To determine the masses and radii of the component stars of \sysname{}
we conducted a joint modeling of the RV and light curves. We used the
JKTEBOP detached eclipsing binary light curve model
\citep{southworth:2004a,southworth:2004b,popper:1981,etzel:1981,nelson:1972},
together with simple Keplerian orbits to describe the RVs. The EBOP model includes the limb darkening, gravity darkening, and ``reflection'' effects, and also accounts for the tidal distortion of the components by approximating the stars as ellipsoids. We do not include Doppler boosting in our light curve model as we estimate an amplitude of $\sim 0.1\%$ \citep[e.g.,][]{shporer:2017}, which is within the photometric noise.

To account for the starspot modulation in the light curves we followed
a method similar to that of \citet{irwin:2011}, where instead of
attempting to model individual spots on the surface of each component,
we instead assume that the full surface brightness of each component
varies following a harmonic series, and use this this to adjust the
$J_{2}/J_{1}$ surface brightness ratio input to the EBOP model at each
time step. 

The assumed Fourier series has the form:
\begin{multline}
\Delta {\rm mag}_{i,j}(t) = a_{0,i,j}\Biggl( \cos\biggl(2\pi\bigl(t/P_{i} + \phi_{0,i}\bigr)\biggr) \\ + \sum_{k=1}^{N_{\rm harm}}b_{k,i}\cos\biggl(2\pi\bigl(t(k+1)/P_{i} + \psi_{k,i} + (k+1)\phi_{0,i}\bigr)\biggr)\Biggr)
\end{multline}
and
\begin{equation}
L_{i,j}(t) = L_{0,i,j}10^{-0.4\Delta {\rm mag}_{i,j}(t)}
\end{equation}
where $L_{i,j}(t)$ is the luminosity of star $i$ (primary or
secondary) in filter $j$ at time $t$, and the average luminosity of
the star is $L_{0,i,j}$. The free parameters in the model are the
rotation period of the star $P_{i}$, the amplitude of the fundamental
mode of the harmonic series $a_{0,i,j}$ in filter $j$, the phase of
the fundamental mode $\phi_{0,i}$, and the relative amplitudes
$b_{k,i}$ and relative phases $\psi_{k,i}$ of the higher harmonics. We
adopt $N_{\rm harm} = 5$ which provides a good fit to the {\em K2}
out-of-eclipse variability. Note that parameterizing it this way,
using the relative amplitudes and phases for the higher harmonics,
ensures that the shape of the signal is the same in all bandpasses,
while the overall amplitude may vary between the bandpasses. It also
has the benefit of removing the large correlation between $\phi_{0,i}$
and $\phi_{k,i} = \psi_{k,i} + (k+1)\phi_{0,i}$ which would be present
if the phases rather than relative phases of the harmonics were used
in the fit. 

We then scale $J_{2}/J_{1}$ in filter $j$ by $L_{2,j}(t)/L_{1,j}(t)$,
and adjust the stellar luminosities and third light contribution input
into EBOP, {\em for that time-step}, as well. Note that the limb
darkening coefficients are kept fixed at each time step in this
analysis, so changes in the central surface brightness of the star
lead to proportial changes in the integrated light from that source.

Our treatment differs from \citet{irwin:2011} in two ways. The first
difference is our use of a higher order Fourier series, compared to a
simple sinusoidal modulation considered by \citet{irwin:2011}. The
higher order series is necessary in our case to represent the more
complicated shape of the {\em K2} out-of-eclipse variations seen in
\sysname{}. The second difference is in the technical details of how we apply
the time-varying surface brightness into the output light curve
model. \citet{irwin:2011} indicate that they modulate the
out-of-eclipse and eclipsed light values computed by JKTEBOP. This
yields the same result as our approach of varying the input
$J_{2}/J_{1}$ values, while also adjusting the total luminosities of
the components.

To allow for the possibility that some of the spotted regions on the
primary are not eclipsed, we added an additional harmonic series to
the flux which does not enter into the eclipse model. Namely we use:
\begin{equation}
L_{\rm tot}(t) = L_{\rm EBOP}(t) + L_{3}(t)
\end{equation}
with 
\begin{multline}
L_{3}(t) = a_{0,3,j}\Biggl( \cos\biggl(2\pi\bigl(t/P_{1} + \phi_{0,3}\bigr)\biggr) \\ + \sum_{k=1}^{N_{\rm harm}}b_{k,3}\cos\biggl(2\pi\bigl(t(k+1)/P_{1} + \psi_{k,3} + (k+1)\phi_{0,3}\bigr)\biggr)\Biggr)
\end{multline}
and using the primary star rotation period $P_{1}$.  This is similar
to the method of \citet{irwin:2011}, except in this case we allow both
the signal shape and amplitude of the uneclipsed modulation to be
independent from the shape and amplitude of the eclipsed
modulation. An additional constant third light parameter is
incorporated as well into the model.

This treatment of the spot modulation clearly is simplistic as it
ignores the detailed inhomogenous brightness distribution on the
surface. However, a more detailed modeling of the spotted surfaces of
the star is beyond the scope of this work. Moreover, as shown by
\citet{irwin:2011}, the simple model is quite flexible in describing
the light curve shape, and accurately captures the resulting
uncertainties in the estimated stellar radii. We also note that we are
assuming here that there is no evolution in time in the dominant
rotational modulation (no significant evolution is seen over the {\em
  K2} observations, but the $r$-band and $i$-band observations were
obtained many years earlier and there may have been significant spot
evolution in the intervening period). This is necessary because the
high precision $i$-band observations do not have significant
out-of-eclipse phase coverage, and we are unable to determine the
shape of the rotational signal independently from those observations.

Each KeplerCam and BOS eclipse light curve was treated as an
independent time-series in the fit, while we also fit each of the four
different HATNet light curves and the full {\em K2} light curve
independently. To account for the long integration time of {\em K2},
we evaluated the full model for the {\em K2} light curve at several
points within a 30\,min time-bin centered on the BJD time of the
observation and averaged these model flux values.

 In order to limit the amplitude of the $i$-band out-of-eclipse
 variability due to spots, we also included a time-binned version of
 the combined KeplerCam light curve shown in
 Fig.~\ref{fig:kepcamrotation}, excluding observations taken during
 the eclipses. Note that this light curve is based on a joint
 reduction of all of the KeplerCam data and thus contains information
 about the relative change in brightness between nights which is not
 included in the independently reduced KeplerCam light curves that we
 use to model the eclipses. The joint KeplerCam light curve also
 includes observations from nights where no eclipse was observed. We
 perform the time-binning to reduce the extent to which we are fitting
 the same information twice (i.e., the out-of-eclipse variations that
 occur at short time-scales). Also note that when we do not include
 this latter light curve in the fit the model converges to a very
 large $i$-band spot variability amplitude of 0.14\,mag, which is
 inconsistent with the out-of-eclipse observations. Most of the
 parameters, however, are insensitive to the inclusion of this light
 curve in the modeling. Exceptions are the $i$-band primary star limb
 darkening coefficients and the $i$-band magnitude difference between
 the primary and secondary stars, which differ by more than $2\sigma$,
 and the orbital inclination and $i$-band surface brightness ratio
 which differ by more than $1\sigma$.

We searched our own KeplerCam observations and the {\em Gaia} catalog for any resolved neighbors which might contaminate the photometry, finding no such neighbors down to $\sim 1\arcsec$ and $G \sim 20$\,mag.

Our method for optimizing the parameters in our fit and estimating the
uncertainties is similar to that used by \citet{irwin:2011}.
We used a Differential Evolution Markov Chain
\citep{terbraak:2006,eastman:2013} to explore the posterior parameter
distributions, assuming uniform priors on the adjusted variables. The
resulting parameters and 1$\sigma$ uncertainties (15.85\% to 84.15\%
confidence region) that we find for \sysname{} are listed in
Table~\ref{tab:sysparam}.  Below we discuss a few additional details
that are relevant to the modeling of this system.

\subsubsection{Trend-Filtering and Error Scaling}

We extended our physical light curve model with a model for systematic
variations due to instrumental or atmospheric effects following the
approach of \citet{bakos:2010:hat11}. The model magnitude for light
curve $j$ at time $t_{i}$ is given by:
\begin{equation}
m_j(t_i) = M_{\rm phys, j}(t_i) + \sum_{k=1}^{N_{\rm EPD, j}} a_{j,k} x_{i,j,k} + \sum_{k=1}^{N_{\rm TFA}} b_{k} y_{i,j,k}
\end{equation}
where $M_{\rm phys, j}$ is the physical model for light curve $j$,
there are $N_{\rm EPD, j}$ sets of external parameters to decorrelate
(EPD) against for light curve $j$, each of which has value
$x_{i,j,k}$, and $N_{\rm TFA}$ template light curves used to model
additional trends (this is the Trend Filtering Algorithm, or TFA, due to \citealp{kovacs:2005:TFA}). These templates have value $y_{i,j,k}$ at
time $i$ for light curve $j$. The parameters $a_{j,k}$ and $b_{k}$ are
linear coefficients which are optimized in our fit. 

For KeplerCam and BOS, the external parameters that we decorrelate
against include the hour angle of the observations (to second order)
and three parameters describing the shape of the point spread function
(each to first order). We use a set of 20 TFA template light curves
for the KeplerCam observations, and a separate set of five template light
curves for the BOS observations. 

For {\em K2} we use a set of 12 vectors which define a 6-harmonic
Fourier series with a period equal to the time-span of the data. This
accounts for any additional variation in the background or other long
time-scale systematics in the EVEREST {\em K2} light curve. While in
principle the pixel-level decorrelation of the variations due to the
spacecraft roll should be applied simultaneously to the fitting of
this large amplitude variable star, the development of a method to
carry out such an analysis is beyond the scope of this paper. We note
that the consistency of the eclipsing system model parameters
estimated solely from the {\em K2} observations with those estimated
solely from the {\em KeplerCam} observations indicates that the {\em
  K2} light curves being analyzed here have not been significantly
distorted by the detrending process. 

We include the HATNet observations in our modeling as well, but in
this case do not include the EPD or TFA terms. For the HATNet
observations we include a dilution factor to account for blending from
poorly resolved neighbors and/or effective dilution due to application
of TFA {\em before} fitting a model.

Because the photometric and RV uncertainties are underestimated, we
include scaling factors applied to the errors. These are varied in the
fit following the method of \citet{gregory:2005}. We note that the
most likely values yield a reduced $\chi^2$ of one, but allowing these
to vary rather than adopting the optimal values ensures that the
uncertainty in these parameters also contributes to the uncertainties
in the other physical parameters.

\subsubsection{Limb Darkening}

For the HATNet $r$-band observations we assume a linear limb darkening
law and fix the coefficients to the values given in
Table~\ref{tab:sysparam}, which are taken from the
\citet{claret:2004} tabulations. For the higher-precision {\em K2} and $i$-band
observations we fit for the limb darkening coefficients. We try both a linear limb darkening law, and a square-root law. For the square-root law we use parameters $c^{\prime}_{1}$ and $c^{\prime}_{2}$, which are related to the traditional parameters $c_{1}$ and $c_{2}$ by:
\begin{eqnarray}
c^{\prime}_{1} & = & c_{1} + 2c_{2}/3 \\
c^{\prime}_{2} & = & c_{1} - 3c_{2}/2 
\end{eqnarray}
for a limb darkening law of the form
\begin{equation}
I_{\mu} = I_{0}(1 - c_{1}(1 - \mu) - c_{2}(1 - \sqrt{\mu}))
\end{equation}
with $\mu = \cos(\theta)$, where $\theta$ is the angle between the
line of sight and the normal to the stellar surface. This
parameterization has the advantage that $c^{\prime}_{1}$ and
$c^{\prime}_{2}$ are not strongly correlated, whereas $c_{1}$ and
$c_{2}$ are. When fitting the square-root law we find that
$c^{\prime}_{1}$ is much more tightly constrained than is
$c^{\prime}_{2}$, for both the {\em K2} and ground-based $i$-band
observations. We also find that there is no significant difference in
$\chi^2$ between the two classes of models, and that differences in
the estimated physical parameters of the binary are unaffected by the
choice of limb darkening law. For this reason our adopted solution is
for the linear limb-darkening law, though we list the fitted
parameters for the square-root law as well. There is no significant
difference in the other parameters between the two classes of model.

\subsubsection{Light Travel-Time Effect}

There are two light travel-time effects that are of potential
significance to the analysis of this system. These are in addition to
the standard barycentric corrections that are applied to the times of
observation and RVs to account for the motion of the Earth about the
center of the Solar System. The first effect is due to the orbits of
the two components of \sysname{} about their own center of mass \citep[see][]{kaplan:2010,kaplan:2014}. The
orbit of the secondary star has a larger semi-major axis about the
center of mass than does that of the primary. As a result, primary
eclipses are observed to occur slightly earlier, and secondary
eclipses slightly later, than would be predicted if the light travel
time were neglected. The observed time difference between primary and
secondary eclipses places a strong constraint on $e \cos \omega$, so
that neglecting the light travel-time may lead to an incorrect
eccentricity measurement. Based on the system parameters determined
for \sysname{}, we expect a correction of $\sim 10$\,s to the
difference between the times of primary and secondary eclipses due to
this effect. Since this is comparable to the precision with which this
time difference is constrained based on our observations (our final
uncertainty on T$_{c}$ is $\sim 2$\,s) we cannot neglect the
effect. We discuss in detail our procedure for accounting for it in
Appendix~\ref{sec:appendixltt}.

The second light travel-time effect of potential significance is a
small correction to the orbital period due to the recessional velocity
of the system. Based on the TRES RVs we measure a systemic velocity of
$\RVgammaBcorr$\,\kms\ which means that \sysname{} recedes from the Solar
System by 29~light seconds every orbital cycle. Correcting for this
effect, the true orbital period of the system is $P =
\periodlttcorr$\,d rather than the observed value $P = \period$\,d. This
0.01\% correction to the period results in proportional corrections to the
masses, radii and semimajor axis. As this is approximately two orders of
magnitude smaller than the uncertainties on these same parameters,
which are currently dominated by uncertainties in the RV
semiamplitudes, we do not apply this correction to the parameters
listed in Table~\ref{tab:sysparam}.

%% --------------------------------------------------------------------
%% Table of astrometric measurements
%%

\ifthenelse{\boolean{emulateapj}}{
    \begin{deluxetable*}{lclc}
}{
    \begin{deluxetable}{lclc}
}
\tablewidth{0pc}
\tabletypesize{\scriptsize}
\tablecaption{
    System parameters for \sysname{}
    \label{tab:sysparam}
}
\tablehead{
    \colhead~~~~~~~~Parameter~~~~~~~~{}  &
    \colhead{Value}                      &              
    \colhead~~~~~~~~Parameter~~~~~~~~{}  &
    \colhead{Value}
}
\startdata
& & & \\
\multicolumn{2}{l}{Light Curve Parameters}   &
\multicolumn{2}{l}{Atmospheric Parameters} \\
& & & \\
~~~$P$ (days) \dotfill                       & $\period$ &
~~~Primary Spectral Type\tablenotemark{e} \dotfill  & $\spectypestarA$ \\
~~~$T_c$ (${\rm BJD}$)\tablenotemark{a,b} \dotfill  & $\jdend$ &
~~~Secondary Spectral Type \dotfill  & $\spectypestarB$ \\
~~~$b$\tablenotemark{c} \dotfill                  & $\bimpact$ &
~~~T$_{\rm eff,A}$\tablenotemark{f} \dotfill  & $\teffstarA$ \\
~~~$i$ (deg) \dotfill                             & $\inclination$ &
~~~T$_{\rm eff,B}$ \dotfill  & $\teffstarB$ \\
~~~$R_{B}/R_{A}$ \dotfill                               & $\rBoverrA$ &
~~~[Fe/H]$_{\rm eff,A}$\tablenotemark{g} \dotfill  & $\fehstarA$ \\
~~~$(R_{A}+R_{B})/a$ \dotfill                           & $\rAplusrBovera$ &
~~~[Fe/H]$_{\rm eff,B}$ \dotfill  & $\fehstarB$ \\
~~~$J_{B}/J_{A}$ (\band{r}) \dotfill                    & $\jBoverjArfilter$ &
 & \\
~~~$J_{B}/J_{A}$ (\band{i}) \dotfill                    & $\jBoverjAifilter$ &
\multicolumn{2}{l}{Orbital Parameters}  \\
~~~$J_{B}/J_{A}$ (\band{Kp}) \dotfill                    & $\jBoverjApfilter$ &
& \\
& &
~~~$K_{A}$ (\kms) \dotfill                           & $\RVKA$ \\
\multicolumn{2}{l}{Assumed Limb Darkening Coefficients\tablenotemark{b}} &
~~~$K_{B}$ (\kms) \dotfill                           & $\RVKB$ \\
& & 
~~~$\sqrt{e}\cos\omega$ \dotfill                  & $\kRVeccen$ \\
\multicolumn{2}{l}{~~~Linear Law} &
~~~$\sqrt{e}\sin\omega$ \dotfill                  & $\hRVeccen$ \\
~~~Primary $\mu$ $r$-band \dotfill & $\starAldcoeffArfilter$ &
~~~$e$ \dotfill                                   & $\RVeccen$ \\
~~~Secondary $\mu$ $r$-band \dotfill & $\starBldcoeffArfilter$ &
~~~$\omega$ \dotfill                              & $\RVomega$ \\
& &
~~~$\gamma$ (\kms)\tablenotemark{h} \dotfill                       & $\RVgammaBcorr$ \\
\multicolumn{2}{l}{Fitted Limb Darkening Coefficients} &
& \\
& &
\multicolumn{2}{l}{Physical Parameters} \\
\multicolumn{2}{l}{~~~Linear Law} &
& \\
~~~Primary $\mu$ $i$-band \dotfill & $\starAldcoeffAifilter$ &
~~~$M_{A}$ (\msunnom) \dotfill                & $\massstarA$ \\
~~~Secondary $\mu$ $i$-band \dotfill & $\starBldcoeffAifilter$ &
~~~$M_{B}$ (\msunnom) \dotfill                & $\massstarB$ \\
~~~Primary $\mu$ $Kp$-band \dotfill & $\starAldcoeffApfilter$ &
~~~$R_{A}$ (\rsunnom) \dotfill                & $\radiusstarA$ \\
~~~Secondary $\mu$ $Kp$-band \dotfill & $\starBldcoeffApfilter$ &
~~~$R_{B}$ (\rsunnom) \dotfill                & $\radiusstarB$ \\
\multicolumn{2}{l}{~~~Square-root Law} &
~~~$\logg_{A}$ (cgs) \dotfill            & $\loggstarA$ \\
~~~Primary $c_{1}+2c_{2}/3$ $i$-band \dotfill & $\starAldcoeffAadjsqrtifilter$ &
~~~$\logg_{B}$ (cgs) \dotfill            & $\loggstarB$ \\
~~~Primary $c_{1}-3c_{2}/2$ $i$-band \dotfill & $\starAldcoeffBadjsqrtifilter$ &
~~~$\rho_{A}$ (\gcmc) \dotfill            & $\densitystarA$ \\
~~~Secondary $c_{1}+2c_{2}/3$ $i$-band \dotfill & $\starBldcoeffAadjsqrtifilter$ &
~~~$\rho_{B}$ (\gcmc) \dotfill            & $\densitystarB$ \\
~~~Secondary $c_{1}-3c_{2}/2$ $i$-band \dotfill & $\starBldcoeffBadjsqrtifilter$ &
~~~$\log L_{A}$ ($\loglsunnom$)\tablenotemark{i} \dotfill & $\logluminositystarA$ \\
~~~Primary $c_{1}+2c_{2}/3$ $Kp$-band \dotfill & $\starAldcoeffAadjsqrtpfilter$ &
~~~$\log L_{B}$ ($\loglsunnom$) \dotfill & $\logluminositystarB$ \\
~~~Primary $c_{1}-3c_{2}/2$ $Kp$-band \dotfill & $\starAldcoeffBadjsqrtpfilter$ &
~~~$a$ (AU) \dotfill                    & $\semimajoraxis$ \\
~~~Secondary $c_{1}+2c_{2}/3$ $Kp$-band \dotfill & $\starBldcoeffAadjsqrtpfilter$ &
~~~$\Delta r$ (mag) \dotfill            & $\deltamagrfilter$ \\
~~~Secondary $c_{1}-3c_{2}/2$ $Kp$-band \dotfill & $\starBldcoeffBadjsqrtpfilter$ &
~~~$\Delta i$ (mag) \dotfill            & $\deltamagifilter$ \\
& &
~~~$\Delta Kp$ (mag) \dotfill            & $\deltamagpfilter$ \\
& &
~~~$d$ (pc)\tablenotemark{j} \dotfill            & $\distance$ \\
\enddata
\tablenotetext{}{We adopt the Nominal Solar conversion constants from IAU 2015 Resolution B3 as listed in \citet{prsa:2016}, using the suggested notation \msunnom, \rsunnom, and \lsunnom\ for these constants. To calculate the bulk density values listed here we assume the CODATA 2014 value of the gravitational constant \citep{mohr:2016}: $G = (6.67408 \pm 0.00031) \times 10^{-11}$\,m$^{3}$\,kg$^{-1}$\,s$^{-2}$.}
\tablenotetext{a}{Times given here, and throughout the paper, are in Barycentric Julian Date (BJD) on the TDB system. Time conversions from UTC to BJD-TDB for the ground-based observations are performed using {\sc Vartools} \citep{hartman:2016:vartools}.}
\tablenotetext{b}{Epoch of primary eclipse.}
\tablenotetext{c}{Impact parameter during primary eclipse relative to the sum of the radii of the two stars.}
\tablenotetext{d}{We fixed the limb darkening coefficients to linear law values from the \citet{claret:2004} tabulations for the HATNet $r$-band light curves. For the higher precision follow-up $i$-band light curves, we allowed the coefficients to vary, trying both a linear and a square-root law. Parameters adopted in the table are for the linear law.}
\tablenotetext{e}{Based on the NIR H$_{2}$O$-$K spectral index using the calibration by \citet{rojasayala:2012}.}
\tablenotetext{f}{Based on cross-correlating NIR spectra against BT~Settl synthetic templates.}
\tablenotetext{g}{Based on averaging the $H$ and $K$-band spectral index-based metallicities using the \citet{terrien:2012} calibration.}
\tablenotetext{h}{$\gamma$ is the
  systemic radial velocity relative to the solar system barycenter
  assuming a RV for Barnard's star of $-110.416\pm0.180$\,\kms\ from
  \citet{chubak:2012}. This velocity has not been corrected for
  gravitational redshifts or convective blueshifts. The radial
  velocity relative to Barnard's star is $\gamma_{\rm rel} =
  \RVgammaB$\,\kms.  
}
\tablenotetext{i}{Calculated from the measured stellar radius and spectroscopically determined effective temperature; assumes a solar effective temperature of T$_{\rm eff,\odot} = \teffsunval$\,K.}
\tablenotetext{j}{Based on the \citet{delfosse:2000} Mass--M$_{K}$ relation for M dwarfs, together with the measured masses of the component stars and the 2MASS $K_{S}$ magnitude of the system; this leads to a lower uncertainty on the distance than using the empirically measured luminosities.}
\ifthenelse{\boolean{emulateapj}}{
    \end{deluxetable*}
}{
    \end{deluxetable}
}
%% --------------------------------------------------------------------

% ####################################################################
\subsubsection{Errors Due to Unmodelled Time-Correlated Variations}
\label{sec:syserr}

The light curve residuals in Figures~\ref{fig:lcpri}--\ref{fig:RVs}
show variations that appear to be correlated in time. These variations
may be due to spots, stellar flares, or instrumental artifacts that
are not accounted for by the trend-filtering model.  In recent years
it has become common practice to account for correlated noise using
the Gaussian Process Regression method, but an inspection of the
moving mean and standard deviation of the light curve residuals shows
that the noise in this case is non-stationary, with sharp variations
in the residuals that are likely due to flares and other stellar
activity phenomena, and would not be well described with commonly used
Gaussian Process kernels. In order to estimate the contributions to
the parameter uncertainties due to correlated variations that are not
accounted for in the model, we carried out a prayer-bead analysis as
follows:

\begin{enumerate}
\item For each light curve and RV curve we subtract the best-fit model to produce residual light curves and RV curves.
\item For each of the residual light curves we determine the time boundaries for all primary and secondary eclipses observed in that light curve. For each eclipse covered in the light curve we then choose at random one of the observed time steps $t_{i}$ in that eclipse and set the residual $r_{0}$, and the observational uncertainty $\sigma_{0}$, for the first time step in that eclipse $t_{0}$ equal to the residual $r_{i}$, and observational uncertainty $\sigma_{i}$, from the chosen time step. We then set $r_{1} = r_{i+1}$, $\sigma_{1} = \sigma_{i+1}$, $r_{2} = r_{i+2}$, $\sigma_{2} = \sigma_{i+2}$ and so on, cycling back to $r_{j} = r_{0}$, $\sigma_{j} = \sigma_{0}$ when $i+j = N$, with $N$ being the number of points in the eclipse. We perform a similar random shift of the residuals for all of the out-of-eclipse observations in that light curve. We treat the eclipses and out-of-eclipse observations independently as we found that systematic variations in the residuals tend to be more pronounced in the eclipses, with differences between the primary and secondary eclipses. While this may introduce an artificial discontinuity in the correlation structure at the edge of the eclipse, treating the eclipses and out-of-eclipse data in a single prayer-bead will tend to produce simulated observations with smaller in-eclipse residuals on average than the actual data, leading to underestimated parameter uncertainties.
\item For the primary and secondary star RV curves we randomly shift the residuals in time in a similar manner. We treat the two stars independently.
\item We then add the best-fit models that were subtracted from the data back to the shifted residual curves and fit a full model to the simulated data using the downhill simplex algorithm to find the maximum likelihood solution.
\end{enumerate}
We repeat this simulation 1000 times. We then take
the standard deviation of the resulting set of parameter values to estimate the
systematic uncertainties due to unmodelled systematic variations in the
light curves and RV measurements. These errors are then added in quadrature to the $1\sigma$ uncertainties based on the MCMC analysis to determine the final parameter uncertainties that are listed in Table~\ref{tab:sysparam}.

% ####################################################################
%% atmospheric parameters
\subsection{Atmospheric Parameters}
\label{sec:nirmodel}

We use the near-IR spectra described in Section~\ref{sec:nirspec} to
determine the metallicity and effective temperature of the components
of \sysname{}. We apply two methods to determine these parameters:
using empirically calibrated spectral indices, and cross-correlating
against theoretical spectral templates. We discuss the results from
each method in turn below, with additional details provided in
Appendices~\ref{sec:spectralindices} and~\ref{sec:ccf}.

\subsubsection{Spectral Indices}

\citet[][hereafter R10 and R12]{rojasayala:2010,rojasayala:2012} and
\citet[][hereafter T12]{terrien:2012} have determined empirical
relations to measure the spectral types and metallicities of M~dwarfs
using K- and H-band spectral indices. We follow these methods, as
described in detail in Appendix~\ref{sec:spectralindices}, to
determine spectral types for the primary and secondary stars of,
\spectypestarA{} and \spectypestarB{}, respectively. These are based
on our disentangled spectra. The spectral type determined for the
primary star based on the three spectra obtained during total eclipse
is \spectypestarAtotal{}. Using the relation between spectral type and
effective temperature given in \citet{bessell:1991}, we estimate
effective temperatures of the component stars of T$_{\rm
  eff,A}=\teffstarbesselA$\,K, and T$_{\rm
  eff,B}=\teffstarbesselB$\,K. We also determined separate
metallicities for the primary and secondary star
[Fe/H]$_{A}=\fehstarA$ and [Fe/H]$_{B}=\fehstarB$, respectively,
which are consistent to within 2$\sigma$. Assuming both components
have the same metallicity, we take the weighted mean of the individual
metallicities to estimate a system metallicity of
[Fe/H]$=\fehsystem$. Systematic uncertainties are included in all of
the parameter uncertainties listed here.

\subsubsection{Cross-Correlation Against Theoretical Spectral Templates}

As an alternative method to determine the stellar atmospheric
parameters we compare our disentangled NIR spectra to model spectra
from the BT-Settl grid \citep{allard:2011} computed using the
\citet{asplund:2009} solar abundances. The details of our method are
described in Appendix~\ref{sec:ccf}. For \sysname{}A and \sysname{}B
we find effective temperatures of T$_{\rm eff,A}=\teffstarbtsettlA$\,K
and T$_{\rm eff,B}=\teffstarbtsettlB$\,K, respectively, from
cross-correlation, and metallicities of [Fe/H]$_{\rm
  A}=\fehstarbtsettlA$ and [Fe/H]$_{\rm B}=\fehstarbtsettlB$,
respectively. Combining the metallicities of the primary and secondary
components yields a metallicity for the system of
[Fe/H]$=\fehsystembtsettl$, which is consistent with the system
metallicity determined from the spectral indices.

The effective temperatures may be combined with the measured stellar
radii to determine the stellar luminosities and bolometric magnitudes
of: L$_{A}=\luminositystarA$\,\lsunnom, L$_{B}=\luminositystarB$\,\lsunnom,
M$_{\rm bol,A}=\bolmagA$\,mag, and M$_{\rm bol,B}=\bolmagB$\,mag,
where we assume a solar effective temperature of T$_{\rm eff,\odot} =
\teffsunval$\,K, and bolometric magnitude of M$_{\rm bol,\odot} =
\mbolsunval$\,mag.

% ####################################################################
%% space velocity
\subsection{Space Velocity}

\begin{figure*}[!ht]
\plotone{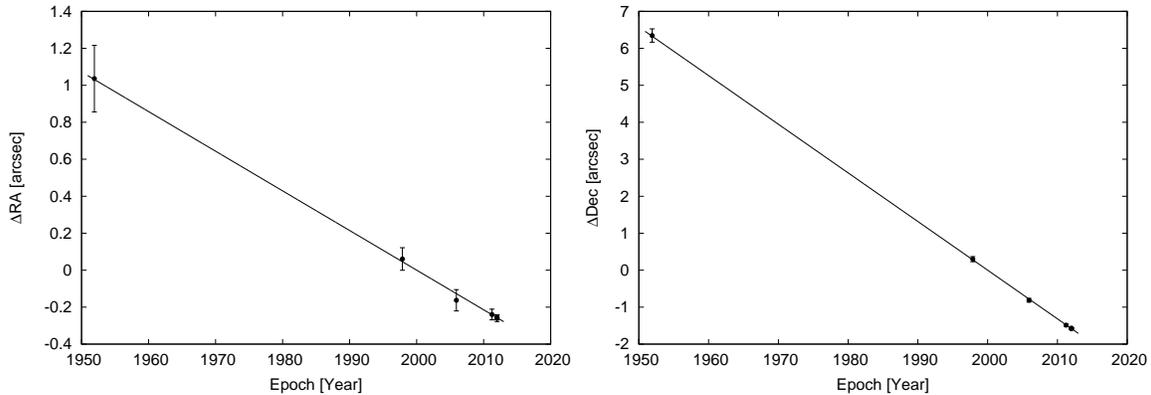}
\caption{
    Right Ascension (Left) and Declination (Right) vs time showing the
    proper motion of \sysname{}. The data is taken from \reftabl{astrom}.
\label{fig:ppm}}
\end{figure*}

\sysname{} does not have a reliable proper motion measurement listed
in any of the available catalogs. We therefore determine its proper
motion using our KeplerCam observations from 2011 and 2012 together
with archival measurements given in the USNO~A2.0 catalog \citep[circa
  1951;][]{monet:1998}, the 2MASS catalog
\citep[1997;][]{skrutskie:2006}, and the SDSS~DR7 primary photometric
catalog \citep[2005;][]{sdss:dr7}. The epochs and measured positions
(in the J2000 equinox) are collected in
\reftabl{astrom}. \reffigl{ppm} shows the change in RA and Dec over
time, together with our best-fit model. From these observations we
measure a proper motion of $\mu_{\rm RA} = \ppmra$\,mas\,yr$^{-1}$ and
$\mu_{\rm Dec} = \ppmdec$\,mas\,yr$^{-1}$, where we follow the
convention that the change in the RA coordinate is given by $\mu_{\rm
  RA}/\cos({\rm Dec})$. We also determine a reference position of RA$
= \sysra$, Dec$ = \sysdec$ (Equinox J2000, Epoch J2000). We note that
the observational precision is insufficient to determine the
parallax. When we try to fit for the parallax we find $\pi = -2 \pm
20$\,mas, whereas the expected value is $\pi = \parallax$\,mas. We
therefore fix the parallax to zero in fitting for the proper motion. We expect the parallax of this system to be determined in the next year with the release of Gaia DR2.

We use the measured proper motion, together with the $\gamma$ RV and
distance from our best-fit model, to determine the UVW space motion of
\sysname{} following \citet{johnson:1987}. We find $U = \velU$\,\kms,
$V = \velV$\,\kms, and $W = \velW$\,\kms, with the convention that $U$
increases toward the Galactic center. Based on the SDSS DR7 kinematic
model for the Galaxy \citep{bond:2010}, 99.5\% of stars with the
Galactic position and velocity of \sysname{} are members of the
disk. The classification of \sysname{} as a member of the Galactic
disk is consistent with the high metallicity measured for the system.

%% --------------------------------------------------------------------
%% Table of astrometric measurements
%%
\ifthenelse{\boolean{emulateapj}}{
    \begin{deluxetable*}{lrrrrr}
}{
    \begin{deluxetable}{lrrrrr}
}
\tablewidth{0pc}
\tabletypesize{\scriptsize}
\tablecaption{
    Astrometric measurements of \sysname{}
    \label{tab:astrom}
}
\tablehead{
    \colhead{JD}  &
    \colhead{RA (deg, J2000)} &
    \colhead{Dec (deg, J2000)} &
    \colhead{RA Error (mas)} &
    \colhead{Dec Error (mas)} &
    \colhead{Source}
}
\startdata
$2433979$ & $132.637612$ & $12.141664$ & $180$ & $180$ & USNO~A2.0 \\
$2450767.9869$ & $132.637341$ & $12.139984$ & $61$ & $68$ & 2MASS \\
$2453710.934$ & $132.637279$ & $12.139675$ & $58$ & $50$ & SDSS \\
$2455651.72551$ & $132.6372576$ & $12.1394879$ & $29$ & $27$ & KeplerCam \\
$2455925.869055$ & $132.6372521$ & $12.1394607$ & $18$ & $20$ & KeplerCam \\
$2455930.89952$ & $132.6372534$ & $12.1394657$ & $14$ & $18$ & KeplerCam \\
\enddata
\ifthenelse{\boolean{emulateapj}}{
    \end{deluxetable*}
}{
    \end{deluxetable}
}
%% --------------------------------------------------------------------

%% ----------------
%% ----------------

\section{Discussion}

\subsection{Significance of Total Eclipses}

The presence of total eclipses in this system provides multiple
benefits. In addition to facilitating the disentanglement of the
spectra for both component stars (Section~\ref{sec:nirspec}), it also
allows many of the light curve parameters to be measured with
significantly better precision than would be possible for a grazing
system. Although a thorough demonstration of this is beyond the scope of this paper, we carried out a few tests based on a preliminary analysis that considered only the ground-based observations (this work was done prior to the {\em K2} observations).  We injected simulated grazing eclipses into the residual light curves from our
best-fit model. For the grazing model we adopted the parameters for
our best-fit model of \sysname{}, except we set the impact parameter
to $b = 0.3$. We then used the DEMC method to fit a model to the
simulated data and to determine the resulting parameter
uncertainties. We found that for the simulated grazing system
$R_{B}/R_{A}$, $(R_{A}+R_{B})/a$, and $i$ had statistical uncertainties that were $4.7$, $3.6$ and $2.1$ times larger than the statistical uncertainties for these same parameters when fitting the actual system with
total eclipses. 
We also found that the limb darkening coefficients for the primary
star could be determined to much better precision when total eclipses
are present. Assuming a linear limb darkening law, the statistical
uncertainty on the limb darkening coefficient for the primary and
secondary stars were $3.6$ and $1.2$ times higher for the grazing
system than for the total eclipsing system.  The only parameter which
we found would be determined with better precision in the grazing
system is $J_{B}/J_{A}$ in the $i$-band, which has a precision for the
grazing system that is $0.90$ times that of the precision for the
total eclipsing system.

While including the {\em K2} observations and accounting for the starspot modulation would affect the relative uncertainties from a grazing vs.\ total-eclipsing system, we still expect that the total eclilpsing system would generally have lower uncertainties than the grazing system.

\subsection{Asynchronous Rotation}

We find that the photometric rotation period of the primary star,
$P_{\rm rot,A} = \starAootvperiod$\,d, is slightly longer than the
orbital period of the system, $P_{\rm orb} = \period$\,d. A similar
slight difference from synchronous rotation was also observed for at
least one of the M dwarf components in the binary LP~661-13 discovered
by \citet{dittmann:2017}. More generally, \citet{balaji:2015} analyzed
a sample of 414 short-period, near-contact and contact binaries
observed by {\em Kepler}, and found that at least 50\% of these
systems exhibit star-spot rotation that is not exactly synchronized
with the orbital periods. One possible explanation for the close, but
not exact, synchronization is differential rotation. As shown by
\citet{scharlemann:1982}, the tidal forces that lead to sychronization
are insufficient to suppress differential rotation within the
star. Stars will have a co-rotating latitude, with rotation that is
slower than the orbital period above that latitude and rotation that
is faster than the orbital period below that latitude. The
observations by \citet{balaji:2015} appear to be consistent with this
picture. They find that the differences between the observed spin
periods of close binary components and the orbital periods are
consistent with being the result of modest differential rotation on
these stars, with a differential rotation parameter that is lower on
average by a factor of $\sim 3$ compared to what is found for isolated
stars. For our observations of \sysname{}A, the relative difference in
frequency between the rotation and orbit $\alpha = (\Omega_{\rm orb} -
\Omega_{\rm rot})/\Omega_{\rm orb} = 0.02$ is well within the range of
$\alpha = (\Omega_{\rm eq} - \Omega_{\rm pole})/\Omega_{\rm eq}$ found
for isolated {\em Kepler} stars by \citet{reinhold:2013}. 

In addition to differential rotation, another potential source of
asychronous rotation is a magnetized wind carrying angular momentum
away from the system. As discussed by \citet{keppens:1997}, magnetized
winds will cause main sequence stars in close binaries to have
rotation periods that are slightly slower than the orbital period. In
effect the wind applies a torque which acts to spin down the star and
may balance the tidal torque which decreases as the system approaches synchronization. The amplitude of this effect depends on
the wind strength, magnetic field strength, their dependence on
rotation period, and the tidal dissipation rate, none of which are
well determined for M dwarf stars.

Interestingly the orbital period of the system is very close to
  the mean of the two additionals periods $P = \starAootvperiod$\,d
  and $P = \starBootvperiod$\,d identified in the {\em K2} light
  curve. As shown in Section~\ref{sec:stellarrotation} we cannot
  determine unambiguously whether the shorter period signal $P =
  \starBootvperiod$\,d arises on the primary or secondary star. If it
  arises on the primary star, then the existence of two rotation
  signals from this star with periods spanning the orbital period
  would be consistent with the differential rotation scenario, where
  the star has two active latitudes above and below the co-rotation
  latitude.

\subsection{Comparison with Other M Dwarf Systems and Theoretical Models}
\label{sec:dartmouthcomparison}

%% ----------------
\begin{figure}[!ht]
\epsscale{1.15} \plotone{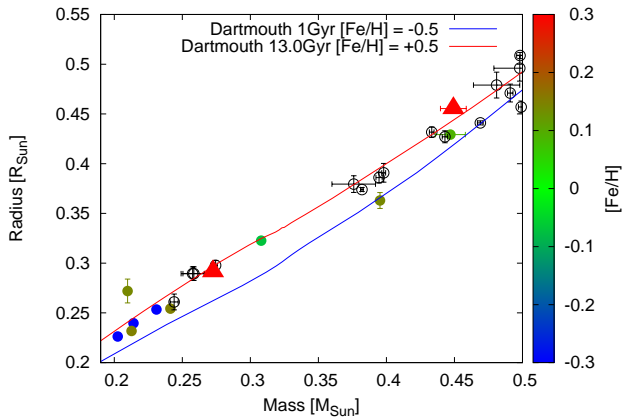}
\caption{ Mass--Radius diagram for M dwarfs with $0.2\,\msunnom < M <
  0.5\,\msunnom$ and masses and radii determined to better than 5\%
  precision. The color-scale of the points indicates the metallicity, if known. Large filled triangles show the components of \sysname{}, smaller circles show other M dwarfs with parameters given
  in Table~\ref{tab:litmebs}. Black open circles are used for systems without a measured metallicity. The lines show theoretical mass--radius
  relations from the Dartmouth
  \citep{dotter:2008} models. Most objects fall between the 1\,Gyr,
        [Fe/H]$=-0.5$ and the 13\,Gyr, [Fe/H]$=+0.5$ Dartmouth models,
        indicating that these models can explain the observations if
        the stars are old and/or metal rich.
\label{fig:massrad}}
\end{figure}
%% ----------------

%% ----------------
\begin{figure}[!ht]
\epsscale{1.15}
\plotone{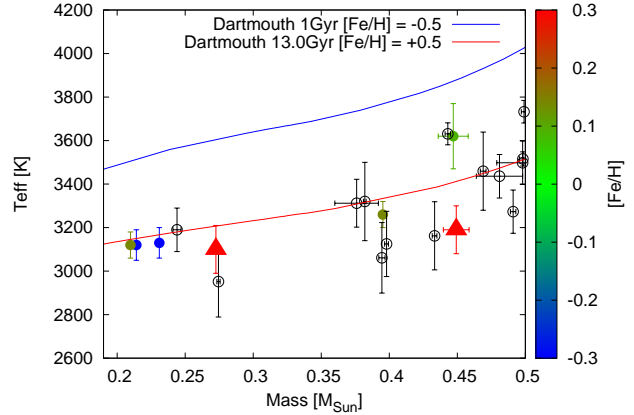}
\caption{Similar to Figure~\ref{fig:massrad}, here we show the
  Mass--T$_{\rm eff}$ diagram. The observed temperatures are
  systematically below the theoretical models, with even the 13\,Gyr,
  [Fe/H]$=+0.5$ model being above most of the observations.
\label{fig:massteff}}
\end{figure}
%% ----------------

%% ----------------
\begin{figure}[!ht]
\epsscale{1.15}
\plotone{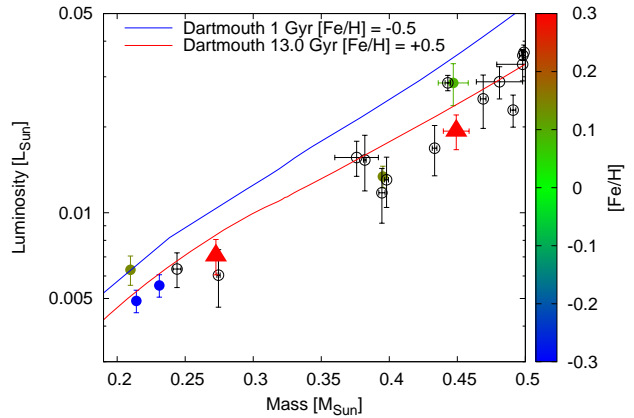}
\caption{Similar to Figure~\ref{fig:massrad}, here we show the
  Mass--Luminosity diagram. The observed luminosities are determined
  from the measured radii and temperatures and are systematically
  below the theoretical models, with even the 13\,Gyr, [Fe/H]$=+0.5$
  model being above most of the observations.
\label{fig:masslum}}
\end{figure}
%% ----------------

There are 22 other M dwarf stars in eclipsing binaries with masses
between $0.2$\,\msunnom\ and $0.5$\,\msunnom, and with masses and
radii measured to better than 5\% precision. These are collected in
Table~\ref{tab:litmebs}. Although \citet{iglesiasmarzoa:2017} report
parameters for the star T-Cyg-12664B that pass these cuts, we exclude
it from this table as \citet{han:2017}, who make use of RVs measured
for both the primary and secondary stars, find higher uncertainty
values that differ substantially from those of
\citet{iglesiasmarzoa:2017}. Figures~\ref{fig:massrad},
\ref{fig:massteff}, and \ref{fig:masslum} show the mass-radius,
mass-T$_{\rm eff}$ and mass-luminosity diagrams for these objects,
respectively.  Overplotted on these are isochrones from the Dartmouth
theoretical stellar evolution models \citep{dotter:2008}. While the
observed radii are generally above the [Fe/H]$=0$, age$=1$\,Gyr
isochrone, they are mostly below, or consistent with, the
[Fe/H]$=+0.5$, age$=13$\,Gyr isochrone. This indicates that most
objects can be explained by these models if they are old and/or metal
rich. While the masses and radii may be consistent with these models,
the observed effective temperatures are systematically lower than the
models. The luminosities, which are computed from the effective
temperatures and radii, are also lower than the models.

For systems with an observed metallicity we can provide a more
quantitative comparison to the models as follows.  To compare the
model to the observed parameters for a system (e.g. the masses, radii,
and metallicity), we make use of the following likelihood function:

\begin{equation}
L \propto \exp{-\frac{1}{2}({\bf X}-{\bf X0})^{\prime}{\bf A}^{-1}({\bf X}-{\bf X0})}
\end{equation}
where ${\bf X}$ is a vector consisting of the observations (e.g.~two
masses, two radii, and the system metallicity), ${\bf X0}$ is a vector
of the mean values for these parameters, and ${\bf A}$ is the
covariance matrix between these parameters (for systems taken from the
literature we assume none of the parameters are correlated). For \sysname{} we determine
this from the MCMC chains produced in fitting the light curves and RV
data, for other objects we assume uncorrelated errors between the
parameters and use the $1\sigma$ errors presented in their papers. To
fit the model to the data we interpolate within a precomputed grid of
isochrones to determine the radii for each trial set of
parameters. The grid is calculated using the Dartmouth web
interface\footnote{http://stellar.dartmouth.edu/models/grid.html}
assuming [$\alpha$/Fe]$=0$ for [Fe/H]$\geq 0$, and
[$\alpha$/Fe]$=+0.2$ for [Fe/H]$< 0$. We restrict the parameter search
to $-2.0\leq$[Fe/H]$\leq+0.5$ and 1\,Gyr$<$age$<$13.8\,Gyr, where the
upper limit on the age is taken to be the age of the universe based on
Planck results \citep{planck:2013}. We carry out a DEMC analysis to
determine the parameters and their uncertainties.

Note that here we often treat directly observed parameters, such as
the metallicity or mass of each star, as free parameters in the model
that are optimized in the process. Doing this accounts for the
uncertainties in the observed values, and the contribution of these
uncertainties to the values predicted by the model for other
parameters, such as the stellar radii or effective temperatures.

We applied this analysis to \sysname, and to each of the six systems in
Table~\ref{tab:litmebs} with measured metallicities. For
PTFEB132.707+19.810 we analyze both the parameters from
\citet{kraus:2017}, and those from \citet{gillen:2017}. The
results are given in Table~\ref{tab:dartmouthmodel} for \sysname{}, CM~Dra, Kepler-16 and LP~661-13, in Table~\ref{tab:dartmouthmodel2} for WOCS~23009 and KOI-126, and in Table~\ref{tab:dartmouthmodel3} for PTFEB132.707+19.810. 

{\em \sysname}: To fit the Dartmouth model to the data for \sysname{}
we vary four parameters: the masses of the two stars, the age of the
system, and the metallicity of the system, fitting these to the
observed masses and radii of the components, and the observed
metallicity of the system. 
The resulting radii are within $\sim
1\sigma$ of the measured values. The best-fit model has $\chi^{2} =
1.3$, and as there is one degree of freedom in this fit, this indicates
an excellent fit. We conclude that the masses, radii, and metallicity
of \sysname{} are consistent with the Dartmouth model. This modelling
yields a 95\% confidence lower limit on the age of $t > 6.6$\,Gyr.
While the Dartmouth model reproduces the masses and radii of the
stars, it predicts somewhat hotter temperatures for both components
(T$_{\rm eff,A}=3487^{+29}_{-22}$\,K, and T$_{\rm
  eff,B}=3255^{+25}_{-21}$\,K) than what we infer from the spectra
(T$_{\rm eff,A}=\teffstarbtsettlA$\,K and T$_{\rm
  eff,B}=\teffstarbtsettlB$\,K). If we include the temperatures as
additional observables to be fit by the model we find a minimum
$\chi^{2} = 5.56$, with three degrees of freedom. In other words the data
and model are still consistent when the effective temperatures are
included in the fit, but the quality of the fit is somewhat poorer. In
this case the 95\% confidence lower limit on the age is $t >
6.6$\,Gyr.

{\em CM~Dra}: Fitting the model without including the temperatures
yields [Fe/H]$=+0.14\pm0.06$ and age$=13.07^{+0.56}_{-1.22}$. A high
metallicity is required to fit the relatively large radii of these
stars, but this is inconsistent with the observed value of
[Fe/H]$=-0.30\pm0.12$. The model radii are $0.9\sigma$ and $1.7\sigma$
smaller (0.7\% and 1\%) than the measured radii of the primary and
secondary, respectively. The resulting $\chi^2$ for the best-fit model
is $16.81$ which has a $4\times10^{-5}$ probability of
occuring by chance when there is one degree of freedom. If the
metallicity is fixed to [Fe/H]$=-0.30$, then the model radii are
$4.5\sigma$ and $6.9\sigma$ (3.4\% and 4.0\%) too small. The model
temperatures ($3270$\,K and $3250$\,K) are also somewhat larger than
the observed values ($3130\pm70$\,K and $3120\pm70$\,K). Including the
temperatures as observables in the fit yields a similar result. We note that \citet{feiden:2014} find that the Darmouth evolution models and observations can be reconciled for CM~Dra by invoking a $\sim 0.2$\,dex $\alpha$-element enhancement, near solar-metallicity, and an age (based partly on the cooling age of the white dwarf companion) of $8.5\pm3.5$\,Gyr.

{\em WOCS~23009}: This long period single-lined binary is a member of
the open cluster NGC~6819. The color-magnitude diagrams for the
cluster, together with multiple eclipsing binaries, enable a precise
determination of the cluster age of $2.62\pm0.25$\,Gyr. Additionally
the metallicity has been precisely determined to be
[Fe/H]$=+0.09\pm0.03$. \cite{sandquist:2013} have previously shown
that the Dartmouth models are consistent with the observed properties
of WOCS~23009. We repeat this comparison but within the framework
presented in this section to allow a fair quantitative comparison with the
other systems. 
In this case we treat the following parameters as
observables to be matched by the model: the effective temperature of
the primary T$_{\rm eff,1}=6320\pm150$\,K, the semi-amplitude of the
primary star's RV orbit $K=6.96\pm0.13$\,\kms, the ratio of the radii
$R_{1}/R_{2} = 4.977\pm0.009$, the sum of the radii relative to the
semimajor axis $(R_{1}+R_{2})/a=0.005836\pm0.000020$, the metallicity
[Fe/H]$=+0.09\pm0.03$, and the age of the cluster
$2.62\pm0.25$\,Gyr. We vary the masses of the two component stars, and
the age and metallicity of the isochrones in our fit. 
We find that the best-fit model has
$\chi^{2}=0.83$, which given that there are two degrees of freedom,
indicates that the model is in excellent agreement with the observations.

{\em KOI-126}: This is a triply eclipsing hierarchical triple system
discovered by {\em Kepler} \citep{carter:2011}. \cite{feiden:2011}
have previously shown that the Dartmouth models are in good agreement
with the observed masses and radii. As for WOCS~23009 we perform our
own modeling of this system using the framework presented here. 
In
this case the observed parameters are the masses and radii of the
three component stars, the temperature of the primary star, and the
metallicity of the system. These parameters are taken from
\citet{carter:2011}.
The free parameters in the model are the
      masses of the three component stars, the metallicity, and the
      age. 
The best-fit model has $\chi^2 =
      2.2$ with three degrees of freedom, indicating a good fit to the
      observations.

{\em Kepler-16}: This is an eclipsing binary system with a transiting
circumbinary planet discovered by \citet{doyle:2011}. The transiting
planet allows the masses and radii of both stars to be determined with
high precision from the light curve alone. Additionally the
temperature and metallicity of the primary star, which dominates the
light of the system, have been determined spectroscopically. 
The
observables that we attempt to fit are: $M_{1} =
0.6897\pm0.0035$\,\msun, $R_{1} = 0.6489\pm0.0014$\,\rsunnom, $M_{2} =
0.20255\pm0.00066$\,\msun, $R_{2} = 0.22623\pm0.00059$\,\rsunnom, T$_{\rm
  eff,1} = 4450 \pm 150$\,K, and [Fe/H]$=$[M/H]$=-0.3\pm0.2$. The free
parameters in the model are the masses of the two stars, and the age
and metallicity of the system. 
The best-fit
model has $\chi^{2} = 17.1$ with two degrees of freedom. The probability
that such a high value of $\chi^2$ is found by chance is only
$2\times10^{-4}$, so the model does not provide a good fit to the
observations within the errors. The model can match the observed
masses and radii of the system to within $1\sigma$, but it requires a
metallicity that is $3.5\sigma$ ($0.69$\,dex) too high to do so. If we
fix the metallicity to $-0.3$ the predicted radius of the primary is
too high by $2.3\sigma$ (0.5\%) while the predicted radius of the
secondary is too low by $8\sigma$ (2.1\%). The model temperature of
the primary is also too high by $2.1\sigma$ (7\%). The high values for
the radius and temperature of the primary are due to the model
choosing a large age of $8.5$\,Gyr to better match the radius of the
secondary. If an age is adopted that fits the primary mass and
radius, then the secondary radius is too large by $19\sigma$ (5\%).

{\em LP~661-13}: This is a double-lined M dwarf eclipsing binary system discovered by \citet{dittmann:2017}. The parameters for the primary component are listed in Table~\ref{tab:litmebs}, while the secondary component has a mass of $0.19400 \pm 0.00034$\,\msun, and a radius of $0.2174 \pm 0.0023$\,\rsunnom. We find that a relatively large age ($12.2\pm1.3$\,Gyr), and super-solar metallicity ($+0.09\pm0.07$\,dex) are required to fit the observed masses and radii. This modelling yields radii for the primary and secondary components that are too small by $2.4\sigma$ (2.4\%) and by $0.4\sigma$ (0.4\%), respectively. The resulting
$\chi^2$ for the best-fit model is 7.7 with one degree of freedom,
indicating a marginally acceptable fit (5\% probability of occurring by chance). When the metallicity is fixed to observed the value of $-0.07$\,dex, then the resulting radii are too small by $3.2\sigma$ (3.2\%) and $1.4\sigma$ (1.5\%), respectively.

{\em PTFEB132.707+19.810}: This is a double-lined M dwarf eclipsing
binary system in the Praesepe cluster discovered by
\citet{kraus:2017}, and independently by \citet{gillen:2017}. We first
fit the parameters from \citet{kraus:2017} for the system in a similar
manner as for CM~Dra, but in this case we fix the age to 1.0\,Gyr, the
minimum age tabulated in the Dartmouth isochrones, given the estimated
age of 600--800\,Myr for the cluster. Fitting the model without
including the temperatures yields a radius for the primary star that
is $1.7\sigma$ larger than the measured radius, and a radius for the
secondary star that is $3.8\sigma$ smaller than the measured
radius. The resulting $\chi^2$ for the best-fit model is 69.3 with two
degrees of freedom, indicating a very poor fit ($8.9\times10^{-16}$
probability of occurring by chance). The model yields temperatures of
$3450$\,K and $3250$\,K for the primary and secondary stars that are
too large by 2.8$\sigma$ and $1.6\sigma$, respectively. While the
1.0\,Gyr Dartmouth isochrone clearly provides a poor fit to the
observed properties of this binary system, we caution that the
secondary component may still be contracting onto the main sequence at
the younger age of the Praesepe cluster, which may explain the
discrepancy. If we instead use the parameters from
\citet{gillen:2017}, which are based on the same {\em K2} light curve, but different spectroscopy, and exclude the effective temperatures, we find
excellent agreement with the models with $\chi^2 = 0.56$ for the
best-fit model. \citet{macdonald:2017} also conclude that the
\citet{gillen:2017} values are in better agreement with models
than the \citet{kraus:2017} parameters.

To summarize the results of our comparison with the Dartmouth models,
we find that the masses, radii, metallicities, and ages (when
independently known) of the stars in \sysname{}, WOCS~23009, and
KOI-126 are well matched by these models, while those of Kepler-16 are
not. For CM~Dra the results are not consistent if we assume a subsolar
[Fe/H] as reported by \citet{terrien:2012} (see however
\citealp{feiden:2014}), while the \citet{kraus:2017} parameters for
PTFEB132.707+19.810 are inconsistent with the models, but the
\citet{gillen:2017} values are in agreement with the models. The
observations of LP~661-13 are in slight disagreement with the models
at the $\sim 2\sigma$ level. We note that the three systems that are
in agreement with the models are older than 1\,Gyr (or at least do not
have independent age determinations indicating that they are younger
than this), and have super-solar metallicities. The other systems
either have sub-solar metallicities (CM~Dra, Kepler-16 and LP~661-13)
or are younger than 1\,Gyr (PTFEB132.707+19.810), and in two cases
have conflicting parameter values, some of which are consistent with
the models, and some of which are not (CM~Dra, PTFEB132.707+19.810).

A similar conclusion that sub-solar metallicity stars are not
well-matched by the Dartmouth models, at least for fully convective
stars, was reached by \citet{feiden:2013} in the context of testing
magnetic models. It is not clear if the agreement with the models for
the higher metallicity systems is fortuitous. Since both increased age
and enhanced metallicity tend to allow for larger radii, if these
binary components are actually inflated due to stellar activity, then
perhaps we should expect to see better agreement with the models for
metal-rich stars, especially when they are allowed to have old
ages. If that is the case, then we should not expect the age inferred
for \sysname{} to be accurate. While the radii of the high metallicity
stars are in agreement with the models, the temperatures are
systematically too low. This is also seen for other stars where the
metallicities have not been determined. While the measured masses and
radii are largely model independent\footnote{That is they depend only
  on very well understood and accepted Keplerian physics.}, the
measured temperatures depend on theoretical atmosphere
models. Therefore we cannot say whether the disagreement between the
measured and expected temperatures is due to errors in the stellar
evolution models, in the atmosphere and spectral synthesis models used
in measuring the temperatures, or both.

\subsection{Summary}

In this paper we have presented the discovery of a new double-lined M
dwarf binary with total secondary eclipses. The results can be summarized as follows:
\begin{enumerate}
\item By combining optical radial velocity measurements for both
  components with photometric observations of the eclipses, we measure
  the masses and radii of both stars to be $M_{A} =
  \massstarA$\,\msunnom, $M_{B} = \massstarB$\,\msunnom, $R_{A}
  = \radiusstarA$\,\rsunnom, and $R_{B} =
  \radiusstarB$\,\rsunnom.
\item We find that the system has a small, but significant, non-zero
eccentricity of $\RVeccen$. 
\item The {\em K2} observations show a strikingly coherent nearly
  sinusoidal variation with a period of $\starAootvperiod$\,d, which
  is slightly longer than the orbital period. We demonstrate that the
  signal is due to the primary star, and interpret it as the rotation
  period of this component. The slight asynchronicity might be due to
  differential rotation, or a magnetized wind which balances the
  torque from tides. Ground-based $r$-band and $i$-band light curves
  obtained many years before {\em K2} show no evidence of this
  variation, with limits on the amplitude that are several times lower
  what was seen in {\em K2} (by nearly a factor of 10 in the case of
  the HATNet $r$-band).
\item The {\em K2} observations show an additional modulation at a period of $\starBootvperiod$\,d (with two peaks per cycle, or a near sinusoidal variation at half this value). We cannot determine whether the signal is due to the primary or secondary component.
\item We obtained near infrared spectra of the system during total
  eclipse, and near both quadrature phases, and used these
  observations to disentangle the spectra of the two components.
\item Based on the disentagled spectra we measure metallicities and
  effective temperatures for the two components of T$_{\rm eff,A} =
  \teffstarA$\,K, T$_{\rm eff,B} = \teffstarB$\,K, [Fe/H]$_{A} =
  \fehstarA$, and [Fe/H]$_{B} = \fehstarB$, or a metallicity of
           [Fe/H]$=\fehsystem$ for the system if we assume the two
           stars have the same abundances. We find consistent results
           when using empricially calibrated spectral indices, and
           when cross-correlating the spectra against BT-Settl
           synthetic templates.
\item The space velocity of the system indicates that it is a member
  of the Galactic disk.
\item We carried out tests which indicate that the total-eclipsing nature of this system
  significantly improves the accuracy with which the parameters may be
  measured.
\item We find that the masses and radii of the stars in this system
  are well-matched by the Dartmouth stellar evolution models for a
  system age of $t > 6.6$\,Gyr. We also find that these same models
  reproduce M dwarfs in two other systems (WOCS~23009B and KOI-126B+C) in the mass range $0.2\,\msunnom < M <
  0.5\,\msunnom$ with well measured masses and radii, and supersolar
  metallicities, but do not match two other systems with subsolar
  metallicity stars (Kepler-16B, LP~661-13A). There are two systems with conflicting sets of
  measured parameters, some of which are in agreement with the
  models, and some of which are not (CM~Dra, PTFEB132.707+19.810).
\end{enumerate}

Further improvement in the precision of the parameter estimates for
this system will require higher precision RV measurements. At $V \sim
16$\,mag, the star is quite faint, and pushes the limits of the
FLWO~1.5\,m telescope used to obtain the RVs presented here. More
precise measurements will require a larger telescope. The treatment of starspots could also be improved. In particular, the {\em K2} light curve may allow spots to be mapped on the surfaces of the component stars.

\acknowledgements 
Partial support for the work reported here was
provided by NASA grants NNX09AB29G, NNX13AJ15G, NNX14AE87G and
NNX17AB61G, and NSF grant AST-1108686. GT acknowledges partial support
for this work from NSF grant AST-1509375. Some of the data presented
in this paper were obtained from the Mikulski Archive for Space
Telescopes (MAST). STScI is operated by the Assocation of Universities
for Research in Astronomy, Inc., under NASA contract
NAS5-26555. Support for MAST for non-HST data is provided by the NASA
Office of Space Science via grant NNX09AF08G and by other grants and
contracts. This paper includes data collected by the Kepler
mission. Funding for the Kepler mission is provided by the NASA
Science Mission directorate.

%% Bibliography

%\bibliographystyle{apj}
%\bibliography{htr318007.bib}

\ifthenelse{\boolean{emulateapj}}{
  \begin{deluxetable*}{lrrrrrrrr}
}{
  \begin{deluxetable}{lrrrrrrrr}
}
\tablewidth{0pc}
\tabletypesize{\footnotesize}
\tablecaption{
  Equivalent Widths and Inferred Metallicities from FIRE/Magellan following \citet{terrien:2012}
  \label{tab:t12param}
}
\tablehead{
  \multicolumn{1}{c}{Target} &
  \multicolumn{1}{c}{EW$_{\rm Ca_{\rm H}}$} &
  \multicolumn{1}{c}{EW$_{\rm K_{\rm H1}}$} &
  \multicolumn{1}{c}{EW$_{\rm Na_{\rm K}}$} &
  \multicolumn{1}{c}{EW$_{\rm Ca_{\rm K}}$} &
  \multicolumn{1}{c}{H$_{2}$O$-$H} &
  \multicolumn{1}{c}{H$_{2}$O$-$K} &
  \multicolumn{1}{c}{[Fe/H]$_{\rm H}$} &
  \multicolumn{1}{c}{[Fe/H]$_{\rm K}$} \\
 }
\startdata
GJ~205 & $3.47\pm0.31$ & $0.727\pm0.069$ & $7.61\pm0.12$ & $5.59\pm0.15$ & $0.910\pm0.015$ & $0.877\pm0.024$ & $0.39\pm0.10$ & $0.498\pm0.019$ \\
GJ~250~B & $2.45\pm0.28$ & $0.481\pm0.068$ & $4.80\pm0.12$ & $4.06\pm0.15$ & $0.944\pm0.014$ & $0.979\pm0.024$ & $-0.05\pm0.10$ & $-0.037\pm0.020$ \\
GJ~283 & $0.61\pm0.28$ & $1.741\pm0.072$ & $4.35\pm0.11$ & $1.07\pm0.14$ & $0.727\pm0.015$ & $0.847\pm0.023$ & $-0.25\pm0.10$ & $-0.294\pm0.020$ \\
GJ~285 & $2.56\pm0.29$ & $1.337\pm0.064$ & $6.95\pm0.11$ & $4.33\pm0.13$ & $0.843\pm0.013$ & $0.857\pm0.024$ & $0.29\pm0.10$ & $0.315\pm0.019$ \\
GJ~3348~B & $1.22\pm0.29$ & $0.920\pm0.065$ & $5.48\pm0.11$ & $3.22\pm0.15$ & $0.838\pm0.014$ & $0.884\pm0.023$ & $-0.33\pm0.10$ & $0.018\pm0.021$ \\
GJ~352 & $2.27\pm0.22$ & $0.738\pm0.055$ & $4.71\pm0.10$ & $3.61\pm0.12$ & $0.875\pm0.012$ & $0.888\pm0.019$ & $-0.03\pm0.08$ & $-0.053\pm0.016$ \\
LHS~2065 & $0.56\pm0.29$ & $2.881\pm0.070$ & $6.80\pm0.11$ & $1.05\pm0.15$ & $0.762\pm0.014$ & $0.797\pm0.024$ & $0.20\pm0.10$ & $0.046\pm0.020$ \\
NLTT~15867 & $0.62\pm0.30$ & $0.693\pm0.067$ & $4.34\pm0.11$ & $1.50\pm0.15$ & $0.816\pm0.014$ & $0.872\pm0.024$ & $-0.64\pm0.11$ & $-0.271\pm0.019$ \\
\sysname{}~A\tablenotemark{a} & $2.72\pm0.22$ & $1.414\pm0.057$ & $7.85\pm0.09$ & $5.17\pm0.12$ & $0.840\pm0.012$ & $0.897\pm0.019$ & $0.39\pm0.08$ & $0.488\pm0.018$ \\
\sysname{}~A\tablenotemark{b} & $2.68\pm0.39$ & $1.066\pm0.086$ & $7.96\pm0.14$ & $5.06\pm0.18$ & $0.839\pm0.019$ & $0.900\pm0.032$ & $0.23\pm0.13$ & $0.493\pm0.024$ \\
\sysname{}~B\tablenotemark{b} & $1.41\pm0.40$ & $2.545\pm0.108$ & $5.91\pm0.19$ & $3.08\pm0.27$ & $0.914\pm0.021$ & $0.871\pm0.034$ & $0.43\pm0.14$ & $0.068\pm0.036$ \\
\enddata
\tablenotetext{a}{Based on the combined spectrum obtained during total secondary eclipse.}
\tablenotetext{b}{Based on the disentangled primary and secondary spectra.}
\ifthenelse{\boolean{emulateapj}}{
  \end{deluxetable*}
}{
  \end{deluxetable}
}

\ifthenelse{\boolean{emulateapj}}{
  \begin{deluxetable*}{lrrrrrr}
}{
  \begin{deluxetable}{lrrrrrr}
}
\tablewidth{0pc}
\tabletypesize{\footnotesize}
\tablecaption{
  Equivalent Widths and Inferred Spectral Types and Metallicities from FIRE/Magellan following \citet{rojasayala:2012}
  \label{tab:r12param}
}
\tablehead{
  \multicolumn{1}{c}{Target} &
  \multicolumn{1}{c}{EW$_{\rm Na_{\rm K}}$} &
  \multicolumn{1}{c}{EW$_{\rm Ca_{\rm K}}$} &
  \multicolumn{1}{c}{H$_{2}$O$-$K} &
  \multicolumn{1}{c}{Spec.~Type} &
  \multicolumn{1}{c}{[Fe/H]} &
  \multicolumn{1}{c}{[M/H]} \\
}
\startdata
GJ~205 & $8.11\pm0.43$ & $6.20\pm0.25$ & $0.967\pm0.010$ & M$1.68\pm0.24$ & $0.496\pm0.060$ & $0.360\pm0.043$ \\
GJ~250~B & $4.74\pm0.43$ & $4.58\pm0.26$ & $0.943\pm0.010$ & M$2.27\pm0.23$ & $-0.005\pm0.060$ & $-0.004\pm0.041$ \\
GJ~283 & $3.64\pm0.48$ & $1.10\pm0.27$ & $0.737\pm0.009$ & M$7.16\pm0.25$ & $-0.408\pm0.073$ & $-0.282\pm0.051$ \\
GJ~285 & $7.04\pm0.43$ & $4.81\pm0.22$ & $0.844\pm0.009$ & M$4.61\pm0.22$ & $0.401\pm0.060$ & $0.283\pm0.041$ \\
GJ~3348~B & $5.49\pm0.45$ & $3.47\pm0.24$ & $0.865\pm0.010$ & M$4.11\pm0.24$ & $0.027\pm0.062$ & $0.022\pm0.046$ \\
GJ~352 & $4.82\pm0.36$ & $3.76\pm0.19$ & $0.930\pm0.009$ & M$2.55\pm0.21$ & $-0.080\pm0.046$ & $-0.053\pm0.035$ \\
LHS~2065 & $6.70\pm0.45$ & $1.16\pm0.29$ & $0.670\pm0.010$ & M$8.76\pm0.23$ & $0.092\pm0.072$ & $0.072\pm0.053$ \\
NLTT~15867 & $4.14\pm0.45$ & $1.84\pm0.25$ & $0.841\pm0.011$ & M$4.72\pm0.24$ & $-0.328\pm0.063$ & $-0.227\pm0.041$ \\
\sysname{}~A\tablenotemark{a} & $8.64\pm0.40$ & $5.77\pm0.22$ & $0.886\pm0.008$ & M$3.62\pm0.18$ & $0.634\pm0.054$ & $0.450\pm0.037$ \\
\sysname{}~A\tablenotemark{b} & $8.34\pm0.60$ & $5.33\pm0.28$ & $0.882\pm0.013$ & M$3.71\pm0.30$ & $0.550\pm0.079$ & $0.394\pm0.057$ \\
\sysname{}~B\tablenotemark{b} & $6.76\pm0.70$ & $2.53\pm0.85$ & $0.828\pm0.016$ & M$5.01\pm0.38$ & $0.077\pm0.143$ & $0.062\pm0.103$ \\
\enddata
\tablenotetext{a}{Based on the combined spectrum obtained during total secondary eclipse.}
\tablenotetext{b}{Based on the disentangled primary and secondary spectra.}
\ifthenelse{\boolean{emulateapj}}{
  \end{deluxetable*}
}{
  \end{deluxetable}
}

\ifthenelse{\boolean{emulateapj}}{
  \begin{deluxetable}{lrr}
}{
  \begin{deluxetable}{lrr}
}
\tablewidth{0pc}
\tabletypesize{\footnotesize}
\tablecaption{
  Atmospheric Parameters from FIRE/Magellan Based on Cross-Correlation with BT-Settl Synthetic Templates
  \label{tab:btsettlparam}
}
\tablehead{
  \multicolumn{1}{c}{Target} &
  \multicolumn{1}{c}{T$_{\rm eff}$} &
  \multicolumn{1}{c}{[Fe/H]} \\
  &
  \multicolumn{1}{c}{(K)} &
  \\
}
\startdata
GJ~205 & $3870 \pm 100$ & $0.24\pm0.15$ \\
GJ~250~B & $3600 \pm 100$ & $-0.13\pm0.15$ \\
GJ~283 & $2690 \pm 100$ & $-0.31\pm0.14$ \\
GJ~285 & $2050 \pm 110$ & $0.27\pm0.14$ \\
GJ~3348~B & $3160 \pm 100$ & $-0.05\pm0.15$ \\
GJ~352 & $3480 \pm 100$ & $-0.03\pm0.15$ \\
LHS~2065 & $2490 \pm 110$ & $0.03\pm0.15$ \\
NLTT~15867 & $2980 \pm 110$ & $0.09\pm0.15$ \\
\sysname~A\tablenotemark{a} & $3200 \pm 100$ & $0.23\pm0.15$ \\
\sysname~A\tablenotemark{b} & $3190 \pm 110$ & $0.25\pm0.13$ \\
\sysname~B\tablenotemark{b} & $3100 \pm 110$ & $0.09\pm0.15$ \\
\enddata
\ifthenelse{\boolean{emulateapj}}{
  \end{deluxetable}
}{
  \end{deluxetable}
}

\ifthenelse{\boolean{emulateapj}}{
  \begin{deluxetable*}{lrrrrl}
}{
  \begin{deluxetable}{lrrrrl}
}
\tablewidth{0pc}
\tabletypesize{\footnotesize}
\tablecaption{
Literature M dwarfs in eclipsing binary systems with masses between $0.2\,\msunnom < M < 0.5 \,\msunnom$, and with masses and radii determined to better than 5\% precision. Except where noted, stars are components of double-lined eclipsing binary systems. We exclude stars with white dwarf binary companions; such systems may have undergone significant mass transfer.
\label{tab:litmebs}
}
\tablehead{
  \multicolumn{1}{c}{Star} &
  \multicolumn{1}{c}{Mass} &
  \multicolumn{1}{c}{Radius} &
  \multicolumn{1}{c}{T$_{\rm eff}$} &
  \multicolumn{1}{c}{[Fe/H]} &
  \multicolumn{1}{c}{Reference(s)} \\
  &
  \multicolumn{1}{c}{(\msunnom)} &
  \multicolumn{1}{c}{(\rsunnom)} &
  \multicolumn{1}{c}{(K)} &
  &
  \\
}
\startdata
MG1-646680~A & $0.499\pm0.002$ & $0.457\pm0.006$ & $3730 \pm 50$ & $\cdots$ & \citep{kraus:2011} \\
NSVS~01031772~B & $0.4982\pm0.0025$ & $0.5087\pm0.0031$ & $3520 \pm 30$ & $\cdots$ & \citep{lopezmorales:2006} \\
WTS~19b-2-01387~A & $0.498\pm0.019$ & $0.496\pm0.013$ & $3498\pm100$ & $\cdots$ & \citep{birkby:2012} \\
MG1-78457~B & $0.491\pm0.002$ & $0.471\pm0.009$ & $3270\pm100$ & $\cdots$ & \citep{kraus:2011} \\
WTS19b-2-01387B & $0.481\pm0.017$ & $0.479\pm0.013$ & $3436\pm100$ & $\cdots$ & \citep{birkby:2012} \\
MG1-2056316A & $0.4690\pm0.0021$ & $0.441\pm0.002$ & $3460\pm180$ & $\cdots$ & \citep{kraus:2011} \\
WOCS~23009~B\tablenotemark{a} & $0.447\pm0.011$ & $0.4292\pm0.0033$ & $3620\pm150$ & $+0.09\pm0.03$ & \citep{sandquist:2013} \\
MG1-646680B & $0.443\pm0.002$ & $0.427\pm0.006$ & $3630\pm50$ & $\cdots$ & \citep{kraus:2011} \\
CU~Cnc~A & $0.4333\pm0.0017$ & $0.4317\pm0.0052$ & $3160\pm150$ & $\cdots$ & \citep{ribas:2003} \\
CU~Cnc~B & $0.3980\pm0.0014$ & $0.3908\pm0.0094$ & $3130\pm150$ & $\cdots$ & \citep{ribas:2003} \\
PTFEB132.707+19.810~A\tablenotemark{b} & $0.3953 \pm 0.0020$ & $0.363 \pm 0.008$ & $3260 \pm 60$ & $+0.14 \pm 0.04$ & \citep{kraus:2017} \\
LSPM~J1112+7626~A & $0.3946\pm0.0023$ & $0.3860\pm0.0055$ & $3060\pm160$ & $\cdots$ & \citep{irwin:2011} \\
MG1-2056316~B & $0.382\pm0.001$ & $0.374\pm0.002$ & $3320\pm180$ & $\cdots$ & \citep{kraus:2011} \\
GJ~3236~A & $0.376\pm0.016$ & $0.3795\pm0.0084$ & $3310\pm110$ & $\cdots$ & \citep{irwin:2009} \\
LP~661-13~A\tablenotemark{c} & $0.30795\pm0.00084$ & $0.3226\pm0.0033$ & $\cdots$ & $-0.07 \pm 0.1$ & \citep{dittmann:2017} \\
LSPM~J1112+7626~B & $0.2745\pm0.0012$ & $0.2978\pm0.0049$ & $2950\pm160$ & $\cdots$ & \citep{irwin:2011} \\
1RXS~J154727.5+450803~B & $0.2585\pm0.0080$ & $0.2895\pm0.0068$ & $\cdots$ & $\cdots$ & \citep{hartman:2011:kmdwarf} \\
1RXS~J154727.5+450803~A & $0.2576\pm0.0085$ & $0.2895\pm0.0068$ & $\cdots$ & $\cdots$ & \citep{hartman:2011:kmdwarf} \\
HATS551-027~A              & $0.244\pm0.003$   & $0.261^{+0.006}_{-0.009}$ & $3190 \pm 100$ & $\cdots$ & \citep{zhou:2015} \\
KOI~126~B\tablenotemark{d} & $0.2413\pm0.0030$ & $0.2543\pm0.0014$ & $\cdots$ & $+0.15\pm0.08$ & \citep{carter:2011} \\
CM~Dra~A\tablenotemark{e} & $0.2310\pm0.0009$ & $0.2534\pm0.0019$ & $3130\pm70$ & $-0.30\pm0.12$ & \citep{morales:2009,terrien:2012} \\
CM~Dra~B\tablenotemark{e} & $0.2141\pm0.0010$ & $0.2396\pm0.0015$ & $3120\pm70$ & $-0.30\pm0.12$ & \citep{morales:2009,terrien:2012} \\
KOI~126~C\tablenotemark{d} & $0.2127\pm0.0026$ & $0.2318\pm0.0013$ & $\cdots$ & $+0.15\pm0.08$ & \citep{carter:2011} \\
PTFEB132.707+19.810~B\tablenotemark{b} & $0.2098 \pm 0.0014$ & $0.272 \pm 0.012$ & $3120 \pm 60$ & $+0.14 \pm 0.04$ & \citep{kraus:2017} \\
Kepler-16~B\tablenotemark{f} & $0.20255\pm0.00066$ & $0.22623\pm0.00059$ & $\cdots$ & $-0.3\pm0.2$ & \citep{doyle:2011} \\
\enddata
\tablenotetext{a}{WOCS~23009~B is the secondary component of a single-lined binary system with a $M=1.468\pm0.030\,\msunnom$ evolved primary star. This binary system is a member of the open cluster NGC~6819. The listed [Fe/H] is the value for the cluster.}
\tablenotetext{b}{PTFEB132.707+19.810 is a member of the Praesepe open cluster, and the adopted metallicity is the value for the cluster. Note that \citet{gillen:2017} independently identified this is a binary, which they label AD~3814. They measure masses of $0.3813\pm0.0074\,\msun$ and $0.2022\pm0.0045\,\msun$, and radii of $0.3610\pm0.0033\,\rsun$ and $0.2256^{+0.0063}_{-0.0049}\,\rsun$ for the primary and secondary stars, respectively.}
\tablenotetext{c}{The metallicity of the LP~661-13 eclipsing binary system was not determined spectroscopically, but was estimated using the absolute $K_{s}$ magnitude and the $MEarth - K_{S}$ broad-band color following \citet{dittmann:2016}.}
\tablenotetext{d}{KOI-126~B and KOI-126~C are components of a triply eclipsing hierarchical triple system. The primary star has a mass of $M=1.347\pm0.032\,\msunnom$. Only light from the primary star has been detected in the spectrum. The listed [Fe/H] is the value determined spectroscopically for the primary. The triple eclipses, together with the RVs for the primary star, enable a determination of the masses and radii of both stars that is independent of stellar evolution models.}
\tablenotetext{e}{\citet{feiden:2014} argue that CM~Dra has [Fe/H]$\sim 0$\,dex and [$\alpha$/Fe]$\ga +0.2$\,dex.}
\tablenotetext{f}{Kepler-16~B is the secondary component of a binary system with a $M=0.6897\pm0.0035\,\msunnom$ primary star. Light from the secondary star has not been detected within the spectrum, however there is a transiting circumbinary planet whose transits around each stellar component, in conjunction with the observed RVs for the primary star, allow a determination of the masses and radii of both stars that is independent of stellar evolution models. The listed [Fe/H] is the [M/H] value determined spectroscopically for the primary.}
\ifthenelse{\boolean{emulateapj}}{
  \end{deluxetable*}
}{
  \end{deluxetable}
}

\ifthenelse{\boolean{emulateapj}}{
  \begin{deluxetable*}{lrrrr}
}{
  \begin{deluxetable}{lrrrr}
}
\tablewidth{0pc}
\tabletypesize{\footnotesize}
\tablecaption{
Results from fitting low-mass eclipsing binary systems with the Dartmouth stellar evolution isochrones. \sysname{}, CM~Dra, Kepler-16, and LP~661-13.
\label{tab:dartmouthmodel}
}
\tablehead{
  \multicolumn{1}{c}{Parameter} &
  \multicolumn{1}{c}{Observed Value} &
  \multicolumn{1}{c}{Model Value\tablenotemark{a}} &
  \multicolumn{1}{c}{$(O-C)/\sigma$\tablenotemark{b}} &
  \multicolumn{1}{c}{$100\% \times (O-C)/O$\tablenotemark{c}}\\
}
\startdata
\cutinhead{\bf \sysname{}} \\
$M_{A}^{\star,\dagger}$ (\msunnom) & $\massstarA{}$ & $0.4665^{+0.0047}_{-0.0040}$ & $-1.68$ & $-4.13\%$ \\
$M_{B}^{\star,\dagger}$ (\msunnom) & $\massstarB{}$ & $0.2732 \pm 0.0029$ & $-0.26$ & $-0.40\%$ \\
$R_{A}^{\dagger}$ (\rsunnom) & $\radiusstarA{}$ & $0.4527^{+0.0035}_{-0.0033}$ & $0.58$ & $0.24\%$ \\
$R_{B}^{\dagger}$ (\rsunnom) & $\radiusstarB{}$ & $0.2906\pm0.0021$ & $0.29$ & $0.24\%$ \\
$T_{\rm eff,A}$ (K) & $\teffstarbtsettlA$ & $3487^{+29}_{-22}$ & $-2.7$ & $-9.31\%$ \\
$T_{\rm eff,B}$ (K) & $\teffstarbtsettlB$ & $3255^{+25}_{-21}$ & $-1.41$ & $-5.00\%$ \\
Age$^{\star}$ (Gyr) & $\cdots$ & $11.4^{+1.7}_{-2.8}$ & $\cdots$ & $\cdots$ \\
$[$Fe/H$]^{\star,\dagger}$ (dex) & $\fehsystem$ & $+0.330^{+0.071}_{-0.080}$ & $-0.40$ & $-10.74\%$ \\
$\chi^2$ (DOF)\tablenotemark{d} & $\cdots$ & $1.3$ (1) & $\cdots$ & $\cdots$ \\
\cutinhead{\bf CM Dra} \\
$M_{A}^{\star,\dagger}$ (\msunnom) & $0.2310 \pm 0.0009$ & $0.23129 \pm 0.00087$ & $-0.32$ & $-0.13\%$ \\
$M_{B}^{\star,\dagger}$ (\msunnom) & $0.2141 \pm 0.0010$ & $0.21515 \pm 0.00092$ & $-1.1$ & $-0.49\%$ \\
$R_{A}^{\dagger}$ (\rsunnom) & $0.2534 \pm 0.0019$ & $0.2516 \pm 0.0013$ & $0.95$ & $0.71\%$ \\
$R_{B}^{\dagger}$ (\rsunnom) & $0.2396 \pm 0.0015$ & $0.2371 \pm 0.0011$ & $1.7$ & $1.04\%$ \\
$T_{\rm eff,A}$ (K) & $3130 \pm 70$ & $3271 \pm 21$ & $-2.0$ & $-4.50\%$ \\
$T_{\rm eff,B}$ (K) & $3120 \pm 70$ & $3253 \pm 20$ & $-1.9$ & $-4.26\%$ \\
Age$^{\star}$ (Gyr) & $\cdots$ & $13.07^{+0.56}_{-1.22}$ & $\cdots$ & $\cdots$ \\
$[$Fe/H$]^{\star,\dagger}$ (dex) & $-0.30 \pm 0.12$ & $+0.14 \pm 0.06$ & $-3.7$ & $147\%$ \\
$\chi^2$ (DOF) & $\cdots$ & $16.81$ (1) & $\cdots$ & $\cdots$ \\
\cutinhead{\bf Kepler-16} \\
$M_{A}^{\star,\dagger}$ (\msunnom) & $0.6897 \pm 0.0035$ & $0.6888^{+0.0035}_{-0.0038}$ & $0.26$ & $0.13\%$ \\
$M_{B}^{\star,\dagger}$ (\msunnom) & $0.20255 \pm 0.00066$ & $0.20296^{+0.00068}_{-0.00061}$ & $-0.62$ & $-0.20\%$ \\
$R_{A}^{\dagger}$ (\rsunnom) & $0.6489 \pm 0.0014$ & $0.6492^{+0.0014}_{-0.0015}$ & $-0.21$ & $-0.05\%$ \\
$R_{B}^{\dagger}$ (\rsunnom) & $0.22623 \pm 0.00059$ & $0.22588^{+0.00059}_{-0.00063}$ & $0.59$ & $0.15\%$ \\
$T_{\rm eff,A}^{\dagger}$ (K) & $4450 \pm 150$ & $4139^{+14}_{-16}$ & $2.1$ & $6.99\%$ \\
$T_{\rm eff,B}$ (K) & $\cdots$ & $3168 \pm 13$ & $\cdots$ & $\cdots$ \\
Age$^{\star}$ (Gyr) & $\cdots$ & $3.90^{+0.85}_{-0.78}$ & $\cdots$ & $\cdots$ \\
$[$Fe/H$]^{\star,\dagger}$ (dex) & $-0.30 \pm 0.20$ & $+0.392 \pm 0.045$ & $-3.5$ & $231\%$ \\
$\chi^2$ (DOF) & $\cdots$ & $17.1$ (2) & $\cdots$ & $\cdots$ \\
\cutinhead{\bf LP 661-13} \\
$M_{A}^{\star,\dagger}$ (\msunnom) & $0.30795 \pm 0.00084$ & $0.30833^{+0.00077}_{-0.00082}$ & $-0.45$ & $-0.12\%$ \\
$M_{B}^{\star,\dagger}$ (\msunnom) & $0.19400 \pm 0.00034$ & $0.19403 \pm 0.00035$ & $-0.088$ & $-0.02\%$ \\
$R_{A}^{\dagger}$ (\rsunnom) & $0.3226 \pm 0.0033$ & $0.3147 \pm 0.0015$ & $2.4$ & $2.45\%$ \\
$R_{B}^{\dagger}$ (\rsunnom) & $0.2174 \pm 0.0023$ & $0.2165 \pm 0.0012$ & $0.39$ & $0.41\%$ \\
$T_{\rm eff,A}$ (K) & $\cdots$ & $3364 \pm 26$ & $\cdots$ & $\cdots$ \\
$T_{\rm eff,B}$ (K) & $\cdots$ & $3243 \pm 21$ & $\cdots$ & $\cdots$ \\
Age$^{\star}$ (Gyr) & $\cdots$ & $12.50^{+0.94}_{-1.7}$ & $\cdots$ & $\cdots$ \\
$[$Fe/H$]^{\star,\dagger}$ (dex) & $-0.07 \pm 0.1$ & $+0.092 \pm 0.072$ & $-1.6$ & $231\%$ \\
$\chi^2$ (DOF) & $\cdots$ & $7.7$ (1) & $\cdots$ & $\cdots$ \\
\enddata
\tablenotetext{}{Parameters marked by a $^{\star}$ are varied in the fit. Parameters marked by a $^{\dagger}$ are treated as observables that are matched to the model in computing the value of $\chi^2$ that is listed.}
\tablenotetext{a}{The optimized value and uncertainty for this parameter that comes from the analysis in Section~\ref{sec:dartmouthcomparison}. The uncertainties do not include any systematic errors in the stellar evolution models.}
\tablenotetext{b}{Difference between the observed parameter value and the model value, divided by the observational uncertainty.}
\tablenotetext{c}{Difference between the observed parameter value and the model value, expressed as a percentage of the observed parameter value.}
\tablenotetext{d}{The $\chi^2$ for the best-fit model. The number of degrees of freedom in the analysis is listed in parentheses following the $\chi^2$ value.}
\ifthenelse{\boolean{emulateapj}}{
  \end{deluxetable*}
}{
  \end{deluxetable}
}

\ifthenelse{\boolean{emulateapj}}{
  \begin{deluxetable*}{lrrrr}
}{
  \begin{deluxetable}{lrrrr}
}
\tablewidth{0pc}
\tabletypesize{\footnotesize}
\tablecaption{
Results from fitting low-mass eclipsing binary systems with the Dartmouth stellar evolution isochrones. WOCS 23009 and KOI-126.
\label{tab:dartmouthmodel2}
}
\tablehead{
  \multicolumn{1}{c}{Parameter} &
  \multicolumn{1}{c}{Observed Value} &
  \multicolumn{1}{c}{Model Value\tablenotemark{a}} &
  \multicolumn{1}{c}{$(O-C)/\sigma$\tablenotemark{b}} &
  \multicolumn{1}{c}{$100\% \times (O-C)/O$\tablenotemark{c}}\\
}
\startdata
\cutinhead{\bf WOCS 23009 \tablenotemark{c}} \\
$M_{A}^{\star}$ (\msunnom) & $\cdots$ & $1.462^{+0.027}_{-0.042}$ & $\cdots$ & $\cdots$ \\
$M_{B}^{\star}$ (\msunnom) & $\cdots$ & $0.4547^{+0.0036}_{-0.0063}$ & $\cdots$ & $\cdots$ \\
$R_{A}$ (\rsunnom) & $\cdots$ & $2.136^{+0.015}_{-0.025}$ & $\cdots$ & $\cdots$ \\
$R_{B}$ (\rsunnom) & $\cdots$ & $0.4292^{+0.0030}_{-0.0050}$ & $\cdots$ & $\cdots$ \\
$T_{\rm eff,A}^{\dagger}$ (K) & $6320\pm150$ & $6295^{+100}_{-130}$ & $0.17$ & $0.40\%$ \\
$T_{\rm eff,B}$ (K) & $\cdots$ & $3551^{+14}_{-18}$ & $\cdots$ & $\cdots$ \\
$K_{A}^{\dagger}$ (\kms) & $6.96\pm0.13$ & $7.075^{+0.052}_{-0.038}$ & $-0.88$ & $-1.65\%$ \\
$R_{A}/R_{B}^{\dagger}$ & $4.977\pm0.009$ & $4.9782^{+0.0086}_{-0.0095}$ & $-0.13$ & $-0.02\%$ \\
$(R_{A}+R_{B})/a^{\dagger}$ & $0.005836\pm0.000020$ & $0.005829\pm0.000021$ & $0.35$ & $0.12\%$ \\
Age$^{\star,\dagger}$ (Gyr) & $2.62\pm0.25$ & $2.68^{+0.34}_{-0.26}$ & $-0.24$ & $-2.29\%$ \\
$[$Fe/H$]^{\star,\dagger}$ (dex) & $+0.09 \pm 0.03$ & $+0.09 \pm 0.03$ & $0.0$ & $0.00\%$ \\
$\chi^2$ (DOF)\tablenotemark{e} & $\cdots$ & $0.83$ (2) & $\cdots$ & $\cdots$ \\
\cutinhead{\bf KOI-126} \\
$M_{A}^{\star,\dagger}$ (\msunnom) & $1.347 \pm 0.032$ & $1.373^{+0.017}_{-0.043}$ & $-0.81$ & $-1.93\%$ \\
$M_{B}^{\star,\dagger}$ (\msunnom) & $0.2413 \pm 0.0030$ & $0.2411\pm0.0019$ & $0.067$ & $0.08\%$ \\
$M_{C}^{\star,\dagger}$ (\msunnom) & $0.2127 \pm 0.0026$ & $0.2139^{+0.0014}_{-0.0016}$ & $-0.46$ & $-0.56\%$ \\
$R_{A}^{\dagger}$ (\rsunnom) & $2.0254 \pm 0.0098$ & $2.0250^{+0.0093}_{-0.0102}$ & $0.041$ & $0.02\%$ \\
$R_{B}^{\dagger}$ (\rsunnom) & $0.2543 \pm 0.0014$ & $0.2543\pm0.0013$ & $0.0$ & $0.00\%$ \\
$R_{C}^{\dagger}$ (\rsunnom) & $0.2318 \pm 0.0013$ & $0.2317\pm0.0010$ & $0.077$ & $0.04\%$ \\
$T_{\rm eff,A}^{\dagger}$ (K) & $5875 \pm 100$ & $5980^{+73}_{-78}$ & $-1.1$ & $-1.79\%$ \\
$T_{\rm eff,B}$ (K) & $\cdots$ & $3255^{+17}_{-19}$ & $\cdots$ & $\cdots$ \\
$T_{\rm eff,C}$ (K) & $\cdots$ & $3227^{+16}_{-18}$ & $\cdots$ & $\cdots$ \\
Age$^{\star}$ (Gyr) & $\cdots$ & $3.71^{+0.41}_{-0.19}$ & $\cdots$ & $\cdots$ \\
$[$Fe/H$]^{\star,\dagger}$ (dex) & $0.15\pm0.08$ & $0.221^{+0.060}_{-0.054}$ & $-0.89$ & $-47\%$ \\
$\chi^2$ (DOF) & $\cdots$ & $2.2$ (3) & $\cdots$ & $\cdots$ \\
\enddata
\tablenotetext{}{Parameters marked by a $^{\star}$ are varied in the fit. Parameters marked by a $^{\dagger}$ are treated as observables that are matched to the model in computing the value of $\chi^2$ that is listed.}
\tablenotetext{a}{The optimized value and uncertainty for this parameter that comes from the analysis in Section~\ref{sec:dartmouthcomparison}. The uncertainties do not include any systematic errors in the stellar evolution models.}
\tablenotetext{b}{Difference between the observed parameter value and the model value, divided by the observational uncertainty.}
\tablenotetext{c}{Difference between the observed parameter value and the model value, expressed as a percentage of the observed parameter value.}
\tablenotetext{d}{WOCS~23009 is a single-lined eclipsing binary system, but a member of the open cluster NGC~6819. The published masses and radii of the components of this system are inferred from the Dartmouth isochrones, so we do not list these as ``observed'' values.}
\tablenotetext{e}{The $\chi^2$ for the best-fit model. The number of degrees of freedom in the analysis is listed in parentheses following the $\chi^2$ value.}
\ifthenelse{\boolean{emulateapj}}{
  \end{deluxetable*}
}{
  \end{deluxetable}
}

\ifthenelse{\boolean{emulateapj}}{
  \begin{deluxetable*}{lrrrr}
}{
  \begin{deluxetable}{lrrrr}
}
\tablewidth{0pc}
\tabletypesize{\footnotesize}
\tablecaption{
Results from fitting low-mass eclipsing binary systems with the Dartmouth stellar evolution isochrones. PTFEB132.707+19.810.
\label{tab:dartmouthmodel3}
}
\tablehead{
  \multicolumn{1}{c}{Parameter} &
  \multicolumn{1}{c}{Observed Value} &
  \multicolumn{1}{c}{Model Value\tablenotemark{a}} &
  \multicolumn{1}{c}{$(O-C)/\sigma$\tablenotemark{b}} &
  \multicolumn{1}{c}{$100\% \times (O-C)/O$\tablenotemark{c}}\\
}
\startdata
\cutinhead{\bf PTFEB132.707+19.810, \citet{kraus:2017} parameters} \\
$M_{A}^{\star}$ (\msunnom) & $0.3953 \pm 0.0020$ & $0.3986 \pm 0.0019$ & $-1.7$ & $\-0.83\%$ \\
$M_{B}^{\star}$ (\msunnom) & $0.2098 \pm 0.0014$ & $0.2110 \pm 0.0015$ & $-0.86$ & $-0.57\%$ \\
$R_{A}$ (\rsunnom) & $0.363 \pm 0.008$ & $0.3764 \pm 0.0015$ & $-1.7$ & $-3.69\%$ \\
$R_{B}$ (\rsunnom) & $0.272 \pm 0.012$ & $0.2260 \pm 0.0012$ & $3.8$ & $16.91\%$ \\
$(M_{B}+M_{A})^{\dagger}$ (\msunnom) & $0.6050 \pm 0.0020$ & $0.6095 \pm 0.0019$ & $-2.3$ & $-0.74\%$ \\
$M_{B}/M_{A}^{\dagger}$ & $0.531 \pm 0.0050$ & $0.5292 \pm 0.0051$ & $0.36$ & $0.34\%$ \\
$(R_{A}+R_{B})^{\dagger}$ (\rsunnom) & $0.635 \pm 0.005$ & $0.6024 \pm 0.0016$ & $6.5$ & $5.13\%$ \\
$R_{B}/R_{A}^{\dagger}$ & $0.75 \pm 0.05$ & $0.6003 \pm 0.0046$ & $3.0$ & $19.96\%$ \\
$T_{\rm eff,A}$ (K) & $3260 \pm 67$ & $3447.6 \pm 6.0$ & $-2.8$ & $-5.75\%$ \\
$T_{\rm eff,B}$ (K) & $3120 \pm 78$ & $3245.1 \pm 4.4$ & $-1.6$ & $-4.01\%$ \\
Age (Gyr) & $0.7 \pm 0.1$ & $1.0$\tablenotemark{d} & $\cdots$ & $\cdots$ \\
$[$Fe/H$]^{\star,\dagger}$ (dex) & $0.14 \pm 0.04$ & $+0.137 \pm 0.014$ & $0.075$ & $2.14\%$ \\
$\chi^2$ (DOF)\tablenotemark{e} & $\cdots$ & $69.3$ (2) & $\cdots$ & $\cdots$ \\
\cutinhead{\bf PTFEB132.707+19.810, \citet{gillen:2017} parameters} \\
$M_{A}^{\star,\dagger}$ (\msunnom) & $0.3813 \pm 0.0074$ & $0.3906 \pm 0.0059$ & $-1.3$ & $\-2.4\%$ \\
$M_{B}^{\star,\dagger}$ (\msunnom) & $0.2022 \pm 0.0045$ & $0.2049 \pm 0.0036$ & $-0.60$ & $-1.3\%$ \\
$R_{A}$ (\rsunnom) & $0.3610 \pm 0.0033$ & $0.3616 \pm 0.0048$ & $-0.18$ & $-0.17\%$ \\
$R_{B}$ (\rsunnom) & $0.2256^{+0.0063}_{-0.0049}$ & $0.2211 \pm 0.0029$ & $0.92$ & $2.0\%$ \\
$(R_{A}+R_{B})^{\dagger}$ (\rsunnom) & $0.5868^{+0.0084}_{-0.0073}$ & $0.5827 \pm 0.0058$ & $0.56$ & $0.70\%$ \\
$R_{B}/R_{A}^{\dagger}$ & $0.624^{+0.017}_{-0.010}$ & $0.611 \pm 0.011$ & $1.3$ & $2.1\%$ \\
$T_{\rm eff,A}$ (K) & $3211^{+54}_{-36}$ & $3428 \pm 12$ & $-4.0$ & $-6.8\%$ \\
$T_{\rm eff,B}$ (K) & $3103^{+53}_{-39}$ & $3240.7 \pm 7.9$ & $-2.6$ & $-4.4\%$ \\
Age (Gyr) & $0.7 \pm 0.1$ & $1.0$\tablenotemark{f} & $\cdots$ & $\cdots$ \\
$[$Fe/H$]^{\star,\dagger}$ (dex) & $0.14 \pm 0.04$ & $+0.131 \pm 0.022$ & $0.23$ & $6.4\%$ \\
$\chi^2$ (DOF) & $\cdots$ & $0.56$ (2) & $\cdots$ & $\cdots$ \\
\enddata
\tablenotetext{}{Parameters marked by a $^{\star}$ are varied in the fit. Parameters marked by a $^{\dagger}$ are treated as observables that are matched to the model in computing the value of $\chi^2$ that is listed.}
\tablenotetext{a}{The optimized value and uncertainty for this parameter that comes from the analysis in Section~\ref{sec:dartmouthcomparison}. The uncertainties do not include any systematic errors in the stellar evolution models.}
\tablenotetext{b}{Difference between the observed parameter value and the model value, divided by the observational uncertainty.}
\tablenotetext{c}{Difference between the observed parameter value and the model value, expressed as a percentage of the observed parameter value.}
\tablenotetext{d}{The age of PTFEB132.707+19.810 was fixed to 1.0\,Gyr in this analysis, which is the youngest age at which the Dartmouth isochrones have been tabulated.}
\tablenotetext{e}{The $\chi^2$ for the best-fit model. The number of degrees of freedom in the analysis is listed in parentheses following the $\chi^2$ value.}
\ifthenelse{\boolean{emulateapj}}{
  \end{deluxetable*}
}{
  \end{deluxetable}
}

\begin{appendix}

\section{A. Correcting for Light Travel-Time Within the \sysname{} system}
\label{sec:appendixltt}

The light travel-time correction for a star to the center-of-mass of
the system is given by
\begin{equation}
\Delta t = z/c
\end{equation}
where $z$ is the line-of-sight component of the star's barycentric
orbit and $c$ is the speed of light. Following \citet{hilditch:2001}
this can be expressed in terms of the eccentric anomaly $E$ via
\begin{equation}
\Delta t = \frac{a}{c} \left( \sqrt{1-e^2} \sin E \cos\omega + (\cos E - e)\sin \omega \right) \sin i
\end{equation}
where $a$ is the semimajor axis of the {\em star's} orbit about the
center-of-mass of the system, and a factor of $-1$ appears in front of
the right-hand-side of the equation for the secondary star. The relation between the time of observation from the Solar-System barycenter $t$ and the appropriate $E$ to use for describing a star's position is given by Kepler's equation corrected for $\Delta t$:
\begin{equation}
E - e \sin E = \frac{2\pi}{P}((t - \Delta t) - T)
\end{equation}
where the reference time $T$ is the time of periastron passage in the
system barycenter frame, and $P$ is the orbital period. Given a value
of $t$ we solve for $E$ using a Newton-Raphson procedure. The radial
velocities for stars 1 and 2 at time $t$ are then determined from
$E_{1}$ and $E_{2}$ using standard formulae.

Given anomalies $E_{1,2}$ the sky-projected $x_{1,2}$ and $y_{1,2}$ positions for star $1$ and $2$ are then determined from:
\begin{eqnarray}
x_{i} & = & (-1)^{i-1}a_{i}(1 - e\cos E_{i})\cos(\theta_{i}+\omega) \\
y_{i} & = & (-1)^{i-1}a_{i}(1 - e\cos E_{i})\sin(\theta_{i}+\omega)\cos i
\end{eqnarray}
where $\theta$ and $E$ are related via
\begin{eqnarray}
\cos\theta_{i} & = & (\cos E_{i} - e)/(1 - e\cos E_{i}) \\
\sin\theta_{i} & = & \sqrt{1-e^2}\sin E_{i}/(1 - e\cos E_{i})
\end{eqnarray}
The sky-projected separation between the two stars at observed time $t$ is then given by
\begin{equation}
\rho = \sqrt{(x_{1}-x_{2})^2 + (y_{1}-y_{2})^2}.
\end{equation}
The observed flux for the system can be determined at time $t$ by
finding time $\tilde{t}$ such that the $\tilde{\rho}(\tilde{t}) =
\rho(t)$ and using $\tilde{t}$ as input to the JKTEBOP routine. Here
$\tilde{\rho}$ is the sky-projected separation calculated without
accounting for intra-system light travel-time. Note that the
time-correction is not properly handled for proximity effects such as
tidal distortion or the reflection effect, however these effects are
negligible for the well-detached \sysname{}
system.

For our analysis we determine the $\tilde{t}$ values one time for all
photometric observations of \sysname{} and then use these as the fixed
times of observation throughout our fitting procedure. While a proper
treatment would determine a new set of $\tilde{t}$ values for each set
of system parameters in the Markov Chain, in practice the
uncertainties on $t-\tilde{t}$ due to uncertainties in the system
parameters are much less than our timing precision for the system, so
this level of complexity, which substantially slows the analysis, is
not required.

\section{B. Details of Spectral Index Calculations}
\label{sec:spectralindices}

\begin{figure}[!ht]
\plotone{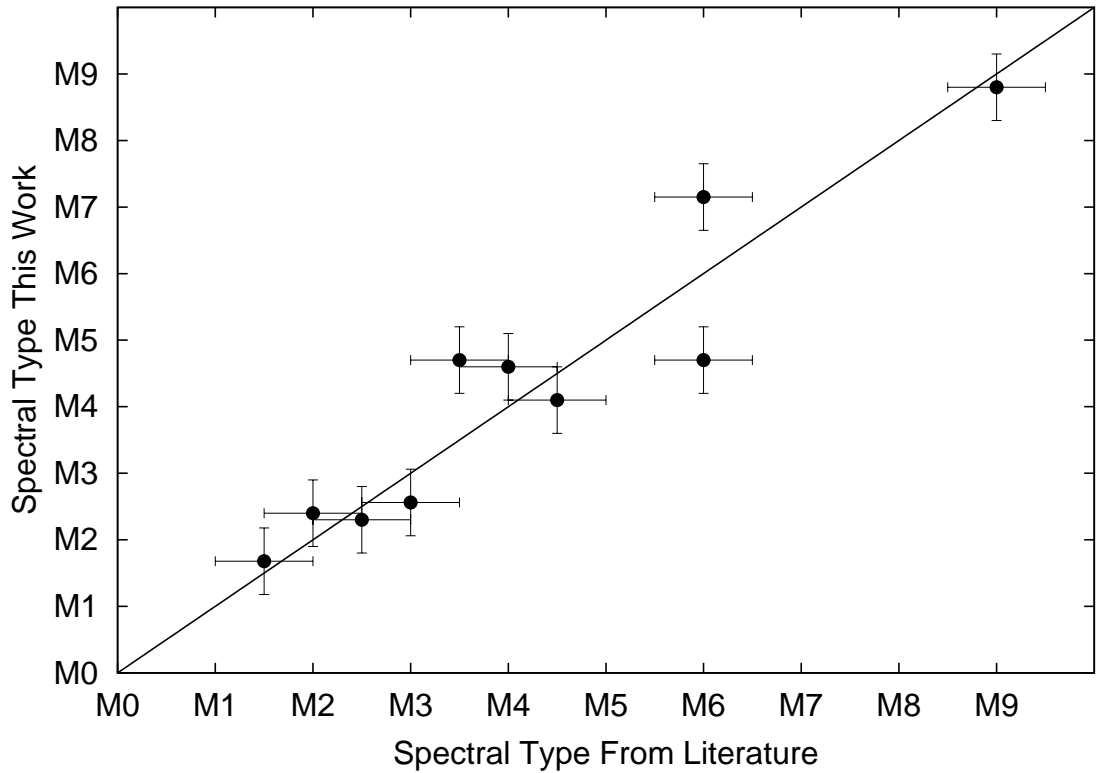}
\caption{ Comparison between spectral types for M dwarf standards
  determined from our FIRE/Magellan observations using the R12
  H$_2$O-K index, and spectral types taken from the literature.
\label{fig:SpecTypeCompare}}
\end{figure}

Spectral types for both components of \sysname{} were determined using
the H$_{2}$O-K index following \citet{rojasayala:2012}. When applied
to the disentangled spectra this yields spectral types of
\spectypestarA{} and \spectypestarB{} for the primary and secondary
stars, respectively. When the index is calculated on the three spectra
obtained during total eclipse, the resulting spectral type is
\spectypestarAtotal{}. The uncertainties here include a systematic
uncertainty of $\pm 0.5$ as given by \citet{rojasayala:2012}, and an
uncertainty in our measurement of the index as described below.  As a
validation of our spectral type estimates, we also obtained FIRE
observations for a number of M dwarf spectral
standards. Figure~\ref{fig:SpecTypeCompare} compares the spectral
types estimated from our FIRE observations of these stars to the
literature values, demonstrating agreement to $\sim \pm 0.5$ spectral
types over the range from M1 to M9. Using the relation between
spectral type and effective temperature given in \citet{bessell:1991},
we estimate effective temperatures of the component stars of T$_{\rm
  eff,A}=\teffstarbesselA$\,K, and T$_{\rm
  eff,B}=\teffstarbesselB$\,K.

Between T12 and R12 there are four separate NIR metallicity
indicators. These include the T12 H-band and K-band [Fe/H] indicators
([Fe/H]$_{H,T12}$ and [Fe/H]$_{K,T12}$, respectively), and the R12
K-band [Fe/H] and [M/H] indicators ([Fe/H]$_{R12}$ and
[M/H]$_{R12}$). The [Fe/H]$_{K,T12}$, [Fe/H]$_{R12}$ and [M/H]$_{R12}$
are not independent indicators in the sense that they make use of the
same spectral features; [Fe/H]$_{H,T12}$, on the other hand, is
independent of the other three indicators. We calculated each of these
metallicity indicators for each of our spectra.

To determine the uncertainties
on these indices we first estimated formal errors for each index by
propagating the uncertinaties in the spectra based on photon
counting statistics through the index calculations. For each index we
then determined a systematic error in precision using a likelihood
function of the form
\begin{equation}
\ln L = \sum_{j}\sum_{i}-\frac{1}{2}(\ln(\alpha^2 + \sigma_{ji}^2)+(X_{ji}-\bar{X_j})^2/(\alpha^2 + \sigma_{ji}^2))
\label{eqn:mcmcL}
\end{equation}
(i.e., we assume a standard Gaussian probability distribution) where
the sum on $j$ is over stars, the sum on $i$ is over individual
measurements for each star, $\alpha$ is the systematic uncertainty for
the index, $\sigma_{ji}$ is the formal uncertainty for the $i$th
observation of the $j$th star, $X_{ji}$ is the measured value of the
index, and $\bar{X_{j}}$ is the estimated value for star $j$. We carry
out an MCMC analysis varying $\alpha$ and $\bar{X_{j}}$ to determine
optimal values and uncertainties for these parameters. The value of
$\alpha$, determined in this manner, represents the excess scatter in
the measurements for an individual star beyond what is expected based
on the formal errors. In addition to this, there are possible errors
in comparing our indices to those given by T12 and R12 (and thus in
using our indices directly in their relations), and there are
additional systematic errors in the relations given by T12 and R12 in
going from spectral metallicity indices to physical metallicities.

\begin{figure}[!ht]
\plotone{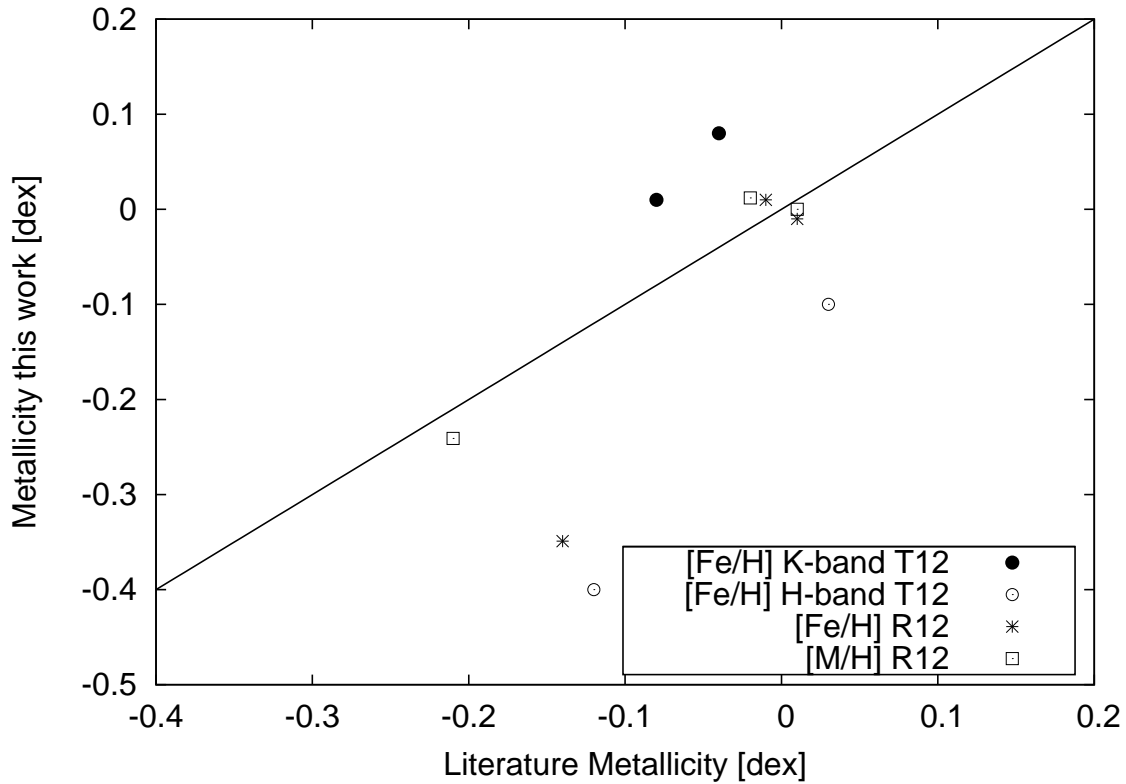}
\caption{ 
Comparison between the metallicities determined from our observations, and metallicities given in the literature for three objects with prior metallicity determinations. We compare each type of metallicity ([Fe/H]$_{H,T12}$, [Fe/H]$_{K,T12}$, [Fe/H]$_{R12}$ and [M/H]$_{R12}$) that is available.
\label{fig:FeHCompare}}
\end{figure}

There are a total of three objects with metallicities given in either of
these catalogs for which we have obtained observations (two of the
objects are in both T12 and R12, while one object is in R12 only). Due
to this small number of calibrators, we do not attempt to derive an
independent metallicity calibration from our
observations. Figure~\ref{fig:FeHCompare} compares the metallicity
indices for these stars from our observations to those given in T12
and R12. We find that an additional scatter in the metallicity indices
of $\pm 0.062$\,dex must be added in quadrature to the formal
uncertainties such that $\chi^2$ per degree of freedom is unity. On
top of this T12 and R12 give estimates of the uncertainties in the
physical metallicities inferred from these indices of $\pm
\fehterrienHsyserr$\,dex for [Fe/H]$_{H,T12}$, $\pm
\fehterrienKsyserr$\,dex for [Fe/H]$_{K,T12}$, $\pm
\fehrojassyserr$\,dex for [Fe/H]$_{R12}$, and $\pm
\mhrojassyserr$\,dex for [M/H]$_{R12}$. 

From our disentangled spectra of \sysname{} we measure
[Fe/H]$_{H,T12}=+\fehterrienHstarAfullerr$,
[Fe/H]$_{K,T12}=+\fehterrienKstarAfullerr$,
[Fe/H]$_{R12}=+\fehrojasstarAfullerr$, and
[M/H]$_{R12}=+\mhrojasstarAfullerr$ for the primary star, and
[Fe/H]$_{H,T12}=+\fehterrienHstarBfullerr$,
[Fe/H]$_{K,T12}=+\fehterrienKstarBfullerr$,
[Fe/H]$_{R12}=+\fehrojasstarBfullerr$, and
[M/H]$_{R12}=+\mhrojasstarBfullerr$ for the secondary star. The error
estimates given here include all the sources of uncertainty discussed
above. For reference, using the four spectra obtained during totality,
we measure [Fe/H]$_{H,T12}=+\fehterrienHstarAtotalfullerr$,
[Fe/H]$_{K,T12}=+\fehterrienKstarAtotalfullerr$,
[Fe/H]$_{R12}=+\fehrojasstarAtotalfullerr$, and
[M/H]$_{R12}=+\mhrojasstarAtotalfullerr$ for the primary. To obtain
final estimates for each star we take the weighted mean of the
[Fe/H]$_{H,T12}$ and [Fe/H]$_{K,T12}$ measurements\footnote{The
  [Fe/H]$_{K,T12}$, [Fe/H]$_{R12}$ and [M/H]$_{R12}$ indices are
  determined from the same spectral features and are thus not
  independent measurements. We adopt the [Fe/H]$_{K,T12}$ index to
  avoid mixing [M/H] and [Fe/H], and because [Fe/H]$_{K,T12}$ has a
  lower uncertainty than [Fe/H]$_{R12}$ for most of our stars.},
finding [Fe/H]$_{A}=\fehstarA$ and [Fe/H]$_{B}=\fehstarB$, which are
consistent to within 2$\sigma$. Assuming both components have the same
metallicity, we take the weighted mean of the individual metallicities
to estimate a system metallicity of [Fe/H]$=\fehsystem$.

%% ----------------
%% ----------------

%% ----------------
%% ----------------

\section{C. Details of Cross-Correlation Against Theoretical Spectral Templates}
\label{sec:ccf}

As an alternative method to determine the stellar atmospheric
parameters we compare our disentangled NIR spectra to model spectra from the
BT-Settl grid \citep{allard:2011} computed using the
\citet{asplund:2009} solar abundances. The models have temperatures
between 2600\,K and 4000\,K in steps of 100\,K, and have [Fe/H]
metallicities between $-4.0$\,dex and $+0.5$\,dex in $0.5$\,dex
increments. A metallicity of [Fe/H]$=+0.3$\,dex is also included. The
models assume solar-scaled abundances with $\alpha$-enhancement for
sub-solar metallicities such that [$\alpha$/Fe]$=+0.2$ for
[Fe/H]$=-0.5$, and [$\alpha$/Fe]$=+0.4$ for [Fe/H]$\leq -1.0$. We only
considered templates with $\log g = 5.0$.

The model spectra are broadened to the resolution of our observations
and then cross-correlated using the XCSAO routine, which is part of
the RVSAO package \citep{kurtz:1998}. We ignore rotational and
turbulent broadening which are both much lower than the instrumental
resolution (the expected projected rotation speeds are
$6.775\pm0.056$\,\kms\ and $4.445\pm0.039$\,\kms\ for the primary and
secondary stars, respectively). The correlation is performed
separately for the $Z+Y$, $J$, $H$, and $K$-bands.

We first determine effective temperatures for each of the stars as
follows. The normalized cross-correlation peak-height $C$ is recorded
for each model in the grid, and we fit a paraboloid to values near the
peak that is a function of [Fe/H] and T$_{\rm eff}$. The T$_{\rm eff}$
value at the peak location is recorded for each band. We then
determine best-estimates of the T$_{\rm eff}$, and uncertainties, for
each star, still analyzing each bandpass separately, using a similar
technique to what was done for the spectroscopic indices. We conducted
an MCMC analysis to explore a likelihood function as in
equation~\ref{eqn:mcmcL}, with $X$ now being the effective
temperature, and replacing $\sigma_{ji}$ with $\alpha_{2}/SN_{ji}$.
Here $\alpha_{2}$ is an additional free parameter and $SN_{ji}$ is the
median S/N ratio for spectrum $i$ of star $j$. This results in T$_{\rm
  eff}$ measurements and $1\sigma$ uncertainties for each star, in
each of the four bandpasses. We then combine the four separate
bandpasses by using another MCMC and a similar likelihood function. In
this case the index $i$ enumerates the different bands, and we use
$\sigma_{ji}$, as the $1\sigma$ uncertainty for band $i$ of star $j$,
rather than $\alpha_{2}/SN_{ji}$. We do this rather than simply taking
the weighted average of the four bandpasses as we found that the
scatter between bands exceeded the measurement uncertainties, and this
is a straightforward method to account for the additional systematic
error. The resulting effective temperatures have uncertainties of
$\sim 100$\,K.

\begin{figure}[!ht]
\plotone{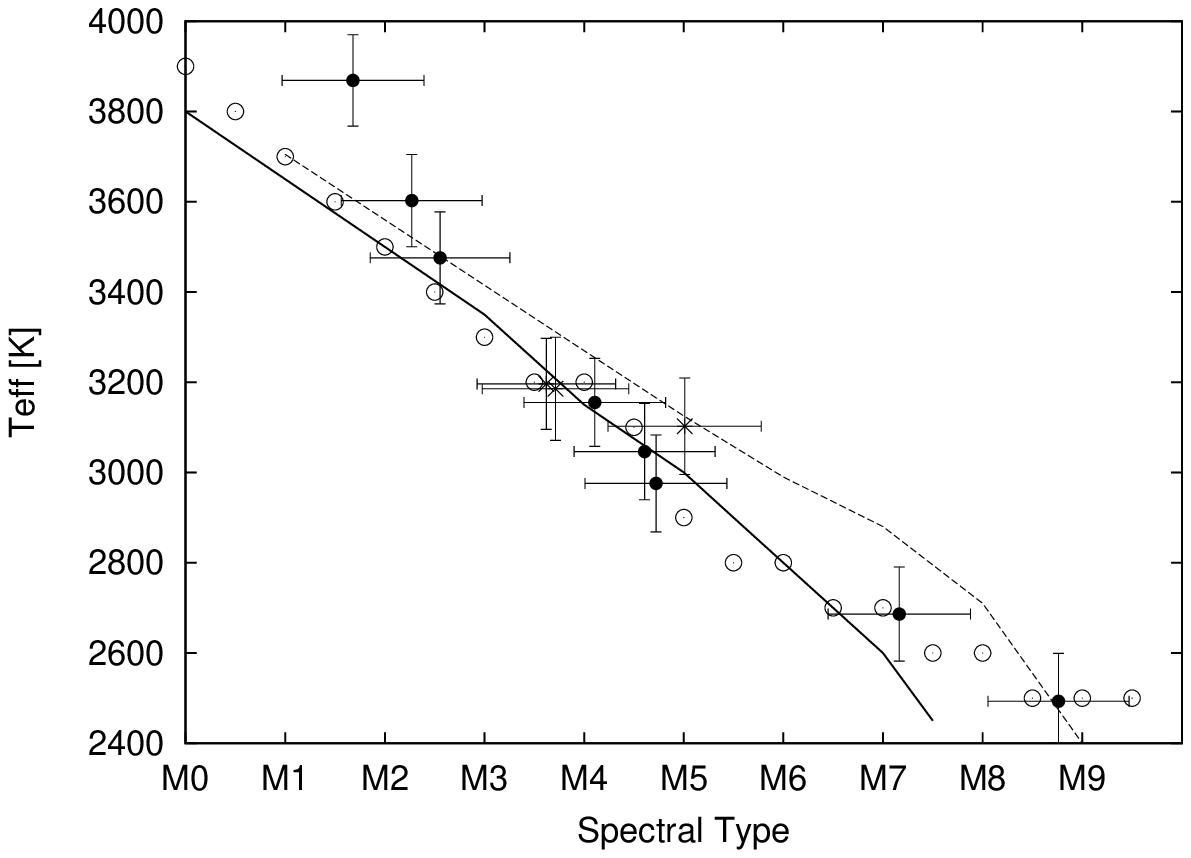}
\caption{ Comparison between the effective temperature determined by
  cross-correlation against BT-Settl synthetic templates and the
  spectral type estimated from the H$_2$O$-$K index and the relation
  from \citet{rojasayala:2012}. The filled circles show standard
  stars, Xs show the disentangled primary and secondary star spectra
  of \sysname{} together with the total eclipse spectrum of
  \sysname{}, the solid line shows the relation from
  \citet{bessell:1991}, the dashed line shows the relation from
  \citet{luhman:2003}, and the open circles show the relation from
  \citet{rajpurohit:2013}. Note that the \citet{luhman:2003} relation
  was calibrated for young stars and is likely not applicable for
  \sysname.
\label{fig:spectypeteff}}
\end{figure}

Figure~\ref{fig:spectypeteff} compares the effective temperatures
estimated in this manner to the spectral types estimated using the
H$_{2}$O$-$K indices. In this plot we also show the relations from
\citet{bessell:1991}, \citet{luhman:2003}, and
\citet{rajpurohit:2013}. We find that our effective temperatures and
spectral types are in good agreement with the \citet{rajpurohit:2013}
relation, in which effective temperatures are determined by
cross-correlating optical spectra against BT-Settl synthetic
templates.

\begin{figure}[!ht]
\plotone{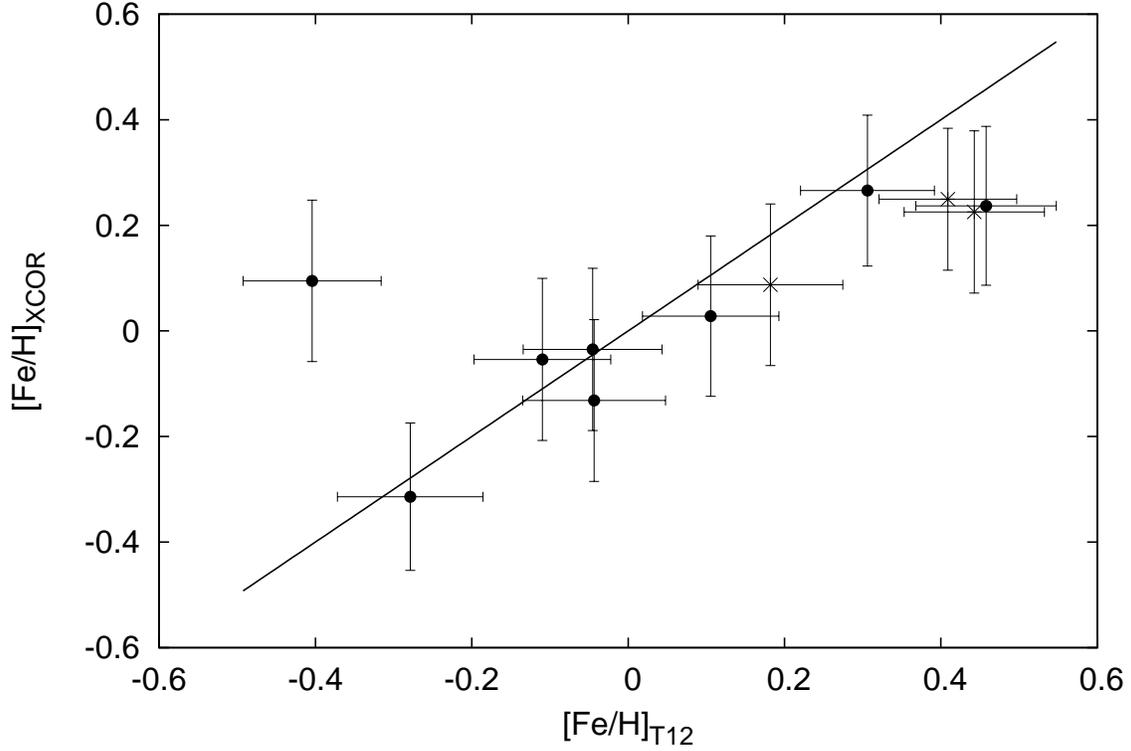}
\caption{ Comparison between the [Fe/H] metallicity determined by
  cross-correlation against BT-Settl synthetic templates (labeled
  [Fe/H]$_{\rm XCOR}$), and the metallicity estimates from the
  combined \citet{terrien:2012} $H$-band and $K$-band indicators
  (labeled [Fe/H]$_{\rm T12}$). We exclude the $Z+Y$-band from the
  correlation which yields systematically lower metallicities than the
  $J$-, $H$-, and $K$-bands. The filled circles show standard stars,
  Xs show the disentangled primary and secondary star spectra of
  \sysname{} together with the total eclipse spectrum of
  \sysname{}. The solid line shows the relation [Fe/H]$_{\rm
    XCOR}=$[Fe/H]$_{\rm T12}$. Excluding NLTT~15867, which is the
  outlier with [Fe/H]$_{\rm T12}<-0.4$, [Fe/H]$_{\rm XCOR}$ is
  tightly correlated with [Fe/H]$_{\rm T12}$, but systematically lower
  by $0.09\pm0.03$\,dex.
\label{fig:fehxcorcomp}}
\end{figure}

Having determined the effective temperature for each star, we next
determine the [Fe/H] metallicity. We do this in a similar manner to
the effective temperatures, except we fix the temperature to the
best-estimated value for each star when finding a [Fe/H] value that
maximizes the correlation for a given spectrum and band. We exclude
the $Z+Y$-band which we found to yield [Fe/H] values that are
systematically lower than the other three bands by $\sim
0.4$\,dex. This band also generally has a lower cross-correlation
peak-height than the other bands indicating systematic differences
between the models and observations in this wavelength range. The
resulting [Fe/H] values have uncertainties of $\sim
0.15$\,dex\footnote{If we do not fix the temperature in finding the
  [Fe/H] values, the results have much larger scatter ($\sim 0.5$ to
  $1$\,dex) and do not correlate with the metallicities based on the
  spectroscopic indices.}. Figure~\ref{fig:fehxcorcomp} compares the
[Fe/H] values so-determined to the values based on the
\citet{terrien:2012} $H$- and $K$-band indices. The two methods yield
metallicities that, aside from one significant outlier in NLTT~15867,
are fairly consistent. If we remove NLTT~15867, then the
cross-correlation-based metallicities are $0.09\pm0.03$\,dex lower
than the T12 metallicities. A difference on this order is not
surprising given the different assumed solar abundance patterns on
which each metallicity system is based. Comparing to metallicities
based on the \citet{rojasayala:2012} indices yields similar results.

For \sysname{}A and \sysname{}B we find effective temperatures of
T$_{\rm eff,A}=\teffstarbtsettlA$\,K and T$_{\rm
  eff,B}=\teffstarbtsettlB$\,K, respectively, from cross-correlation,
and metallicities of [Fe/H]$_{\rm A}=\fehstarbtsettlA$ and
[Fe/H]$_{\rm B}=\fehstarbtsettlB$, respectively. Combining the
metallicities of the primary and secondary components yields a
metallicity for the system of [Fe/H]$=\fehsystembtsettl$, which is
consistent with the system metallicity determined from the T12
indicators.

%% ----------------
%% ----------------

%% ----------------
%% ----------------

%% ----------------
\begin{figure*}[!ht]
\plottwo{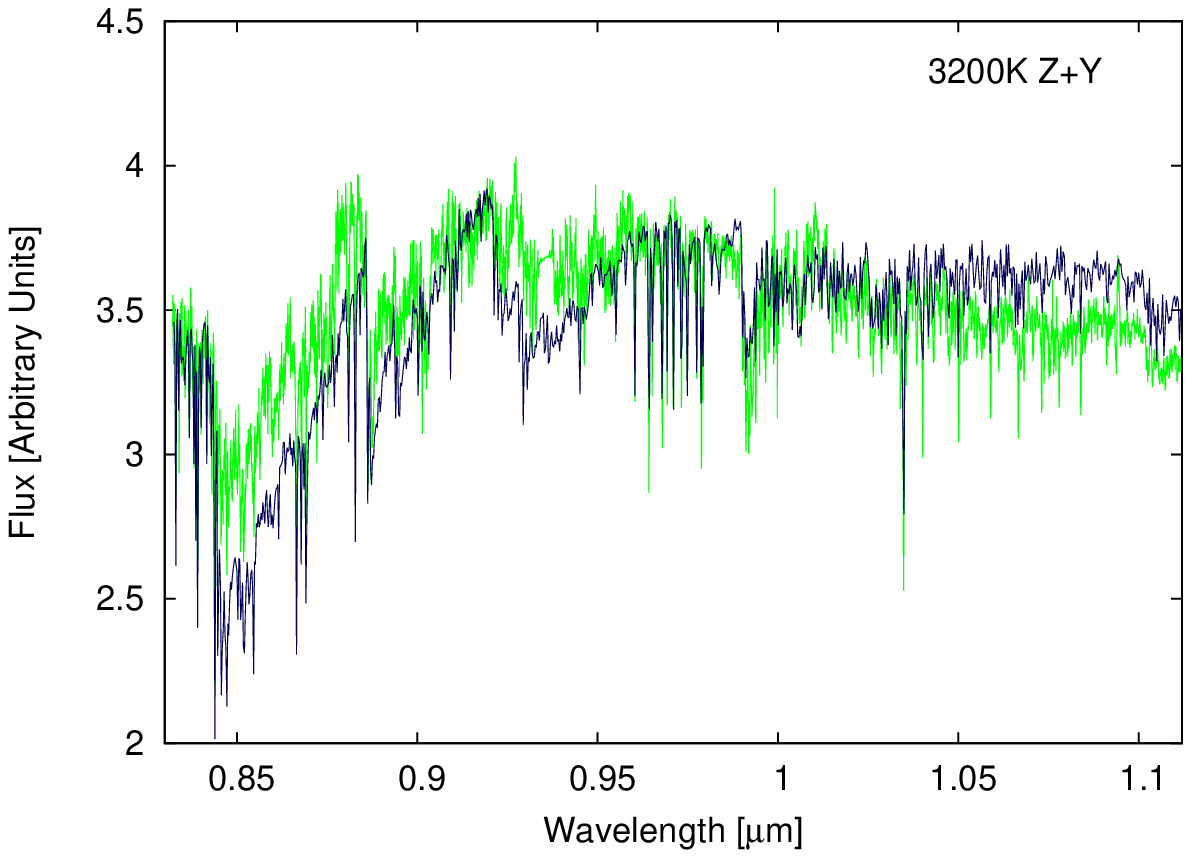}{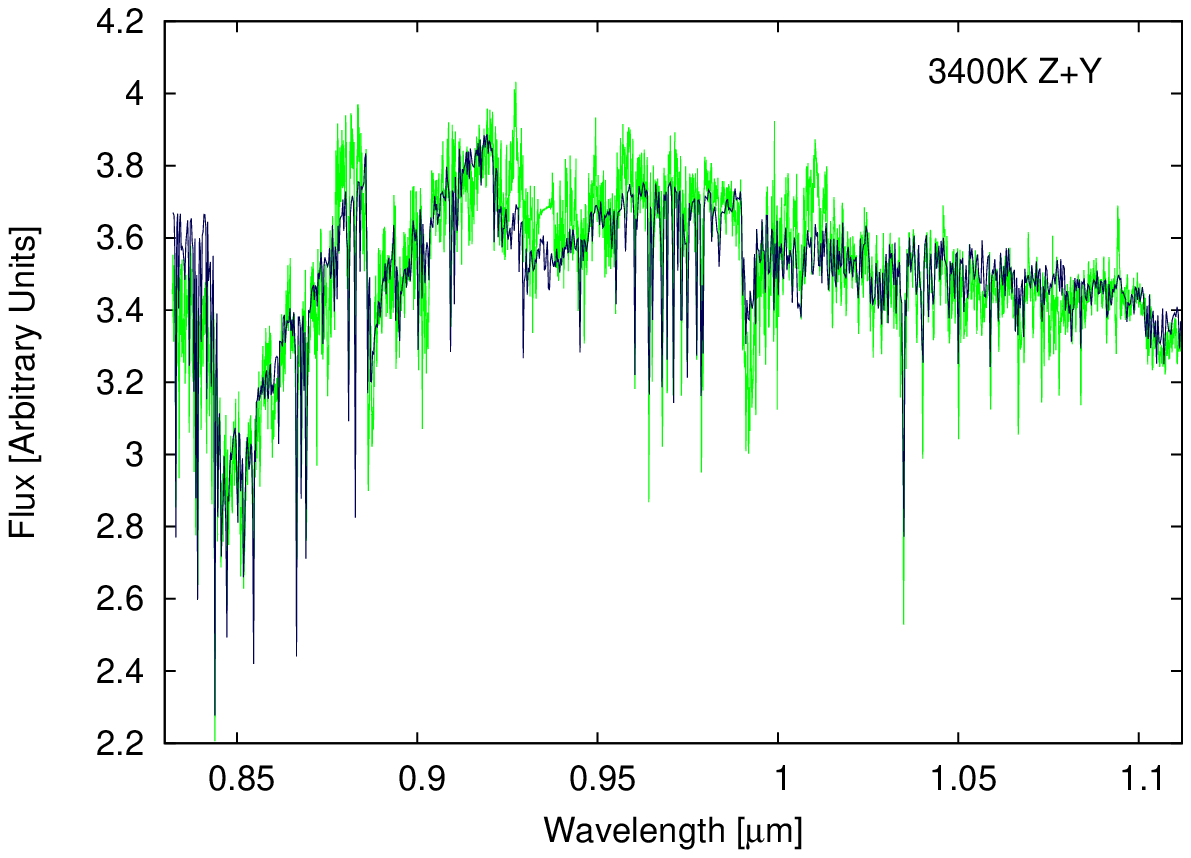}
\plottwo{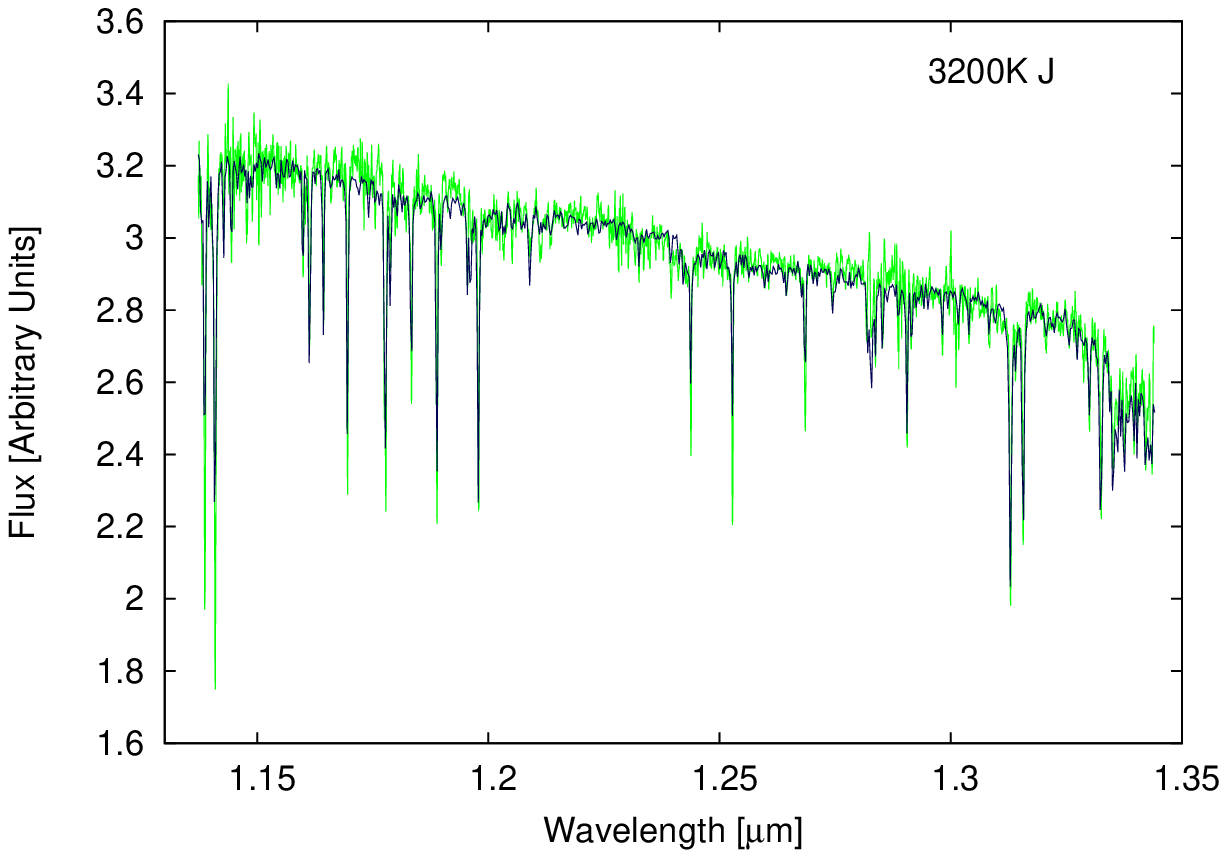}{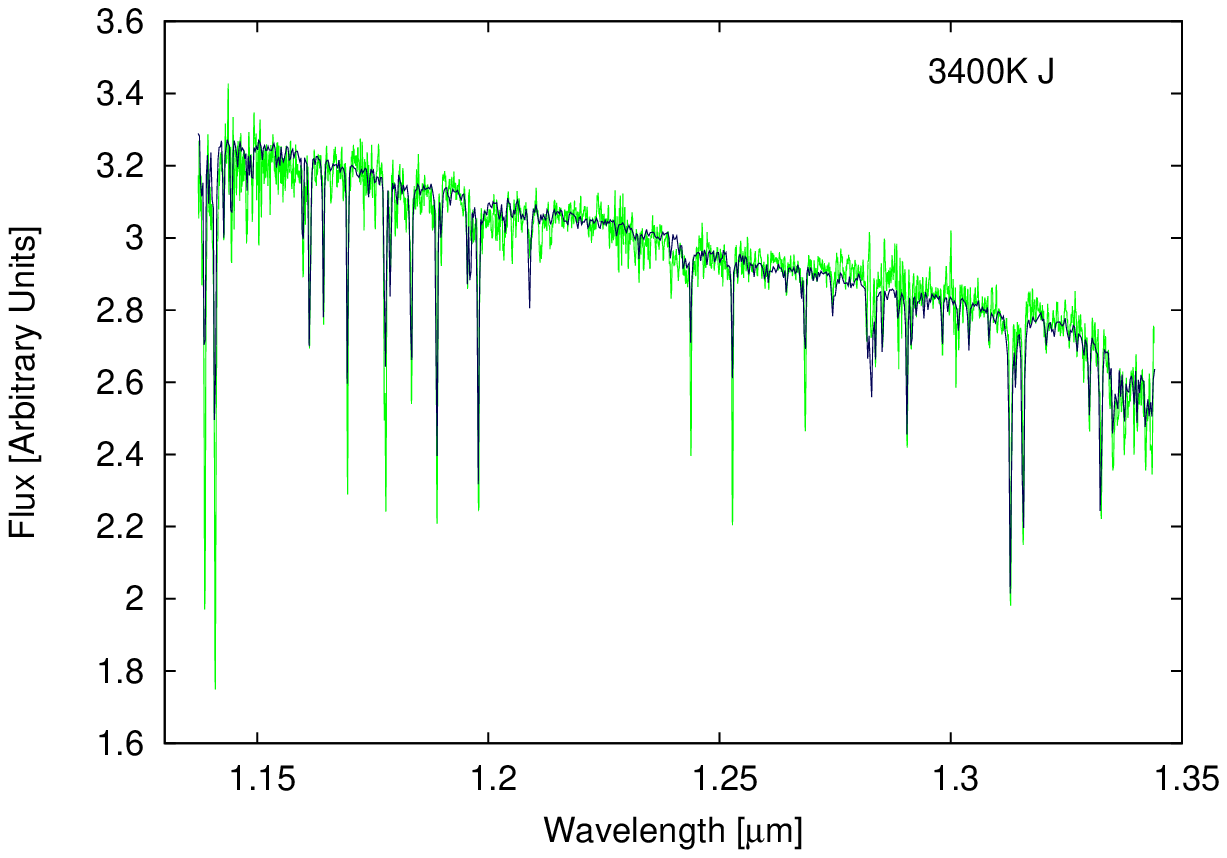}
\plottwo{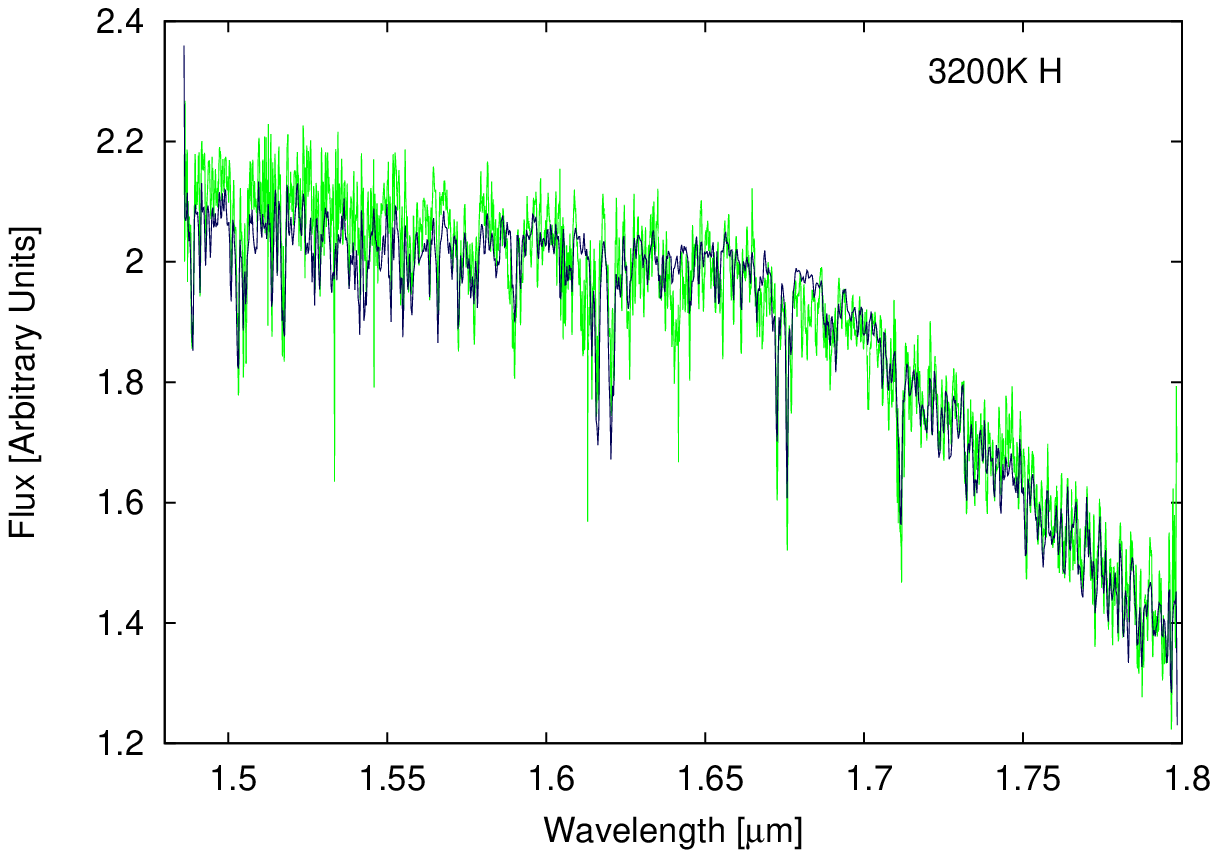}{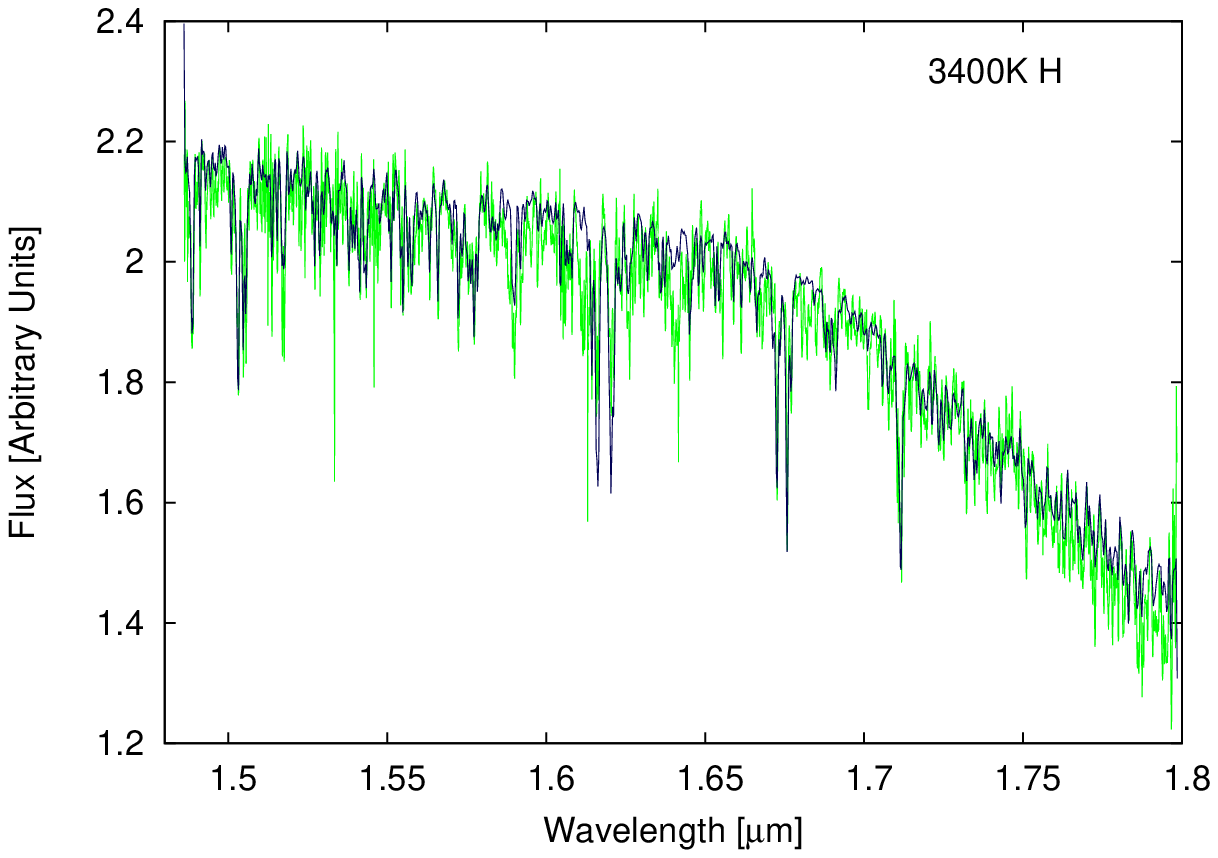}
\plottwo{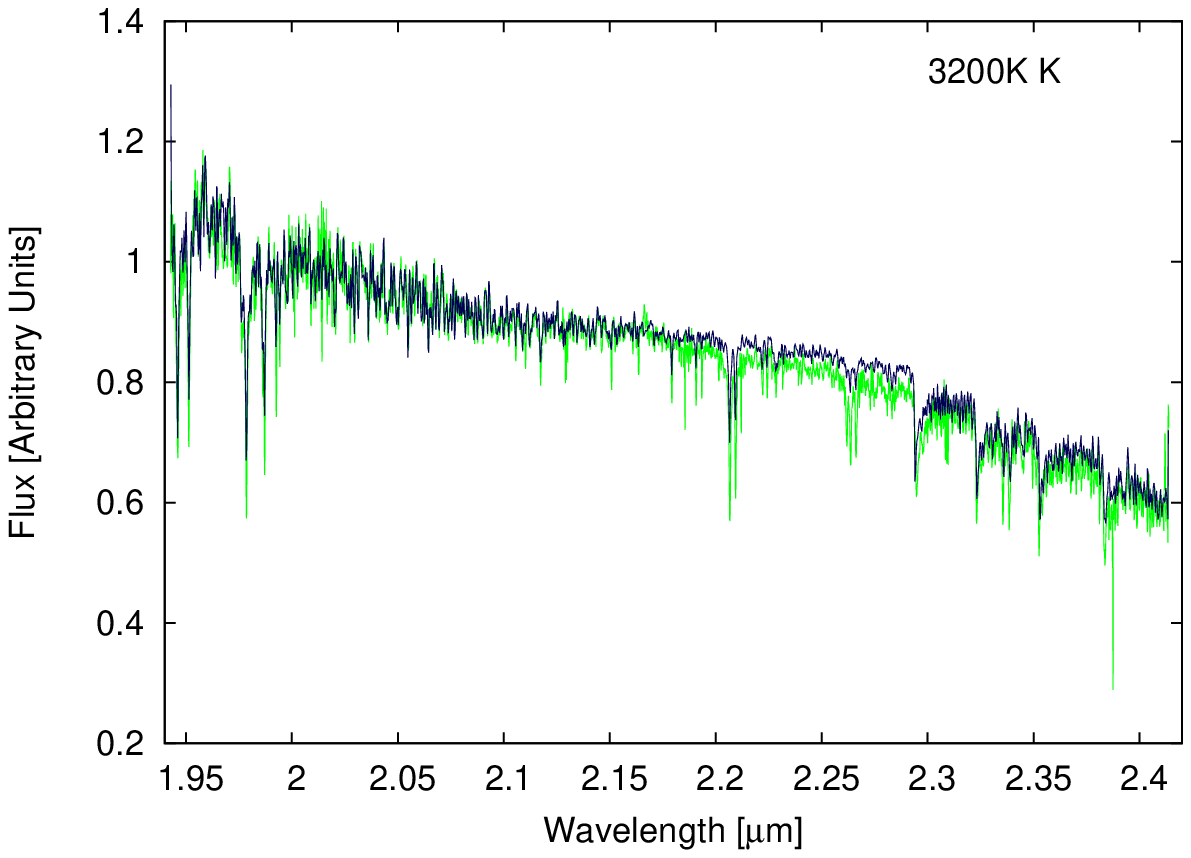}{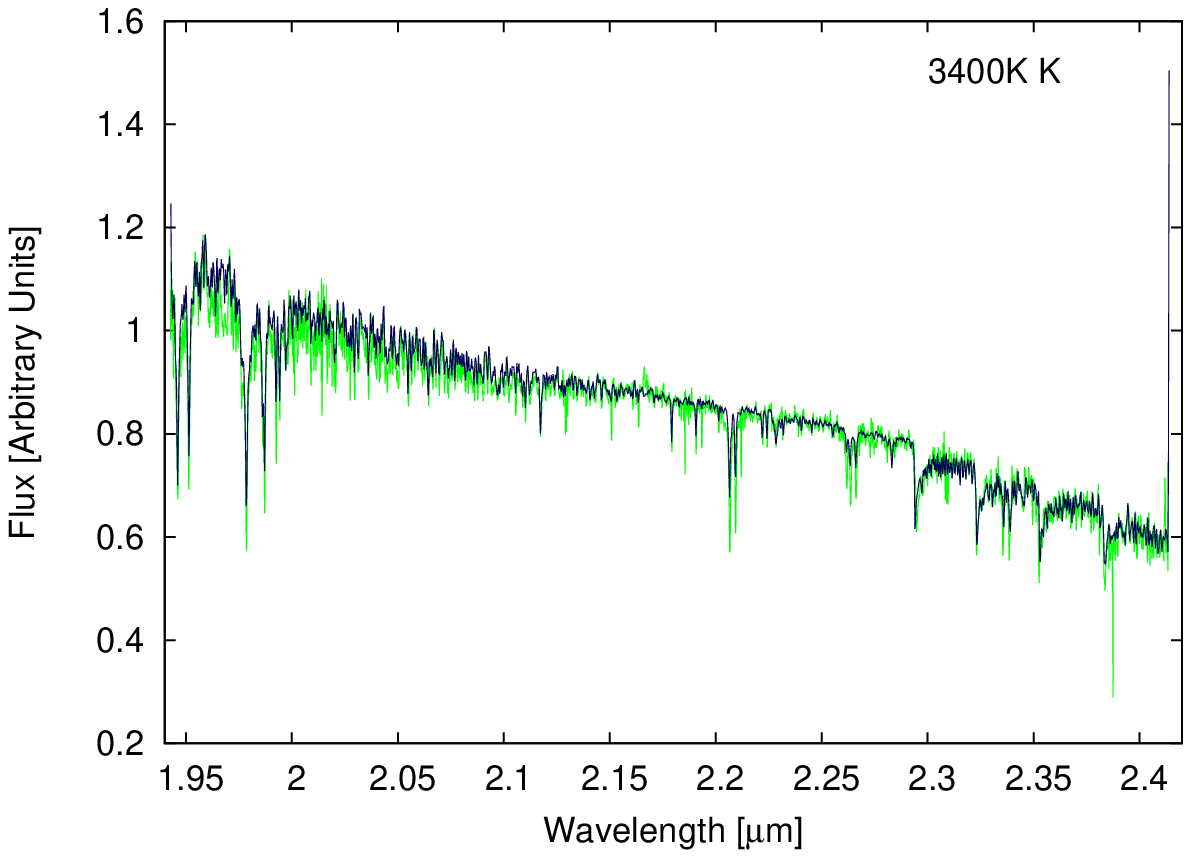}
\caption{
 NIR spectra for primary star of \sysname{} from Magellan/FIRE, showing the $Z+Y$, $J$, $H$ and $K$ bands separately (green lines). On the left-hand-side we overlay the BT-Settl template with the highest cross-correlation (blue lines). This template has T$_{\rm eff} = 3200$\,K and [Fe/H]$=0.3$. On the right-hand-side we overlay a T$_{\rm eff}=3400$\,K, [Fe/H]$=0.3$ template (blue lines) which provides a better match to the bluest ($Z+Y$) band as well as some of the features in the $K$-band. A separate vertical scaling is applied to the templates in each band to provide the best match to the observed spectra.
\label{fig:NIRprimaryspecbands}}
\end{figure*}
%% ----------------

%% ----------------
\begin{figure*}[!ht]
\plottwo{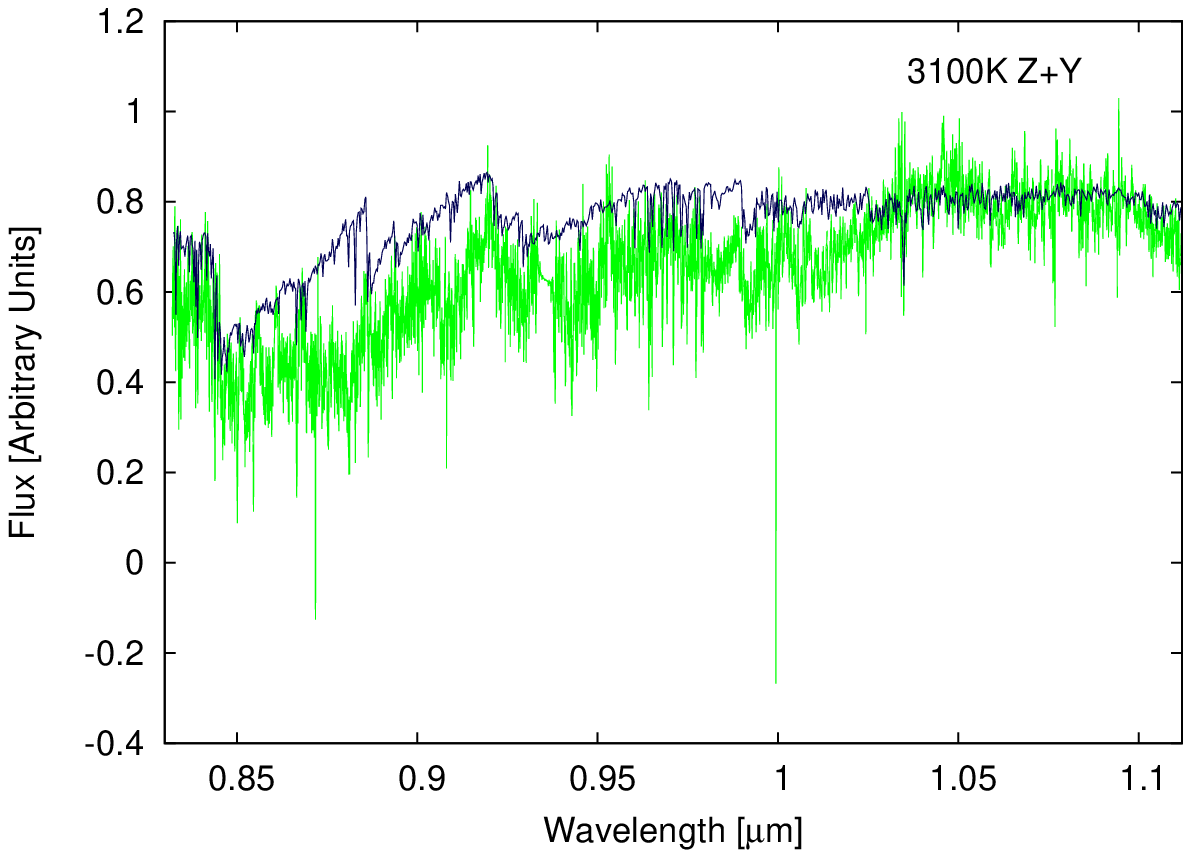}{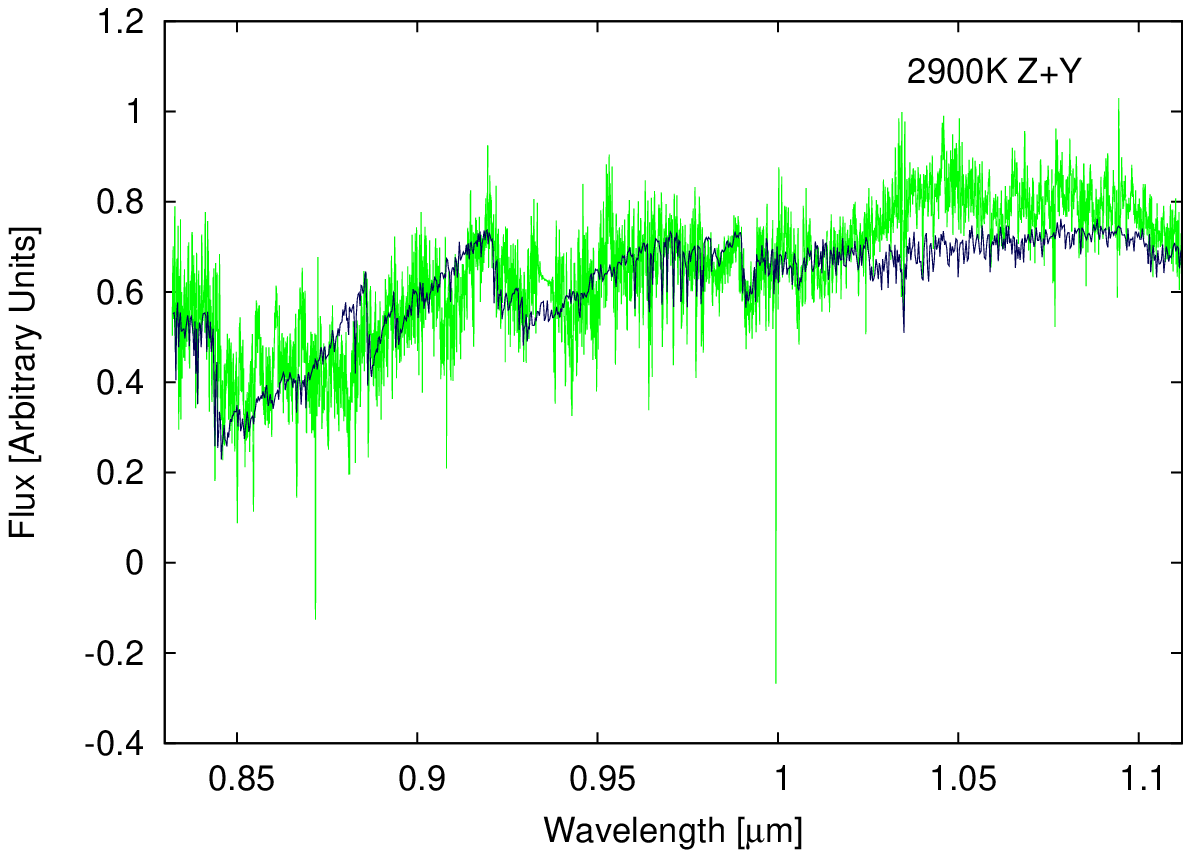}
\plottwo{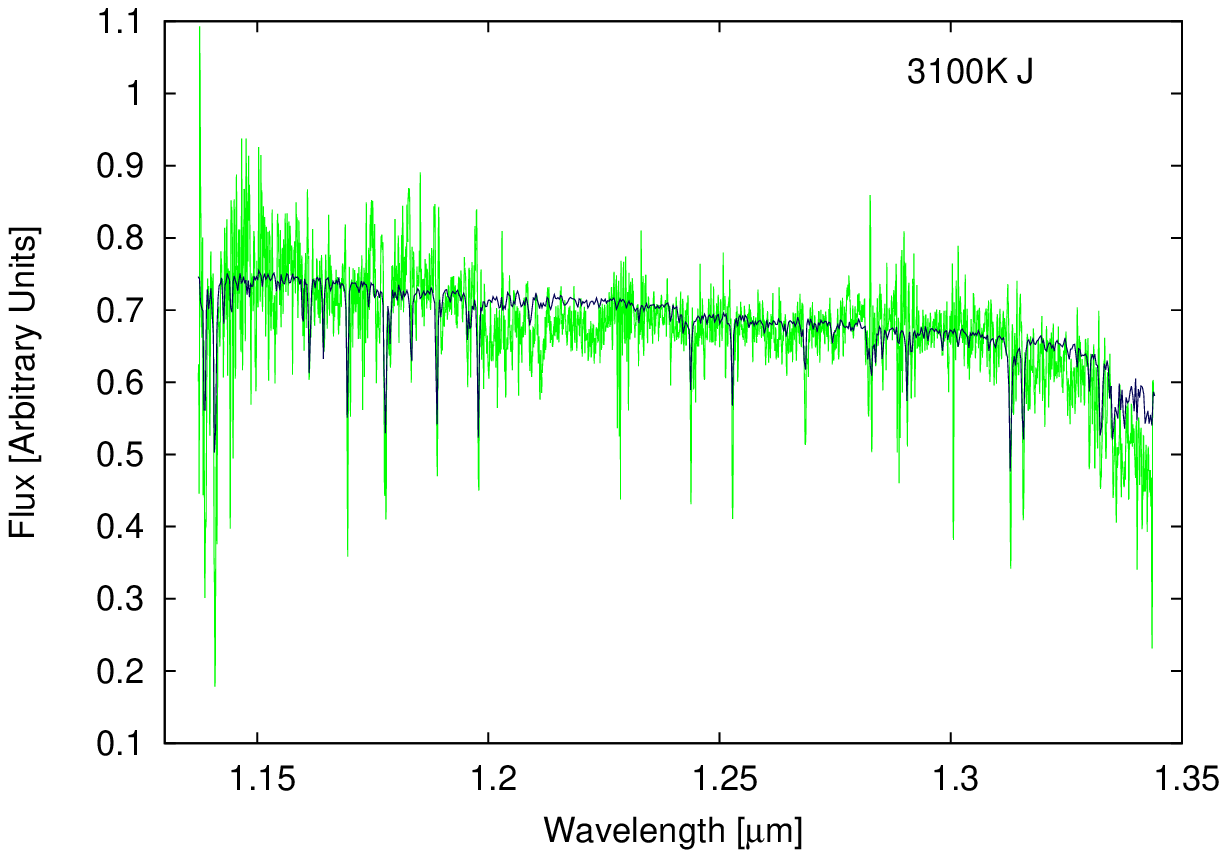}{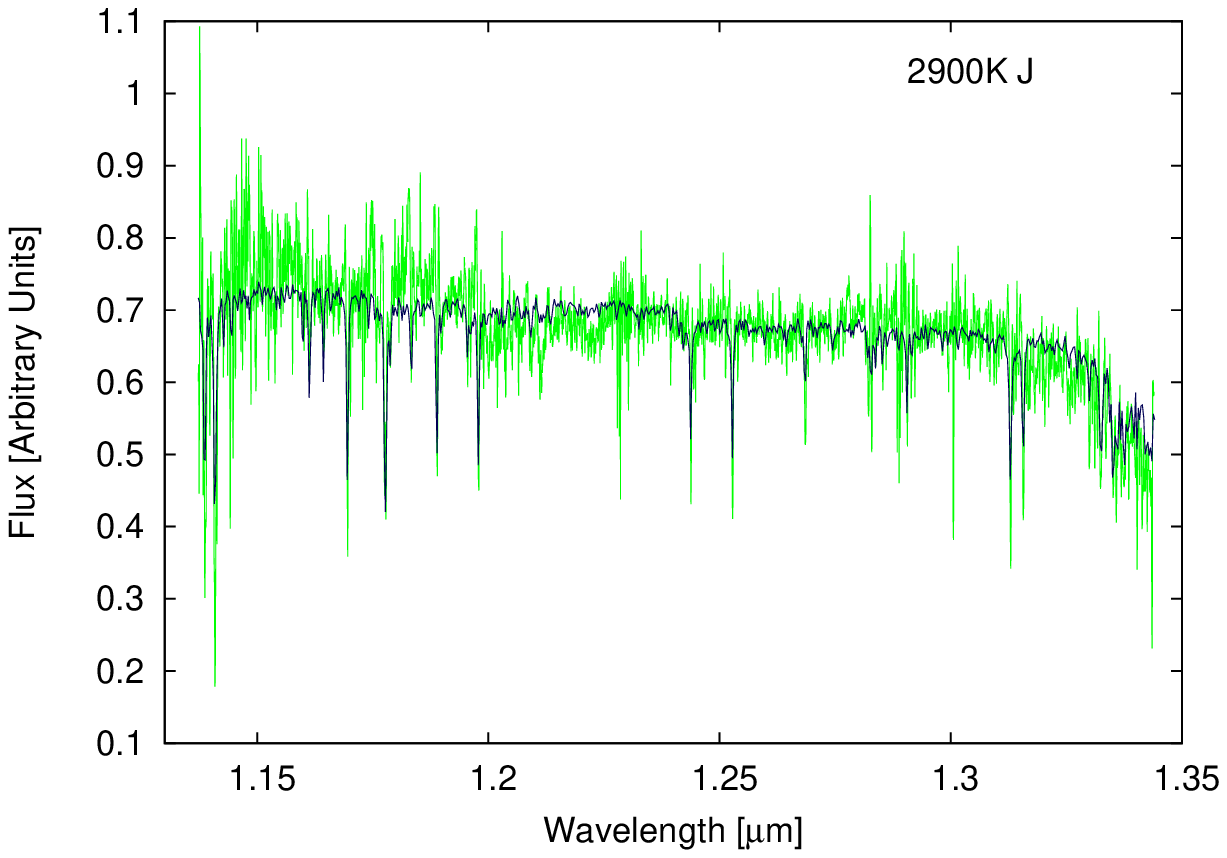}
\plottwo{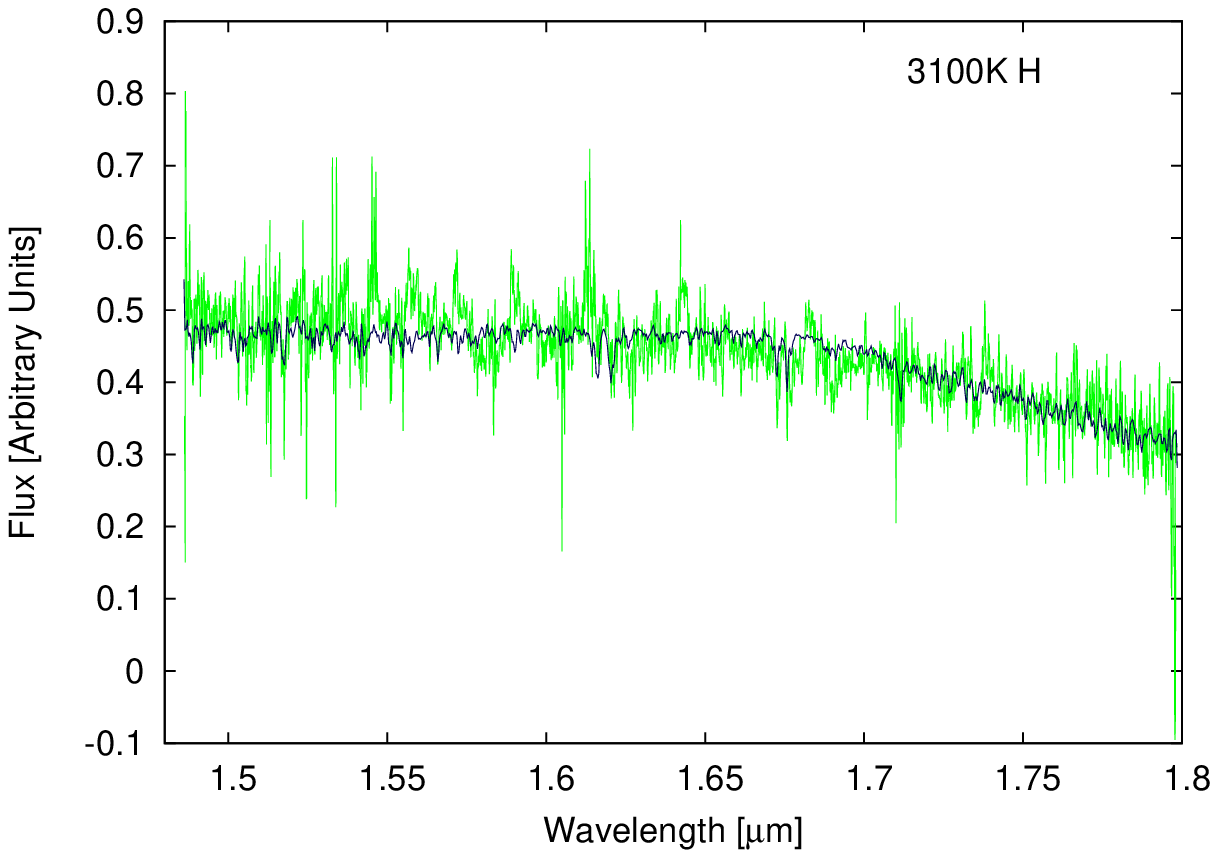}{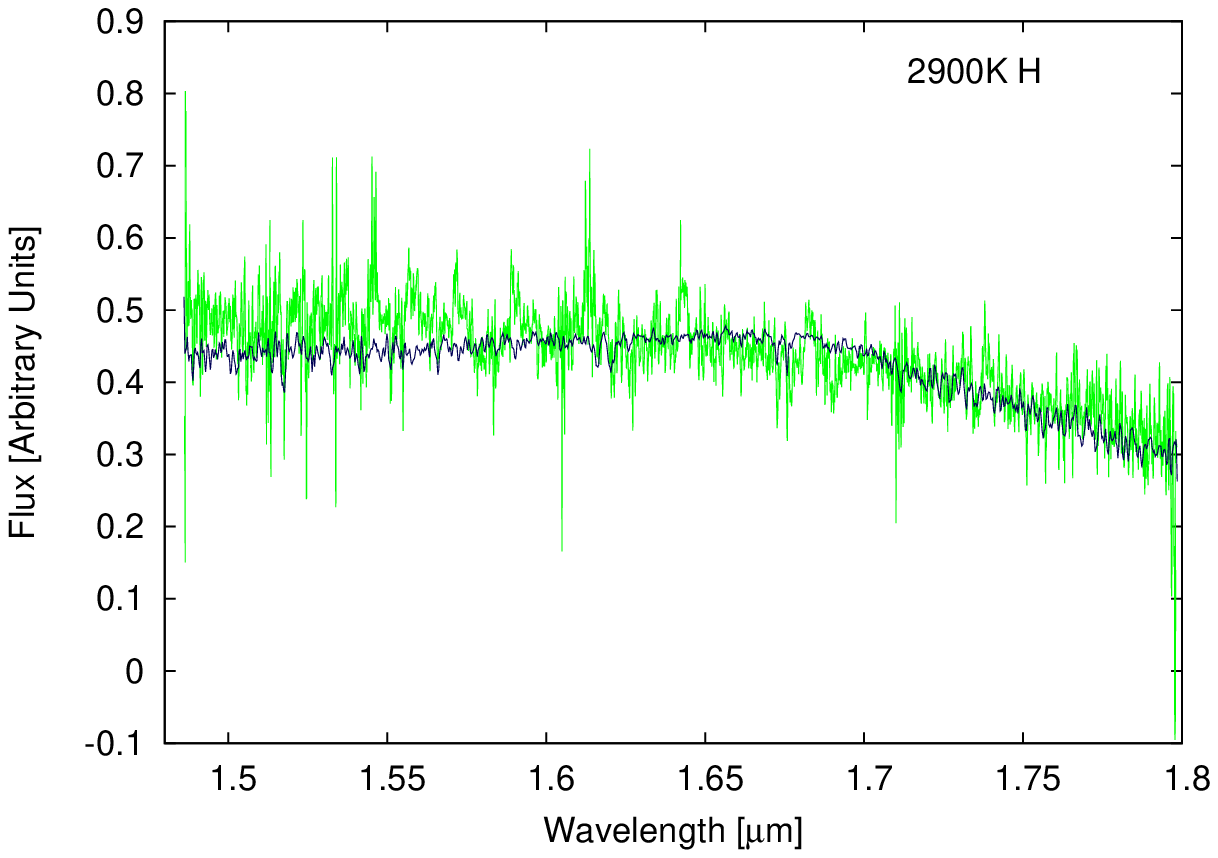}
\plottwo{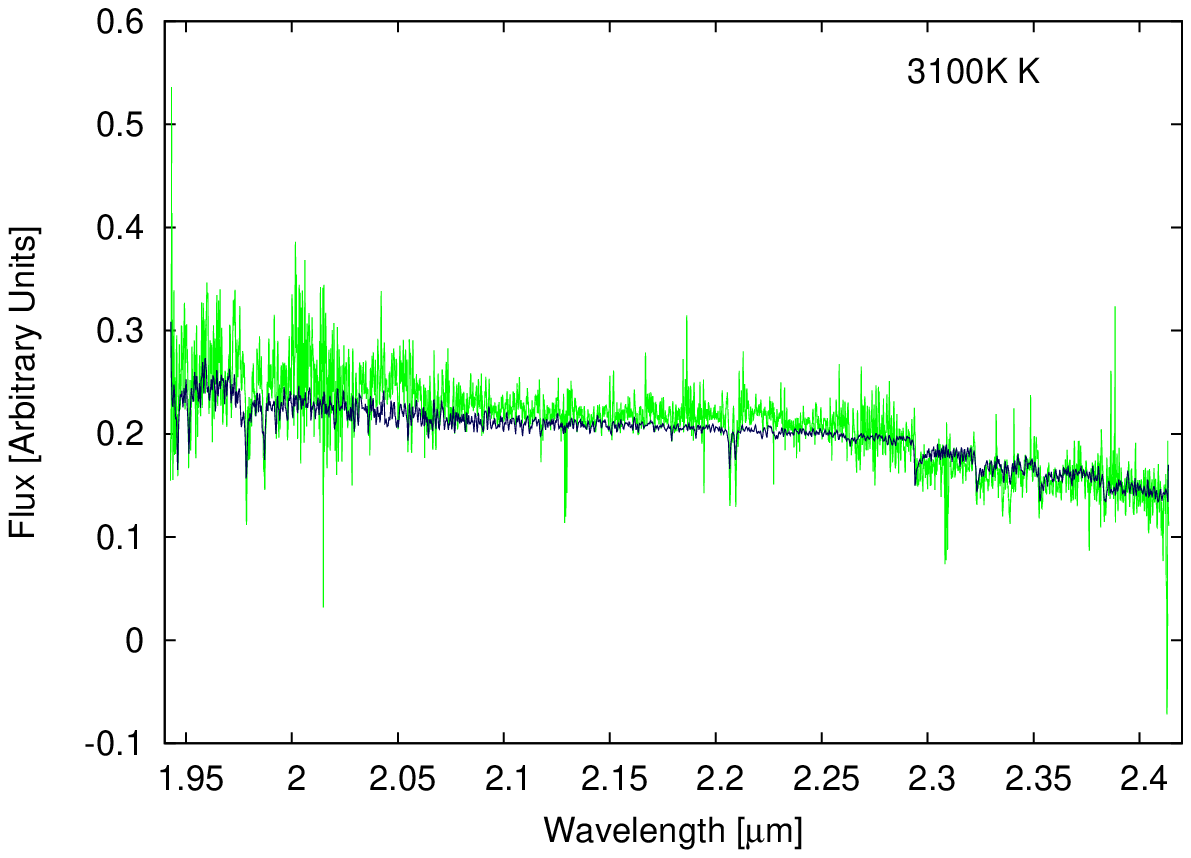}{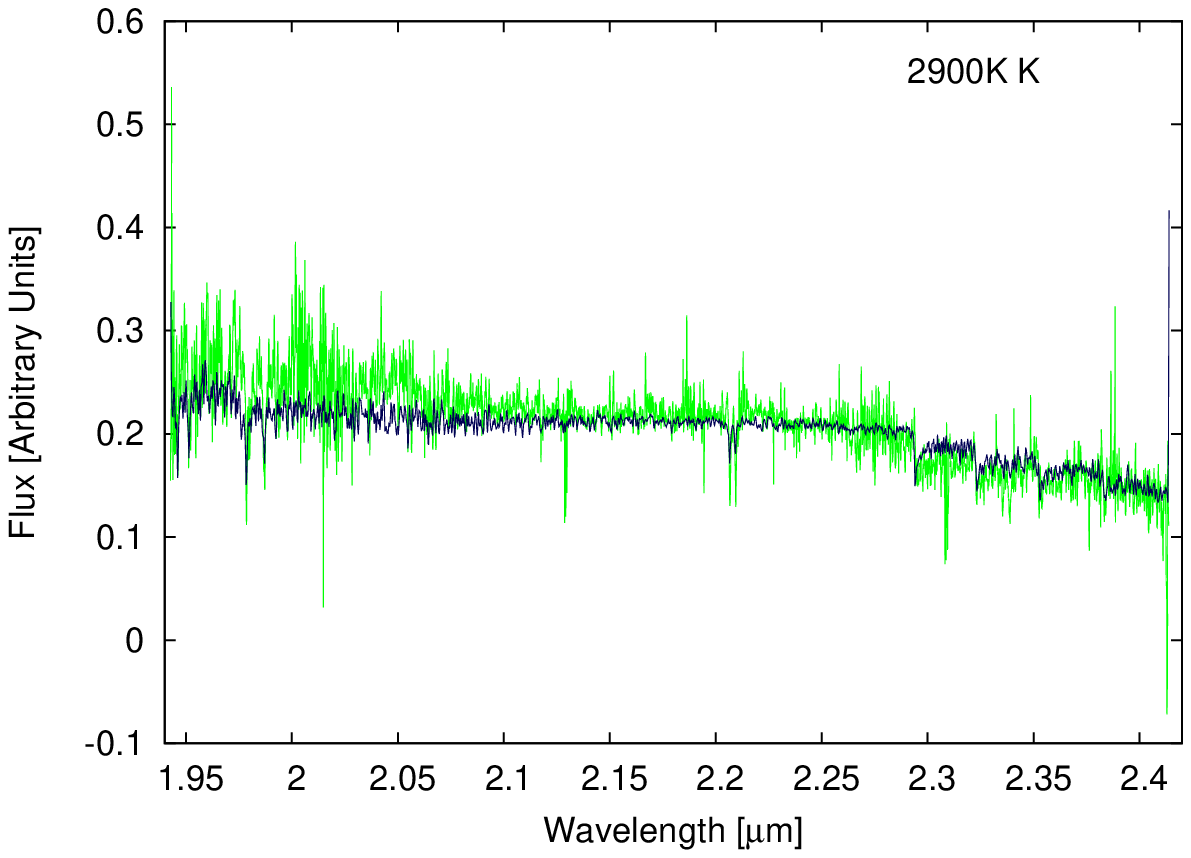}
\caption{
Same as Figure~\ref{fig:NIRprimaryspecbands}, here we show the secondary component. The overlayed template on the left-hand-side has T$_{\rm eff}=3100$\,K, [Fe/H]$=0.3$, while on the right-hand-side it has T$_{\rm eff}=2900$\,K, [Fe/H]$=0.3$.
\label{fig:NIRsecondaryspecbands}}
\end{figure*}
%% ----------------

\end{appendix}

\end{document}